\documentclass[11pt]{article}
\pdfoutput=1 
\usepackage{amssymb}
\usepackage{amsmath}
\usepackage{amstext}
\usepackage{graphicx,epsfig}
\usepackage{epsfig}
\usepackage{verbatim} 
\usepackage{fancybox}
\usepackage{color}
\usepackage{ulem}
\usepackage{enumitem}
\usepackage{subfigure}
\usepackage{bbm}
\usepackage{parskip}
\usepackage[numbers,sort&compress]{natbib}
\usepackage{ytableau}

\usepackage{tikz}
\usetikzlibrary{decorations}
\pgfdeclaredecoration{complete sines}{initial}
{
    \state{initial}[
        width=+0pt,
        next state=upsine,
        persistent precomputation={\pgfmathsetmacro\matchinglength{
            \pgfdecoratedinputsegmentlength / int(\pgfdecoratedinputsegmentlength/\pgfdecorationsegmentlength)}
            \setlength{\pgfdecorationsegmentlength}{\matchinglength pt}
        }] {}
    \state{upsine}[width=\pgfdecorationsegmentlength,next state=downsine]{
        \pgfpathsine{\pgfpoint{0.25\pgfdecorationsegmentlength}{0.5\pgfdecorationsegmentamplitude}}
        \pgfpathcosine{\pgfpoint{0.25\pgfdecorationsegmentlength}{-0.5\pgfdecorationsegmentamplitude}}
    }
    \state{downsine}[width=\pgfdecorationsegmentlength,next state=upsine]{
        \pgfpathsine{\pgfpoint{0.25\pgfdecorationsegmentlength}{-0.5\pgfdecorationsegmentamplitude}}
        \pgfpathcosine{\pgfpoint{0.25\pgfdecorationsegmentlength}{0.5\pgfdecorationsegmentamplitude}}
}
    \state{final}{}
}

\newcommand{\Comment}[1]{{}}
\definecolor{darkblue}{rgb}{0.15,0.35,0.55}
\definecolor{reddish}{rgb}{0.65, 0.2, 0.2}
\usepackage[linktocpage=true]{hyperref}
\hypersetup{
colorlinks=true,
citecolor=darkblue,
linkcolor=reddish,
urlcolor=darkblue,
pdfauthor={},
pdftitle={},
pdfsubject={}
}

\newcommand{\be}{\begin{equation}}
\newcommand{\ee}{\end{equation}}
\newcommand{\bea}{\begin{eqnarray}}
\newcommand{\eea}{\end{eqnarray}}
\newcommand{\beas}{\begin{eqnarray*}}
\newcommand{\eeas}{\end{eqnarray*}}
\newcommand{\nn}{\nonumber}

\def\({\left(}
\def\){\right)}

\newcommand{\qb}{\mathbf{q}}

\newcommand{\kb}{\mathbf{k}}

\newcommand{\rd}{{\rm d}}

\newcommand{\IA}{{\cal I}}
\newcommand{\JA}{{\cal J}}

\newcommand{\IB}{{\tilde{\cal I}}}
\newcommand{\JB}{{\tilde{\cal J}}}

\newcommand{\ti}{{\tilde i}}
\newcommand{\tj}{{\tilde j}}

\def\gsim{ \lower .75ex \hbox{$\sim$} \llap{\raise .27ex \hbox{$>$}} }
\def\lsim{ \lower .75ex \hbox{$\sim$} \llap{\raise .27ex \hbox{$<$}} }

\def\xyma{\xymatrix@M.7em}
\def\xymas{\xymatrix@M.1em}
\newcommand{\ba}{\begin{eqnarray}}
\newcommand{\ea}{\end{eqnarray}}

\title{}
\author{}

\numberwithin{equation}{section}

\begin{document}
%
%\maketitle
\renewcommand{\thefootnote}{\fnsymbol{footnote}}
~
\vspace{1.75truecm}
\begin{center}
{\LARGE \bf{Massive Spin-2 Scattering and}}\\ \vspace{.2cm}
{\LARGE \bf{Asymptotic Superluminality}}
\end{center} 

\vspace{1truecm}
\thispagestyle{empty}
\centerline{{\Large Kurt Hinterbichler,${}^{\rm a,}$\footnote{\href{mailto:kurt.hinterbichler@case.edu}{\texttt{kurt.hinterbichler@case.edu}}} Austin Joyce,${}^{\rm b,}$\footnote{\href{mailto:austin.joyce@columbia.edu}{\texttt{austin.joyce@columbia.edu}}} and Rachel A. Rosen${}^{\rm b,}$\footnote{\href{mailto:rar2172@columbia.edu}{\texttt{rar2172@columbia.edu}}}}}
\vspace{.5cm}
 
\centerline{{\it ${}^{\rm a}$CERCA, Department of Physics,}}
 \centerline{{\it Case Western Reserve University, 10900 Euclid Ave, Cleveland, OH 44106}} 
 \vspace{.25cm}
 
 \centerline{{\it ${}^{\rm b}$Center for Theoretical Physics, Department of Physics,}}
 \centerline{{\it Columbia University, New York, NY 10027}} 
 \vspace{.25cm}

 \vspace{.8cm}
\begin{abstract}
\noindent
We place model-independent constraints on theories of massive spin-2 particles by considering the positivity of the phase shift in eikonal scattering.  The phase shift is an asymptotic $S$-matrix observable, related to the time delay/advance experienced by a particle during scattering.  Demanding the absence of a time advance leads to constraints on the cubic vertices present in the theory.  We find that, in theories with massive spin-2 particles, requiring no time advance means that either: (i) the cubic vertices must appear as a particular linear combination of the Einstein--Hilbert cubic vertex and an $h_{\mu\nu}^3$ potential term or (ii) new degrees of freedom or strong coupling must enter at parametrically the mass of the massive spin-2 field.  
These conclusions have implications for a variety of situations.  Applied to theories of large-$N$ QCD, this indicates that any spectrum with an isolated massive spin-2 at the bottom must have these particular cubic self-couplings.  Applied to de Rham--Gabadadze--Tolley massive gravity, the constraint is in accord with results obtained from a shockwave calculation: of the two free dimensionless parameters in the theory there is a one parameter line consistent with a subluminal phase shift. 

\end{abstract}

\newpage

\setcounter{tocdepth}{2}
\tableofcontents
\newpage
\renewcommand*{\thefootnote}{\arabic{footnote}}
\setcounter{footnote}{0}

\section{Introduction and Summary}
A central outstanding problem in the study of massive high-spin particles  ($s\geq 2$) is to construct an ultraviolet (UV) complete theory which has an infrared (IR) effective description in terms of an isolated massive higher spin, or even to construct an effective theory with a strong coupling scale that does not go to zero with the mass of the particle.  There are many UV complete examples---both theoretically and in nature---where massive higher-spin states arise, {\it e.g.}, as mesons in confining gauge theories, in Kaluza--Klein theories, or in string theory.   However, in these examples there is always a tower of other high-spin states with parametrically the same mass. Accordingly, it is of great interest to derive model-independent constraints on low energy theories of massive high-spin particles and their UV completions. 

In this paper we will study IR constraints on massive spin-2 particles, both because it is the first high-spin massive particle and because of recent interest in its possible relevance to gravity. There is a consistent effective field theory (EFT) description of massive spin-2 fields with a cutoff scale parametrically larger than its mass~\cite{ArkaniHamed:2002sp,Creminelli:2005qk}, but  it is not known whether this type of theory can arise as a Higgs-like phase of Einstein gravity or some other local, Lorentz-invariant ultraviolet completion.\footnote{If such a Higgs mechanism exists, it is thought that it should realize the symmetry breaking pattern ${\rm ISO}(d,1)_{\rm local}\times{\rm Diff}(d,1)\to{\rm SO}(d,1)_{\rm diagonal}$~\cite{Goon:2014paa,Torabian:2017bqu}. In~\cite{Porrati:2001db} an interesting construction is presented where the graviton gets a Schwinger-type mass in AdS, and recently a non-local UV-extension of massive gravity has been proposed~\cite{Gabadadze:2017jom}.}

Within a generic low-energy theory, there are precious few observable quantities which reveal information about possible UV completion.  One such infrared constraint comes from dispersion relations, which constrain forward scattering amplitudes and signal obstructions to UV completion by a local Lorentz-invariant quantum field theory with an $S$-matrix satisfying typical analyticity requirements~\cite{Adams:2006sv,deRham:2017avq,deRham:2017zjm}. This has recently been applied to massive spin-2 particles~\cite{Cheung:2016yqr,Bonifacio:2016wcb}. Another traditional constraint placed on low-energy theories is absence of superluminality. Often these constraints are derived by looking for some classical background solution to the effective theory, considering fluctuations around this solution and demanding that the fluctuations be subluminal.  For example, the classic Velo--Zwanziger problems \cite{Velo:1969bt,Velo:1970ur} and more recent constraints on de Rham--Gabdadze--Tolley massive gravity \cite{Deser:2012qx,Deser:2014fta,Deser:2014hga,Deser:2015wta} are of this type.  However, these bounds are less robust than the sharp $S$-matrix analyticity constraints; there are always questions about whether the backgrounds in question can be reached dynamically within the regime of validity of the effective theory~\cite{Burrage:2011cr,Hassan:2017ugh}, or whether the superluminality itself is visible within the effective theory~\cite{Goon:2016une}.

A better indicator of superluminality would be a sharply-defined $S$-matrix observable, one which does not depend on the existence of or choice of a nontrivial classical background solution.  An example of such a quantity is the phase shift in eikonal scattering amplitudes.  The phase shift in eikonal scattering has long been known to be related to the asymptotic time delay or advance that particles experience when traversing a shockwave geometry~\cite{tHooft:1987vrq,Kabat:1992tb}.  In the case of an asymptotic time advance, a succession of eikonal scattering events or, equivalently, a particle crossing multiple shockwaves, would accumulate a large enough time advance to propagate outside the lightcone used to define the theory.  Provided that the time advance for a single scattering event is measurable within the regime of validity of the EFT, this would indicate that the given theory is superluminal.

The study of eikonal scattering was recently reinvigorated by the analysis of~\cite{Camanho:2014apa} which related the phase shift to on-shell three-point scattering amplitudes. Demanding that the theory does not possess an asymptotic time advance places constraints on the coefficients of cubic terms in the theory or, conversely, the presence of an asymptotic time advance at some scale implies that new physics must enter there.   The fact that the asymptotic phase shift depends only on the on-shell 3-point amplitudes and does not depend on a specific classical background or choice of off-shell Lagrangian makes it a robust and model-independent constraint.  The three-point amplitudes themselves are fixed by Lorentz invariance up to a finite number of constants, and the constraints then apply to these constants independent of the structure of the rest of the theory.  

In~\cite{Camanho:2014apa}, these constraints were worked out in the massless spin-2 case.  An example where their constraints apply is an effective theory consisting of Einstein gravity plus higher-curvature corrections, schematically of the form
\be
S = M_{\rm Pl}^{D-2}\int\rd^Dx\sqrt{-g}\left(R+\frac{1}{\Lambda^2}R_{\rm GB}^2+\cdots\right),
\label{eq:RplusR2}
\ee
where $R_{\rm GB}^2$ stands for the ghost-free Gauss--Bonnet contractions of the Riemann tensor, and the scale suppressing it is parametrically below the Planck scale, $\Lambda \ll M_{\rm Pl}$, so that the contributions to scattering from the $R^2$ terms can become important relative to the Einstein--Hilbert vertices. If we expand out the curvature terms and canonically normalize the fluctuations $h_{\mu\nu}$ around the flat solution, we have schematically,
\be
S = \int\rd^Dx\left(h\partial^2 h+\frac{1}{M_{\rm Pl}^\frac{D-2}{2}}h^2\partial^2  h+\cdots+\frac{1}{\Lambda^2M_{\rm Pl}^\frac{D-2}{2}} h^2\partial^4 h+\cdots\right).
\ee
If we imagine that any higher-curvature invariants in \eqref{eq:RplusR2} are suppressed by $M_{\rm Pl}$, then the lowest scale suppressing interactions in this theory is $\Lambda_{\rm c} = \Lambda^{4/(D+2)} M_{\rm Pl}^{(D-2)/(D+2)}$,
which is where we should expect tree level partial-wave unitarity to break down and the theory to become strongly coupled. Therefore, naively we should expect that~\eqref{eq:RplusR2} is a well-defined effective field theory up until the scale $\Lambda_{\rm c}$. However, in~\cite{Camanho:2014apa} it was shown that the eikonal phase can become negative for some polarizations at impact parameters $b\sim \Lambda^{-1}$, signaling some kind of superluminal propagation at the scale $\Lambda\ll \Lambda_{\rm c}$. Under the assumption that the ultraviolet completion of the theory does not have such superluminality, something new has to happen at the scale $\Lambda$, which is parametrically lower than the strong-coupling scale, to fix this problem. Indeed, this happens in explicit examples. For example, in bosonic string theory, where such higher-curvature corrections appear for the graviton~\cite{Zwiebach:1985uq}---and the scale $\Lambda\sim\ell_{\rm s}^{-1}$ is the string scale---new states enter at precisely this scale in order to make the total Shapiro time delay positive~\cite{DAppollonio:2015fly}.\footnote{However, in this case higher-curvature operators are also suppressed by the string scale, so $\Lambda_{\rm c}$ gets brought down to the string scale as more and more higher-derivative operators are considered.}

In this paper we perform a similar analysis for massive spin-2. That is, we use the positivity of the phase shift in eikonal scattering amplitudes as an IR constraint on the possible three-point structures that can appear in a theory of a single massive spin-2 particle.\footnote{A similar study was done for the case of $D=3$ massive gravity in~\cite{Edelstein:2016nml}.}  For a single massive spin-2 field, Lorentz invariance fixes the cubic vertices to be one of five structures in generic dimension.
We find that for a generic choice of cubic interactions in the theory, a time advance would be detectable at impact parameters of order the inverse graviton mass, $b\sim m^{-1}$. This can be avoided with only one particular choice of cubic vertices, leaving only the freedom to rescale the Planck mass.  Away from this choice of parameters, either interactions which give rise to this time advance have greatly suppressed coefficients so that they are effectively absent from the theory, or new particles or strong coupling enter at the scale $m$.  Note that this calculation is well within the regime of validity of the effective theory, and in fact is independent of the precise strong coupling scale of the effective theory, which is generally parametrically larger than the graviton mass. A generic effective field theory of a massive spin-2 in four dimensions has as its cutoff the scale $\Lambda_5 = (m^4M_{\rm Pl})^{1/5}$~\cite{ArkaniHamed:2002sp,Creminelli:2005qk}, but by carefully tuning interactions, the cutoff can be raised to $\Lambda_3 = (m^2M_{\rm Pl})^{1/3}$. This can be done in two ways.  The first is by using the nonlinear Einstein--Hilbert kinetic term and tuning potential interactions.  This leads to the de Rham--Gabadadze--Tolley (dRGT) theory of massive gravity~\cite{deRham:2010ik,deRham:2010kj} (see \cite{Hinterbichler:2011tt,deRham:2014zqa} for reviews).  The second is to keep the linear theory kinetic term and tune interactions.  This leads to the pseudo-linear interacting theory which can have particular derivative interactions in addition to potential terms~\cite{Folkerts:2011ev,Hinterbichler:2013eza}. The precise cutoff of the effective theory therefore depends on the full structure of interactions in the theory.  Our analysis on the other hand is only sensitive to the cubic structure of the theory, and  applies equally well regardless of the cutoff or the choice of potential beyond cubic order. In this sense, it is completely model-independent.

There are at least two immediate applications of the eikonal constraints: to theories of massive gravity and to theories of large-$N$ QCD.  For the two-parameter dRGT massive gravity, our result leaves a one-parameter family consistent with positivity of the eikonal phase. This agrees qualitatively with the results of~\cite{Camanho:2016opx}, found by considering scattering off of shockwave backgrounds of the theory.\footnote{In the revised version of~\cite{Camanho:2016opx} the authors have changed their conclusions to argue that the eikonal phase is negative for all points in the parameter space they consider. Our results adapted to the situation they consider differ in some details. We find agreement between the eikonal scattering computation we do and the shockwave analysis we perform in Sec.~\ref{sec:shockwaveamp}.} Additionally, this one-parameter family intersects the two dimensional compact blob-like region of~\cite{Cheung:2016yqr} consistent with dispersion relations following from $S$-matrix analyticity, leaving a finite size one-dimensional line in parameter space with no obstructions from either analysis.   Regarding QCD, the constraints apply to a confining gauge theory with a large number of colors, $N$.  Such a theory can be thought of as a weakly-coupled theory of interacting higher spins.  The massive higher spin particles are the hadrons and glueballs of the theory, and some positive power of $1/N$ serves as a coupling constant governing their interaction strengths.  Constraints on massive higher spins thus serve as constraints on the possible spectra and interactions of large-$N$ QCD.  Our results indicate that if some large-$N$ QCD-like theory has an isolated massive spin-2 excitation, then its cubic self-couplings must appear in a specific combination.

The eikonal positivity constraint is typically enforced because of its apparent connection to causality~\cite{Dray:1984ha,Camanho:2014apa}.\footnote{It is worth noting that causality, even in this asymptotic sense, is quite difficult to define rigorously in General Relativity, essentially because there is no unique way to identify a perturbed spacetime with Minkowski space~\cite{Tipler:1980aq,Olum:1998mu}. This difficulty is partially addressed in~\cite{Gao:2000ga}. These difficulties are mostly absent in the theory of a massive spin-2 because it requires an underlying reference spacetime to define the theory in the first place.}
However, this connection is somewhat tenuous. The usual argument that is given against the type of asymptotic time advance that eikonal scattering captures is that it would allow one to build a time machine (see {\it e.g.},~\cite{Adams:2006sv,Camanho:2014apa,Camanho:2016opx}).
It is somewhat unclear that this is the case though; it is perfectly possible for the theory to be causal, but on a widened lightcone compared to the one used to define the theory~\cite{Babichev:2007dw,Geroch:2010da}. Further, it was argued in~\cite{Papallo:2015rna} that this type of time machine cannot arise as the Cauchy evolution of some hyperbolic set of equations.\footnote{This is somewhat similar to the arguments against Gott time machines~\cite{Gott:1990zr} in General Relativity~\cite{Carroll:1991nr}.} Nevertheless, even if such an asymptotic time advance does not violate causality, it would seem to violate microcausality, {\it i.e.}, make it possible to find two local operators which do not commute outside the lightcone used to define the theory, which should not happen in the IR description of a local Lorentz-invariant UV theory~\cite{Dubovsky:2007ac}.
However, this connection, as far as we know, has not been rigorously demonstrated.   We will keep these issues in mind when imposing absence of this type of asymptotic superluminality in the infrared EFT.

It is also worthwhile to pause and describe the assumptions implicit in imposing the constraint that the eikonal phase is positive in the IR theory. Foremost, we are assuming that whatever UV completes the theory of interest is devoid of time advances. While this condition is satisfied in known examples, there is no theorem that this must be the case. Further we are assuming that the leading eikonal phase is accurately captured by resummation of ladder graphs---as we will describe. There is no proof that this works to all orders, and in fact is known to fail for spin-0 and spin-1 exchange. We are also assuming that the $S$-matrix is a well-defined observable quantity in the theory of interest. In cosmological applications, the spacetime is not asymptotically flat, so it not clear that constraints from scattering need apply. Our viewpoint is that it is important to bear these limitations in mind, but it is nevertheless an interesting question to ask what constraints positivity of the eikonal phase place on the IR effective theory, and it is in this spirit that we proceed.

In what follows we will first review scattering in the eikonal approximation and then apply these techniques to compute the asymptotic time advance felt in a generic theory of a massive spin-2. We then compare these results to those obtained by explicitly solving for the propagation in a shockwave background. Finally, we comment on the implications of our results and future directions.

\noindent{\bf Conventions:} We work with mostly plus metric signature, with the curvature conventions of \cite{Carroll:2004st}.  We denote the spacetime dimension by $D$ and we restrict to $D>3$.  Symmetrization and anti-symmetrization is done with weight one.

%%%%%%%
\section{Eikonal Scattering Amplitudes}
\label{sec:eikonalSstuff}
We first review the eikonal approximation and its relation to on-shell cubic structures. We then describe the kinematics of eikonal scattering and a well-adapted basis of polarization tensors.

\subsection{The Eikonal Phase}

The eikonal regime of $2\rightarrow 2$ scattering corresponds to large center of mass energy and relatively large impact parameter~\cite{Cheng:1969eh,Levy:1969cr,Abarbanel:1969ek}. At large center of mass energy, the eikonal approximation gives the leading contribution to forward scattering $(t/s \to 0)$. In this kinematic limit, scattering occurs between two highly boosted particles through the exchange of many soft modes. The eikonal approximation is expected to correspond to summing all ladder and crossed ladder diagrams of the form
\vspace{-.5cm}
 \be
 \label{eq:eikonalpicture}
 \begin{tikzpicture}[line width=1.7 pt,baseline={([yshift=-3ex]current bounding box.center)},vertex/.style={anchor=base,
    circle,fill=black!25,minimum size=18pt,inner sep=2pt}]
\draw[style] (-1.2,0) -- (1.2,0);
\draw[style] (-1.2,1) -- (1.2,1);
\draw[style={decorate,decoration=complete sines},line width=1] (-.04,0) -- (-.04,2);
\draw[style={decorate,decoration=complete sines},line width=1] (.04,0) -- (.04,2);
\node[scale=1] at (1.75, .5) {$\boldsymbol+$};
\end{tikzpicture}
 \begin{tikzpicture}[line width=1.7 pt,baseline={([yshift=-3ex]current bounding box.center)},vertex/.style={anchor=base,
    circle,fill=black!25,minimum size=18pt,inner sep=2pt}]
\draw[style] (-1.2,0) -- (1.2,0);
\draw[style] (-1.2,1) -- (1.2,1);
\draw[style={decorate,decoration=complete sines},line width=1] (-.54,0) -- (-.54,2);
\draw[style={decorate,decoration=complete sines},line width=1] (-.46,0) -- (-.46,2);
\draw[style={decorate,decoration=complete sines},line width=1] (.46,0) -- (.46,2);
\draw[style={decorate,decoration=complete sines},line width=1] (.54,0) -- (.54,2);
\node[scale=1] at (3.7, .5) {$\boldsymbol+~~\boldsymbol\cdots\boldsymbol+{\rm crossed~ladders}$,};
\end{tikzpicture}
\ee
in the limit of small momentum transfer. In this limit, the scattered particles are nearly on-shell on the top and bottom sides (rails) of the ladder. 

With these approximations, the ladder and crossed-ladder diagrams re-sum into an exponential form in impact parameter space~\cite{Cheng:1969eh,Levy:1969cr,Abarbanel:1969ek,Kabat:1992tb}
\be
\label{eq:eikonal}
i{\cal M}_{\rm eik}(s, t) = 2s\int\rd^{D-2}\vec b\,e^{i\vec q\cdot\vec b}\left(e^{i\delta(s,\vec b)}-1\right),
\ee
where $\vec b$ is the impact parameter, the Fourier conjugate variable to the momentum transfer, $\vec q$, where $-\vec q\,{}^2 = t$ (we will make the eikonal kinematics and spin dependence explicit in Section \ref{sec:eikonlkin}). The eikonal phase, $\delta$, is given by
\be
\label{eq:eikphase}
\delta(s,\vec b) = \frac{1}{2s}\int\frac{\rd^{D-2}\vec q}{(2\pi)^{D-2}}\,e^{-i\vec q\cdot\vec b}{\cal M}_{4}(s, \vec q\,{}^2),
\ee
where ${\cal M}_{4}$ is the part of the tree-level amplitude given by the $t$-channel graph with eikonal kinematics,
\vspace{-1cm}
\be
\begin{tikzpicture}[line width=1.7 pt,baseline={([yshift=-3ex]current bounding box.center)},vertex/.style={anchor=base,
    circle,fill=black!25,minimum size=18pt,inner sep=2pt}]
\node[scale=1] at (-2.25, .5) {${\cal M}_{4}(s, \vec q\,{}^2)=$};
\draw[style] (-.8,0) -- (.8,0);
\draw[style] (-.8,1) -- (.8,1);
\draw[style={decorate,decoration=complete sines},line width=1] (-.04,0) -- (-.04,2);
\draw[style={decorate,decoration=complete sines},line width=1] (.04,0) -- (.04,2);
\node[scale=1] at (1.2, .2) {$.$};
\end{tikzpicture}
\ee

The rough idea behind eikonal re-summation is that the particles on the rails of the ladder diagrams are always very nearly on-shell, so that ladder and crossed ladder diagrams essentially factorize into a product of tree level amplitudes. The relevant combinatorial factors associated with permuting momenta in the rungs conspire to give the exponential in~\eqref{eq:eikonal}~\cite{Giudice:2001ce,Giddings:2011xs}. This is worked out in detail for scalars in~\cite{Levy:1969cr} and in Appendix~\ref{app:eikonal} we demonstrate the eikonal exponentiation for external and exchanged particles of arbitrary spin.

The exponentiation of ladder graphs into a phase is very robust, but a general proof that these diagrams accurately capture the leading behavior of the eikonal limit ($t/s\to 0$) of the full scattering amplitude is so far missing. In fact, it is known that for exchange of spin-0 or spin-1 particles, there are non-ladder diagrams which contribute at the same order as the ladder diagams~\cite{Tiktopoulos:1971hi,Cheng:1987ga,Kabat:1992pz}. It is believed that for exchange of particles with spin $\geq 2$ that the ladder approximation accurately captures the leading effects, and sub-leading corrections have been checked to be small in some cases~\cite{Akhoury:2013yua,Bjerrum-Bohr:2016hpa}. In what follows we assume that the ladder and crossed ladders capture the leading eikonal amplitude, but it would be interesting to return to examine sub-leading corrections in the future.

The eikonal phase~\eqref{eq:eikphase} is related to the delay in lightcone coordinate time, $\Delta x^{\scriptscriptstyle -}$, experienced by the particle moving in the $x^{\scriptscriptstyle +}$ direction after interacting with the other particle moving in the $x^{\scriptscriptstyle -}$ direction~\cite{Kabat:1992tb,Camanho:2014apa}\footnote{The relation between this and the time delay experienced by a classical particle traversing a shockwave is made fairly explicit in~\cite{Saotome:2012vy} by considering the eikonalized gravitons as mediating an interaction between a probe particle and a classical gravitational background. The eikonal amplitude is reproduced by considering repeated interaction with a  background field of the Aichelburg--Sexl form~\cite{Aichelburg:1970dh}.}
\be
\Delta x^{\scriptscriptstyle -} = \frac{1}{\lvert p^{\scriptscriptstyle -}\rvert}\delta(s, b).
\ee
We are therefore interested in computing the sign of the eikonal phase, $\delta$, to ascertain whether interactions lead to an asymptotic time delay or time advance.  A positive phase shift, $\delta >0$, leads to a time delay (Shapiro delay), a negative phase shift, $\delta <0$, leads to a time advance (superluminality).

The problem of computing the time advance/delay experienced by a particle thus reduces to a 4-point tree-level scattering amplitude computation in impact parameter space with eikonal kinematics. In~\cite{Camanho:2014apa}, it was pointed out that the leading eikonal phase, $\delta$, can actually be computed knowing only the on-shell three-point scattering amplitudes in the theory. The argument relies on a complex momentum shift and is somewhat reminiscent of arguments used to derive the BCFW recursion relations~\cite{Britto:2005fq} and related $S$-matrix constraints~\cite{Benincasa:2007xk,Schuster:2008nh}. Consider the eikonal phase~\eqref{eq:eikphase} and analytically continue the first component of the transverse momentum, $\vec q$, by the complex shift
\be
q^1 \mapsto q^1-i\kappa\, .
\ee
Then imagine making $\kappa$ arbitrarily large and real, {\it i.e.}, deform the integration contour in the integral~\eqref{eq:eikphase} in the complex plane by pushing it out towards infinity.  Choosing $\vec b$ to point along the first direction so that $\kappa b^1 > 0$, this leads to exponential suppression so the contour at infinity will not contribute to the $q^1$ integral (assuming amplitudes are polynomially bounded, which they always are in effective field theories), but any poles in the lower complex momentum plane {\it will}. The residues of these poles in the scattering amplitude at complex momenta are precisely the product of on-shell three-point functions~\cite{Schuster:2008nh,Camanho:2014apa}, so we have
\be
\int\frac{\rd^{D-2}\vec q}{(2\pi)^{D-2}}\,e^{-i\vec q\cdot\vec b}{\cal M}_{4}(s, \vec q) = \int\frac{\rd^{D-2}\vec q}{(2\pi)^{D-2}}\,e^{-i\vec q\cdot\vec b}{\cal M}^I_3(p_1,p_3,q)\frac{N_{IJ}}{\vec q\,{}^2 + m^2}{\cal M}^J_3(q,p_2,p_4)\, ,
\ee
where $N_{IJ}$ stands for the tensor structure of the propagator of the exchanged particle, which comes from summing over all possible intermediate polarization states. Though we have written ${\cal M}_3$ as a function of three arguments, it should be thought of as an on-shell object, and therefore only has two independent momentum arguments.

We can then trade the factors of $\vec q$ in the on-shell amplitudes for derivatives with respect to the impact parameter, $\vec b$, in order to write~\cite{Camanho:2014apa}
\bea
\delta(s,b) &=& \frac{\sum_I{\cal M}_3^{13 I}(i\partial_{\vec b}){\cal M}_3^{I24}(i\partial_{\vec b})}{2s}\int\frac{\rd^{D-2}\vec q}{(2\pi)^{D-2}}\frac{e^{-i\vec q\cdot\vec b}}{\vec q^2 + m^2} \nn\\
&=& \frac{\sum_I{\cal M}_3^{13 I}(i\partial_{\vec b}){\cal M}_3^{I24}(i\partial_{\vec b})}{2s}\left[ \frac{1}{2\pi^{\frac{D-2}{2}}}\left(\frac{m}{b}\right)^\frac{D-4}{2}K_\frac{D-4}{2}(mb)\right]
\, ,
\eea
where the sum is over all the possible internal polarization states.

In the cases we are interested in, the exchanged particles in the amplitude are massive spin-2s, so what we get is
\be
\delta(s,b)  = \frac{{\cal M}_3^{13,\alpha\beta}(i\partial_{\vec b})N_{\alpha\beta\mu\nu}{\cal M}_3^{\mu\nu,24 }(i\partial_{\vec b})}{2s}\frac{1}{2\pi^{\frac{D-2}{2}}}\left(\frac{m}{b}\right)^\frac{D-4}{2}K_\frac{D-4}{2}(mb)\, ,
\label{eq:phaseshiftmassivespin2}
\ee
where ${\cal M}_3^{13 \alpha\beta}(i\partial_{\vec b})$ is an on-shell amplitude with an external massive spin-2 and $N_{\alpha\beta\mu\nu}$ is the numerator of the massive graviton propagator
\be
N_{\alpha\beta\mu\nu} = \frac{1}{2}(P_{\alpha\mu}P_{\beta\nu}+P_{\alpha\nu}P_{\beta\mu})-\frac{1}{D-1}P_{\alpha\beta}P_{\mu\nu} ~~~~{\rm with}~~P_{\mu\nu} = \eta_{\mu\nu}+\frac{1}{m^2}p_\mu p_\nu\, ,
\ee
which comes from the completeness relation of massive graviton polarizations:
\be
\sum_I \epsilon_{\alpha\beta}^I\epsilon_{\mu\nu}^{*I} = N_{\alpha\beta\mu\nu}.
\ee

The problem of computing the time delay in an arbitrary theory of massive spin-2s has been essentially reduced to the concrete problem of enumerating the possible on-shell 3-point vertices and using them to compute the operator ${\cal M}_3^{13,\alpha\beta}(i\partial_{\vec b})N_{\alpha\beta\mu\nu}{\cal M}_3^{\mu\nu,24 }(i\partial_{\vec b})$. In what follows we will describe how to do this and how to extract the time delay.

\subsection{Eikonal Kinematics}
\label{sec:eikonlkin}
We will now make the eikonal kinematics explicit.
We are interested in $2\rightarrow 2$ scattering where particle $A$ with mass $m_A$ scatters off of particle $B$ with mass $m_B$.  Particle $A$ has incoming momentum $p_1^\mu$ and outgoing momentum $p_3^\mu$, particle $B$ has incoming momentum $p_2^\mu$ and outgoing momentum $p_4^\mu$.

Throughout we work in lightcone coordinates $(x^{\scriptscriptstyle -},x^{\scriptscriptstyle +},x^i)$,
\be x^{\scriptscriptstyle \pm}={1\over \sqrt{2}}\left(x^0\pm x^1\right)\, , \ee
where the Minkowski metric takes the form,
\be \eta_{\mu\nu}=\left(\begin{array}{ccc}0 & -1 & 0 \\-1 & 0 & 0 \\0 & 0 & \delta_{ij}\end{array}\right),\ee
with $i,j,\ldots=2,\ldots,D-1$ running over the transverse directions.

We want the amplitude with the following kinematics, 
\begin{align}
p_1^\mu&=\left({1\over 2p^{\scriptscriptstyle +}}\left({{\vec q\ }^2\over 4}+m_A^2\right),p^{\scriptscriptstyle +} ,{ q^i \over 2}\right)\, , & p_3^\mu&=\left({1\over 2 p^{\scriptscriptstyle +}}\left({{\vec q\ }^2\over 4}+m_A^2\right),p^{\scriptscriptstyle +},-{ q^i \over 2}\right)\, , \\
p_2^\mu&=\left(p^{\scriptscriptstyle -},{1\over 2 p^{\scriptscriptstyle -}}\left({{\vec q\ }^2\over 4}+m_B^2\right), -{ q^i\over 2 }\right)\, , & p_4^\mu &=\left(p^{\scriptscriptstyle -},{1\over 2 p^{\scriptscriptstyle -}}\left( {{\vec q\ }^2\over 4}+m_B^2\right),{ q^i\over 2 }\right)\,.
\end{align}
These are exactly on-shell: $p_1^2=p_3^2= -m_A^2$ and  $p_2^2=p_4^2=-m_B^2$, $p_1^\mu+p_2^\mu=p_3^\mu+p_4^\mu$.  The independent Mandelstam invariants are
\bea &&s=-(p_1+p_2)^2={(m_A^2+2p^{\scriptscriptstyle +}p^{\scriptscriptstyle -})(m_B^2+2p^{\scriptscriptstyle +}p^{\scriptscriptstyle -})\over 2p^{\scriptscriptstyle +}p^{\scriptscriptstyle -}}+{m_A^2+m_B^2+4p^{\scriptscriptstyle +}p^{\scriptscriptstyle -}\over 2 p^{\scriptscriptstyle +}p^{\scriptscriptstyle -}}{{\vec q\ }^2\over 4}+{1\over 2 p^{\scriptscriptstyle +} p^{\scriptscriptstyle -}}{{\vec q\ }^4\over 16},\ \ \nn\\
&&t=-(p_1-p_3)^2=-{\vec q\ }^2\, .\eea

We construct polarization tensors for the massive particles out of the following transverse (labeled by $T$) and longitudinal (labeled by $L$) massive spin-1 polarization tensors:
\begin{align}
 \epsilon_T^\mu(p_1)&=\left({{\vec q \ }\cdot {\vec e}_1 \over 2p^{\scriptscriptstyle +}},0 ,{  e_1^i  }\right)\, , &  \epsilon_L^\mu(p_1)&=\left({1\over 2m_Ap^{\scriptscriptstyle +}}\left({{{\vec q \ }}^2\over 4}-m_A^2\right),{p^{\scriptscriptstyle +}\over m_A} ,{ q^i \over 2m_A}\right)\, ,  \nn\\
 \epsilon_T^\mu(p_2)&=\left(0,-{{\vec q \ }\cdot  {\vec e}_2 \over 2p^{\scriptscriptstyle -}},{  e_2 ^i   }\right)\, ,&  \epsilon_L^\mu(p_2)&=\left({p^{\scriptscriptstyle -}\over m_B} ,{1\over 2m_Bp^{\scriptscriptstyle -}}\left({{{\vec q \ }}^2\over 4}-m_B^2\right),-{ q^i \over 2m_B}\right)\, ,  \nn\\
\epsilon_T^\mu(p_3)&=\left(-{{\vec q \ }\cdot  {\vec e}_3 \over 2p^{\scriptscriptstyle +}},0 ,{  e_3^i   }\right)\, ,  &\epsilon_L^\mu(p_3)&=\left({1\over 2m_Ap^{\scriptscriptstyle +}}\left({{{\vec q \ }}^2\over 4}-m_A^2\right),{p^{\scriptscriptstyle +}\over m_A} ,-{ q^i \over 2m_A}\right)\, ,  \nn\\
 \epsilon_T^\mu(p_4)&=\left(0,{{\vec q \ }\cdot  {\vec e}_4 \over 2p^{\scriptscriptstyle -}} ,{  e_4^i   }\right)\, , &  \epsilon_L^\mu(p_4)&=\left({p^{\scriptscriptstyle -}\over m_B},{1\over 2m_Bp^{\scriptscriptstyle -}}\left({{{\vec q \ }}^2\over 4}-m_B^2\right) ,{ q^i \over 2m_B}\right)\, .\label{spi1polse}
\end{align}
Here the $\vec e$ 's are normalized vectors that live in the $(D-2)$-plane transverse to $x^{\scriptscriptstyle +},x^{\scriptscriptstyle -}$; there are $D-2$ independent such vectors, and so there are $D-2$ independent $T$ polarization vectors.  Thus the transverse polarizations actually come along with an additional label $\lambda=1,2,\cdots,D-2$ which indexes an orthonormal basis of the transverse space $e_\lambda^i$,
\be  
e_{\lambda i}e_{\lambda'}^i=\delta_{\lambda\lambda'}\, ,~~~~~~~~~~~~~
\sum_\lambda e_\lambda^i e_\lambda^j=\delta^{ij}\, .
\ee
For example, we will usually choose the standard basis of linear polarization vectors,
\be 
e_\lambda^i=\delta_\lambda^i\, .
\ee
The polarization vectors \eqref{spi1polse} are exactly transverse, orthonormal, and complete,
\begin{align}
& p_{a\mu}\epsilon_T^\mu(p_a)=p_{a\mu}\epsilon_L^\mu(p_a)=0, \nn\\
& \epsilon_{T,\lambda \ \mu}(p_a)^\ast \epsilon_{T,\lambda'}^\mu(p_a)=\delta_{\lambda\lambda'},~~~~~~~\epsilon_{L\mu}(p_a)^\ast \epsilon_L^\mu(p_a)=1,~~~~~~~\epsilon_{T\mu}(p_a)^\ast \epsilon_L^\mu(p_a)=0\, ,\nn\\
&\epsilon_{L}^\mu (p_a) \epsilon_L^\nu(p_a)^\ast+\sum_\lambda \epsilon_{T,\lambda}^\mu (p_a) \epsilon_{T,\lambda}^\nu(p_a)^\ast=\eta^{\mu\nu}-{1\over p_a^2}p_a^\mu p_a^\nu\, .
\end{align}
where $a=1,2,3,4$ labels the momenta.

The polarization tensors for a massive spin-2 are constructed out of these as follows:
\begin{align}
\epsilon_T^{\mu\nu}(p_a) &=\epsilon_T^{\mu}(p_a)\epsilon_T^{\nu}(p_a)\, ,   \nn\\
\epsilon_V^{\mu\nu}(p_a) &={i\over \sqrt{2}}\left(\epsilon_T^{\mu}(p_a)\epsilon_L^{\nu}(p_a) +\epsilon_L^{\mu}(p_a)\epsilon_T^{\nu}(p_a) \right)\, , \nn\\
 \epsilon_S^{\mu\nu}(p_a) &= \sqrt{D-1\over D-2}\left[ \epsilon_L^{\mu}(p_a) \epsilon_L^{\nu}(p_a) -{1\over D-1}\left(\eta^{\mu\nu} -{1\over p_a^2}p_a^\mu p_a^\nu  \right) \right] \, . \label{spin2polse}
\end{align}
Here $T$, $V$, $S$ stand for tensor, vector and scalar polarizations, respectively.
In the expression for $\epsilon_T^{\mu\nu}(p_a)$, it is understood that we replace $e_ie_j\mapsto e_{ij}$ with $e_{ij}$, which is symmetric and traceless.

As in the spin-1 case, the $\epsilon_V^{\mu\nu}$ depend on a transverse vector $e_i$ and so it comes with an additional label $\lambda=1,2,\cdots,D-2$ running over a basis of these transverse vectors.   The $\epsilon_T^{\mu\nu}$ depend on a transverse, symmetric, and traceless tensor $e_{ij}$, and so it depends on an additional label $\tilde\lambda=1,2,\cdots {(D-2)(D-1)\over 2}-1$ indexing a basis $e_{\tilde\lambda}^{ij}$ of transverse, symmetric, and traceless tensors,
\be e_{\tilde \lambda \ ij}e_{\tilde \lambda'}^{ij}=\delta_{\tilde\lambda\tilde\lambda'},~~~~~~~\sum_{\tilde \lambda} e_{\tilde \lambda}^{ij} e_{\tilde\lambda}^{kl}={1\over 2}\left(\delta^{ik}\delta^{jl}+\delta^{jk}\delta^{il}-{2\over D-2}\delta^{ij}\delta^{kl}\right).\ee  

We can construct an explicit basis in the following way (see {\it e.g.}~\cite{Benakli:2015qlh} for a similar construction of a non-orthonormal basis)
\begin{itemize}
\item The $\oplus$ polarizations:  ${\bf e}_{\oplus_{j}}$, $j=1,\cdots,D-3$ are diagonal with $1$s along the diagonal from the $11$-th entry to the $jj$-th entry and $-j$ in the $j+1,j+1$-th entry, with an overall normalization factor:
\be
{\bf e}_{\oplus_{j}}=\frac{1}{\sqrt{j(j+1)}}\left(
\begin{array}{cccc}
1&0&0&\cdots\\
0&\ddots&0&\cdots\\
0& 0& -j& \cdots\\
\vdots & \vdots & \vdots & \ddots
\end{array}
\right).
\ee
There are $D-3$ independent polarizations of this type, and they form a basis of traceless diagonal matrices.  
\item The $\otimes$ polarizations:  ${\bf e}_{\otimes_{ij}}$, $i,j=1,\cdots,D-2$, $i<j$ are off-diagonal with $1/\sqrt{2}$ in the $ij$-th entry and  $1/\sqrt{2}$ in the $ji$-th entry.  For example  
\be
{\bf e}_{\otimes_{12}}=\frac{1}{\sqrt 2}\left(
\begin{array}{cccc}
0&1&0&\cdots\\
1&0&0&\cdots\\
0& 0& 0& \cdots\\
\vdots & \vdots & \vdots & \ddots
\end{array}
\right)\,.
\ee
There are $\frac{(D-2)(D-3)}{2}$ independent polarizations of this type, and they form a basis of off-diagonal traceless symmetric tensors.
\end{itemize}
Together there are $\frac{(D-2)(D-3)}{2}+D-3 = \frac{(D-2)(D-1)}{2}-1$ independent tensor polarizations. This is the correct number of helicity-2 polarizations in $D$ dimensions. Combining these tensor polarizations with the $D-2$ vector polarizations and the single scalar polarization, we find a total of $\frac{(D+1)(D-2)}{2}$ polarizations, the correct number for a massive spin-2 in $D$-dimensions.

The spin-2 polarization tensors \eqref{spin2polse} are all properly transverse, orthonormal, and complete:
\begin{align}
\nonumber
&p_{a\mu}\epsilon_T^{\mu\nu}(p_a)=p_{a\mu}\epsilon_V^{\mu\nu}(p_a)=p_{a\mu}\epsilon_S^{\mu\nu}(p_a)=0, \\
 &\epsilon_{T,\tilde\lambda \ \mu\nu}(p_a)^\ast \epsilon_{T,\tilde\lambda'}^{\mu\nu}(p_a)=\delta_{\tilde\lambda\tilde\lambda'},~~~~~~~\epsilon_{V,\lambda \ \mu\nu}(p_a)^\ast \epsilon_{V,\lambda'}^{\mu\nu}(p_a)=\delta_{\lambda\lambda'},~~~~~~~\epsilon_{S\mu\nu}(p_a)^\ast \epsilon_S^{\mu\nu}(p_a)=1, \\\nonumber
& \epsilon_{T\mu\nu}(p_a)^\ast \epsilon_V^{\mu\nu}(p_a)= \epsilon_{T\mu\nu}(p_a)^\ast \epsilon_S^{\mu\nu}(p_a)= \epsilon_{V\mu\nu}(p_a)^\ast \epsilon_S^{\mu\nu}(p_a)=0\, ,\\\nonumber
& \epsilon_{S}^{\mu\nu} (p_a) \epsilon_S^{\alpha\beta}(p_a)^\ast+\sum_\lambda \epsilon_{V,\lambda}^{\mu\nu} (p_a) \epsilon_{V,\lambda}^{\alpha\beta}(p_a)^\ast+\sum_{\tilde\lambda}\epsilon_{T,\tilde \lambda}^{\mu\nu} (p_a) \epsilon_{T,\tilde\lambda}^{\alpha\beta}(p_a)^\ast ={1\over 2}\left( P^{\mu\alpha} P^{\nu\beta}+P^{\nu\alpha} P^{\mu\beta}-{2\over D-1}P^{\mu\nu} P^{\alpha\beta}\right),
\end{align}
where $P^{\alpha\beta}\equiv  \eta^{\mu\nu}-{1\over p_a^2}p_a^\mu p_a^\nu$.

We can now proceed to evaluate amplitudes using these kinematics.  The eikonal limit is the limit where $p^{\scriptscriptstyle +},p^{\scriptscriptstyle -}$ is taken to be large compared to everything else. 
In this limit, 
\be 
s\rightarrow 2p^{\scriptscriptstyle +} p^{\scriptscriptstyle -}\, ,
\ee
and in all our later expressions we will use $s$ and $2p^{\scriptscriptstyle +} p^{\scriptscriptstyle -}$ interchangeably.

\section{On-Shell Cubic Amplitudes for Massive Spin-2}

The eikonal amplitude depends only on the on-shell 3-point functions of the theory, so we will first enumerate the possible on-shell cubic vertices that can appear in the computation.

\subsection{General Construction of On-Shell Cubic Vertices}
\label{sec:onshell3ptvert}

Lorentz invariance places strong constraints on on-shell three particle scattering amplitudes.\footnote{With real external momenta, all such amplitudes vanish kinematically. However, as we saw in Section~\ref{sec:eikonalSstuff} the calculation is sensitive to the on-shell amplitudes at complex momenta, which do not have to vanish.}  Given a set of three particles, there are only a finite number of three-point structures consistent with Lorentz invariance.  A useful description of the construction of these amplitudes can be found in~\cite{Costa:2011mg}, which we review here briefly.

We want momentum-space on-shell 3-point scattering amplitudes involving 3 particles of spins $s_1,s_2,s_3$ and masses $m_1,m_2,m_3$.  The amplitudes are a polynomial depending on the three momenta $p_1^\mu,p_2^\mu,p_3^\mu$, and three polarization tensors $\epsilon_1^{\mu_1\cdots \mu_{s_1}}$, $\epsilon_2^{\mu_1\cdots \mu_{s_2}}$, $\epsilon_3^{\mu_1\cdots \mu_{s_3}}$.  The momenta are all ingoing and on-shell: $p_a^2=-m_a^2$ ($a=1,2,3$), $\sum_{a=1}^3 p_a^\mu=0$.  The polarization tensors are all symmetric, transverse and traceless: $\epsilon_a^{(\mu_1\cdots \mu_{s_a})}=\epsilon_a^{\mu_1\cdots \mu_{s_a}}$, $p_{a\mu}\epsilon_a^{\mu\mu_2\cdots \mu_{s_a}}=0$, $\epsilon_{a\mu}^{\ \ \mu\mu_3\cdots \mu_{s_a}}$=0 ($a=1,2,3$).

 Since the degrees of freedom are carried by transverse, traceless tensors, it is most convenient to work in an index-free notation where we introduce auxiliary polarization vector variables, $z_a^\mu$, which are null, $z_a^2 = 0$, and transverse, $ p_a\cdot z_a = 0$.  We then write amplitudes in terms of the $z_a$ and make the identification
 \be z_a^{\mu_1}\cdots z_a^{\mu_{s_a}}\longleftrightarrow \epsilon_a^{\mu_1\cdots \mu_{s_a}}.\ee

The scattering amplitude with spins $\{s_1,s_2,s_3\}$ must be homogeneous of order $s_a$ in each of the $z_a$. There are two types of contractions involving the $z$'s: we can either dot them with themselves or with the $p_a$. Taking into account that $z_a^2=0$ and $z_a\cdot p_a=0$, we have 9 independent contractions:
\begin{align}
\label{eq:structurelist1} 
& z_1\cdot z_2\, , & &z_1\cdot z_3\, ,&  &z_2\cdot z_3 \, ,\\
&z_1\cdot p_2\, ,& & z_1\cdot p_3\, ,& &z_2\cdot p_1\, ,\nn\\
&z_2\cdot p_3\, ,& &z_3\cdot p_1\, ,& &z_3\cdot p_2 \, .\nn
\label{eq:structurelist4}
\end{align}
This number is further reduced by taking into account momentum conservation, $p_1^\mu+p_2^\mu+p_3^\mu = 0,$
to eliminate 3 of the dot products. We choose to eliminate as follows
\begin{align}
 z_1\cdot p_3 &\mapsto - z_1\cdot p_2\, ,\nn\\
 z_2\cdot p_1 &\mapsto - z_2\cdot p_3\, ,\nn\\
 z_3\cdot p_2 &\mapsto- z_3\cdot p_1\, .
\end{align}
Finally, there are no independent dot products among the $p_a$'s themselves because the on shell conditions can be used to reduce them all to functions of the masses.

The most general on-shell three-point scattering amplitude is thus a scalar function of the six independent dot products $z_1\cdot z_2$, $z_1\cdot z_3$, $z_2\cdot z_3$, $z_1\cdot p_2$, $z_2\cdot p_3$, $z_3\cdot p_1$, and takes the form
\be
{\cal M}_{s_1s_2s_3}(p_1,p_2,p_3) =  c_{s_1s_2s_3}\, (z_1\cdot z_2)^{n_{12}}( z_1\cdot z_3)^{n_{13}}( z_2\cdot z_3)^{n_{23}}( z_1\cdot p_2)^{m_{12}}( z_2\cdot p_3)^{m_{23}}( z_3\cdot p_1)^{m_{31}},
\label{eq:onshell3ptansatz}
\ee
where $c_{s_1s_2s_3}$ is an overall constant, and due to the requirement that the amplitude be must be  order $s_a$ in each of the $z_a$, the various (non-negative) powers are restricted to satisfy the system of equations
\begin{align}
n_{12}+n_{13}+m_{12} &= s_1\, ,\\
n_{12}+n_{23}+m_{23} &= s_2\, ,\\
n_{13}+n_{23}+m_{31} &= s_3.
\end{align}
These equations have a finite number of solutions in non-negative integers (enumerated in \cite{Costa:2011mg}), and each solution gives an independent scattering amplitude.

In the case of massless representations with spin $\geq 1$, we must also impose gauge invariance, which in this context amounts to the scattering amplitude being invariant under the shift
\be
z_a\mapsto z_a + \epsilon p_a\, ,
\ee
for arbitrary $\epsilon$.
 This condition further restricts the number of allowed structures, but since we will be interested only in massive particles we will not have to deal with it.

In situations where some of the particles are identical, we can decompose the amplitudes further into irreducible representations of the symmetric group of particle interchange, and only those structures invariant under interchange are allowed.  

Finally, there can be additional parity violating amplitudes in some dimensions.  We will not consider these cases here, though it would be interesting to come back to them in the future. 

\subsection{Massive Spin-2 Three-Point Structures}
\label{sec:3ptstructures}
Using the construction in Section~\ref{sec:onshell3ptvert}, we can find a basis for the allowed on-shell three-point amplitudes for a single massive spin-2 particle. We want to consider a theory with a single massive spin-2 of mass $m$, so we demand that they are totally symmetric in the external particles. This leads to 5 different structures, a basis for which is the following: 

 \begin{itemize}
\item 0-derivative structure
\be 
{\cal A}_1={m^2\over M_{\rm Pl}^\frac{D-2}{2}}\,  (z_1\cdot z_2)( z_2\cdot z_3)(z_3\cdot z_1).
\ee
\item 2-derivative structures
\begin{align}
{\cal A}_2&={1\over M_{\rm Pl}^\frac{D-2}{2}}\, \left[ (p_1\cdot  z_3)^2 (z_1\cdot  z_2)^2 + (p_3\cdot  z_2)^2 (z_1
\cdot  z_3)^2 + (p_2\cdot  z_1)^2 (z_2\cdot  z_3)^2 \right],\\\nonumber
{\cal A}_3&={1\over M_{\rm Pl}^\frac{D-2}{2}}\, \Big[ (p_1\cdot  z_3)(p_3\cdot  z_2)( z_1\cdot  z_2)( z_1\cdot  z_3) + 
 (p_1\cdot  z_3)(p_2\cdot  z_1)(z_1\cdot  z_2)(z_2\cdot  z_3) \\
 &~~~~~~~~~~~~
 ~~~~~~~~~~~~~~~~~~~~~~~~~~~~~~~~~~~~~~~~~~~~~
 +
  (p_2\cdot  z_1)(p_3\cdot  z_2)(z_1\cdot  z_3)( z_2\cdot  z_3) \Big].
\end{align}
\item 4-derivative structure
\be
{\cal A}_4= {1\over M_{\rm Pl}^\frac{D-2}{2}m^2}(p_1\cdot  z_3)(p_2\cdot  z_1)(p_3\cdot  z_2 )\Big[(p_1\cdot  z_3)(
z_1\cdot  z_2) + (p_3\cdot  z_2)(z_1\cdot  z_3) + 
(p_2\cdot  z_1)(z_2\cdot z_3)\Big].
\ee
\item 6-derivative structure
\be  
{\cal A}_5={1\over M_{\rm Pl}^\frac{D-2}{2}m^4}\left(p_1\cdot z_3\right)^2\left(p_2\cdot z_1\right)^2\left(p_3\cdot z_2\right)^2. 
\ee
\end{itemize}

Here we have chosen all the amplitudes to scale with a power of some Planck mass, $M_{\rm Pl}$, and momenta to scale with powers of $1/m$.  There is no loss of generality in these assignments because in a general cubic amplitude which is a linear combination of these,
\be \sum_{i=1}^5 a_i {\cal A}_i \,,\ee
where the $a_i$ are dimensionless coefficients, any different choice of scalings can be absorbed into the $a_i$.

This counting of independent structures can be understood from the fact that there are five possible cubic terms in the action for a massive spin-2 field which cannot be field redefined away or into each other:
\begin{align}
{\cal L}_1 &\sim h_{\mu\nu}^3,\\
{\cal L}_2 &\sim\sqrt{-g} \left. R\right\rvert_{(3)}, \\
{\cal L}_3& \sim  \delta^{[\mu_1}_{\nu_1} \delta^{\mu_2}_{\nu_2}\delta^{\mu_3}_{\nu_3} \delta^{\mu_4]}_{\nu_4}\partial_{\mu_1}\partial^{\nu_1}h_{\mu_2}^{\ \nu_2}h_{\mu_3}^{\ \nu_3} h_{\mu_4}^{\ \nu_4},\\
{\cal L}_4 &\sim\sqrt{-g} \left. \left(R_{\mu\nu\rho\sigma}^2-4R_{\mu\nu}^2+R^2 \right)\right\rvert_{(3)},\\
{\cal L}_5 &\sim\sqrt{-g} \left. R^{\mu\nu}_{\ \ \rho\sigma}R^{\rho\sigma}_{\ \ \alpha\beta}R^{\alpha\beta}_{\ \ \mu\nu}\right\rvert_{(3)},
\end{align}
Three of these are familiar from the massless case: ${\cal L}_2$ is the cubic part of the Einstein--Hilbert action, ${\cal L}_4$ is the cubic part of the Gauss--Bonnet term (which is trivial in $D=4$) and ${\cal L}_5$ is the cubic part of the Riemann cubed term (which is the same on-shell as Weyl cubed).  
The other two Lagrangians appear only in the massive case because they are not diffeomorphism invariant:  
${\cal L}_1$ is the cubic part of the potential in dRGT massive gravity which survives on-shell ($h_{\mu}^{\ \mu}=0)$, and ${\cal L}_3$ is the two-derivative pseudo-linear term of \cite{Folkerts:2011ev,Hinterbichler:2013eza}.

Note that the on-shell amplitudes stemming from these Lagrangians are {not} given by the ${\cal A}_i$ above in a direct manner, but instead are linear combinations of the ${\cal A}_i$. 
In detail, the amplitudes we get from the canonically-normalized expansion $g_{\mu\nu} = \eta_{\mu\nu} +2M_{\rm Pl}^{2-D\over 2}h_{\mu\nu}$ and the usual Feynman rules are
\begin{itemize}
\item $h^3$:
\begin{align}
{\cal L}_1&= 
 {m^2\over 3M_{\rm Pl}^\frac{D-2}{2}}h_{\mu\nu}^3\,  ,  \nonumber\\
 {\cal B}_1&= {2m^2\over M_{\rm Pl}^\frac{D-2}{2}}\,  z_1\cdot z_2 \ \,  z_2\cdot z_3 \ \,   z_3\cdot z_1 \, .
\end{align}
\item Einstein--Hilbert:
\begin{align} 
\nonumber
{\cal L}_2&= { M_{\rm Pl}^{D-2}\over 2}\sqrt{-g} R\big\rvert_{(3)}\, , \\
{\cal B}_2&={2\over M_{\rm Pl}^\frac{D-2}{2}}\left(  p_1\cdot z_3 \ \,  z_1\cdot z_2+ p_3\cdot z_2 \ \,  z_1\cdot z_3+ p_2\cdot z_1 \ \,  z_2\cdot z_3 \right)^2 -3 {\cal B}_1  \, .
\end{align}

\item Pseudo-linear:
\begin{align}
{\cal L}_3&= {4! \over M_{\rm Pl}^{D-2}}  \delta^{[\mu_1}_{\nu_1} \delta^{\mu_2}_{\nu_2}\delta^{\mu_3}_{\nu_3} \delta^{\mu_4]}_{\nu_4}\partial_{\mu_1}\partial^{\nu_1}h_{\mu_2}^{\ \nu_2}h_{\mu_3}^{\ \nu_3} h_{\mu_4}^{\ \nu_4} \, , \\\nonumber
{\cal B}_3&=-{1\over M_{\rm Pl}^\frac{D-2}{2}}\left[  \left( p_1\cdot z_3\right)^2\left( z_1\cdot z_2\right)^2 + \left( p_3\cdot z_2\right)^2\left( z_1\cdot z_3\right)^2+ \left( p_2\cdot z_1\right)^2\left( z_2\cdot z_3\right)^2 \right] -{1\over 2}{\cal B}_2+{3\over 2} {\cal B}_1  \, .
\end{align} 
\item Gauss--Bonnet:
\begin{align}
{\cal L}_4&= { M_{\rm Pl}^{D-2}\over m^2}\sqrt{-g} \left(R_{\mu\nu\rho\sigma}^2-4R_{\mu\nu}^2+R^2 \right)\Big|_{(3)} \, ,\\\nonumber
{\cal B}_4&= -{80\over M_{\rm Pl}^\frac{D-2}{2}m^2} p_1\cdot  z_3\  p_2\cdot  z_1\  p_3\cdot  z_2 \ \left(p_1\cdot  z_3 \
z_1\cdot  z_2 + p_3\cdot  z_2\  z_1\cdot  z_3 + 
   p_2\cdot  z_1\  z_2\cdot  z_3\right)  -20 {\cal B}_3 +30 {\cal B}_1 \, .
\end{align}
\item $R^3$:
\begin{align}
{\cal L}_5 &= { M_{\rm Pl}^{D-2}\over m^4}\sqrt{-g} \left. R^{\mu\nu}_{\ \ \rho\sigma}R^{\rho\sigma}_{\ \ \alpha\beta}R^{\alpha\beta}_{\ \ \mu\nu}\right|_{(3)} \, , \nonumber\\
{\cal B}_5 &= {48\over M_{\rm Pl}^\frac{D-2}{2}m^4}\left(p_1\cdot z_3\right)^2\left(p_2\cdot z_1\right)^2\left(p_3\cdot z_2\right)^2 -{3\over 10} {\cal B}_4+6 {\cal B}_2+12 {\cal B}_1\, .
\end{align}
\end{itemize}
The higher-derivative Lagrangian terms come with amplitudes that have lower derivative ``tails" stemming from the on-shell conditions $\square\rightarrow m^2$.
We can write the Lagrangian amplitudes as linear combinations of the structure amplitudes,
\begin{align}
{\cal B}_1&=2{\cal A}_1 \, ,\nonumber\\
{\cal B}_2&= 2{\cal A}_2+4 {\cal A}_3-6 {\cal A}_1 \, , \nonumber\\
{\cal B}_3&= -2 {\cal A}_2-2{\cal A}_3+6 {\cal A}_1 \, ,\nonumber\\
{\cal B}_4&= -80 {\cal A}_4+40 {\cal A}_2+40 {\cal A}_3-60 {\cal A}_1\, ,\nonumber\\
{\cal B}_5&=48 {\cal A}_5+24 {\cal A}_4+12 {\cal A}_3+6 {\cal A}_1\, ,
\end{align}
and, inversely, the structures amplitudes in terms of the Lagrangian amplitudes,
\begin{align}
 {\cal A}_1&={ {\cal B}_1\over 2} \, ,\nonumber\\
{\cal A}_2&=  -{
   {\cal B}_2\over 2} + {3 {\cal B}_1\over 2} - 
  {\cal B}_3 \, ,\nonumber\\
{\cal A}_3&= {{\cal B}_2\over 2 }+ {{\cal B}_3\over 
  2} \, ,\nonumber\\
{\cal A}_4&= -{{\cal B}_4\over 80} + {3 {\cal B}_1\over 8 }- 
 { {\cal B}_3\over 4}\, ,\nonumber\\
{\cal A}_5&= -{{\cal B}_2\over 8} + 
  {{\cal B}_4\over 160 }- {{\cal B}_1\over 4 }+ {{\cal B}_5\over 48}\, .
\end{align}
If we write a general three-point amplitude as a linear combination of these objects with dimensionless coefficients,
\be 
{\cal A}=\sum_{i=1}^5 b_i {\cal B}_i=\sum_{i=1}^5 a_i {\cal A}_i \, ,
\ee
then the coefficients, $a_i$, of the structure basis can be written in terms of the coefficients, $b_i$, of the Lagrangian basis as
\begin{align}
a_1 &= 2 b_1 - 6 b_2 + 6 b_3 - 60 b_4 + 
  6 b_5\, , \nonumber\\
  a_2 &= 2 b_2 - 2 b_3 + 40 b_4\, , \nonumber\\
  a_3 &=  4 b_2 - 2 b_3 + 40 b_4 + 
  12 b_5 \, ,\nonumber\\
  a_4 &= -80 b_4 + 24 b_5 \, , \nonumber\\ 
  a_5 &= 48 b_5\, ,
\end{align}
 and inversely:
\begin{align}
 b_1 &= 
 {a_1\over 2} + {3 a_2\over 2} + {3 a_4\over 8} - {a_5\over 4}\, ,\nonumber\\
 b_2 &= -{a_2\over 2} + {a_3\over 2} -{ a_5\over 8}\, , \nonumber\\
 b_3 &= -a_2 + {a_3\over 2 }-{ a_4\over 4} \, ,\nonumber\\
 b_4 &= -{a_4\over 80} +{ a_5\over 
  160} \, ,\nonumber\\
  b_5 &= {a_5\over 48 }\, .
\end{align}

We have enumerated only the completely symmetric structures relevant to a single massive spin-2, but in the case of multiple spin-2's there are additional structures which are not completely symmetric under interchange.  It would be interesting to revisit these in the future.

\subsection{Coupling to Scalar Particles}
We will also be interested in the eikonal scattering amplitude between a massive spin-2 and a scalar particle.   For this we will we need the possible cubic vertices between a single massive spin-2 and two identical scalars.  Using the construction of Section~\ref{sec:onshell3ptvert}, there is a unique such cubic vertex that is symmetric under interchanging the scalars, and it takes the form
\be
{\cal A}_{s} = -\frac{2}{M_{\rm Pl}^\frac{D-2}{2}}(z_2\cdot p_1)(z_2\cdot p_3)\,.
\label{eq:scalarscalarspin2vertex}
\ee

This structure, and the normalization we have chosen for it, corresponds to the amplitude obtained from the cubic part of the minimal coupling of a canonically-normalized scalar of mass $M$,
\be {\cal L}_s= -{1\over 2}\sqrt{-g}\Big((\partial\phi)^2+M^2\phi^2\Big)\Big\rvert_{(3)}\, .\ee
The only cubic coupling of a massive graviton to matter is therefore also diffeomorphism invariant and gives nothing new beyond what is familiar from ordinary General Relativity.\footnote{In particular, this implies that for scalars the doubly coupled matter scenarios in massive gravity \cite{deRham:2014naa,Hinterbichler:2015yaa} do not give anything new beyond minimal coupling at the on-shell 3-point level.}

\section{Spin-2--Spin-2 Eikonal Scattering}

We start with the case of pure massive spin-2 scattering.  We compute the following $t$-channel tree diagram in the eikonal limit, using the kinematics of Section~\ref{sec:eikonlkin}:

\begin{center}
\includegraphics[height=1.1in,width=1.7in]{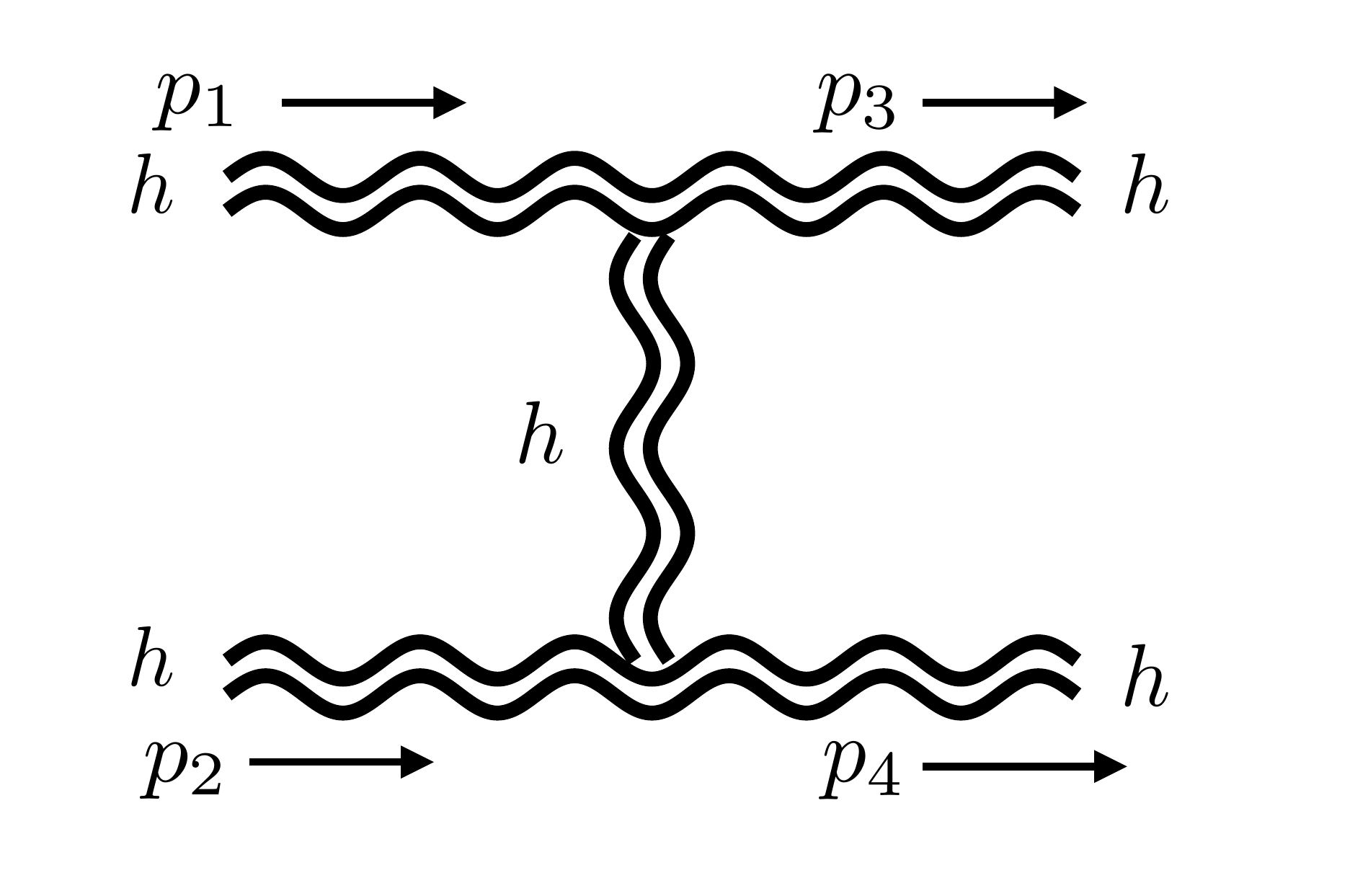}
\end{center}

For the vertices, we allow a general linear combination of all 5 three-point amplitudes to be present in the theory, so that both cubic interactions are a generic sum of all the possible three-point structures enumerated in Section~\ref{sec:3ptstructures},
\be
\label{eq:spin2vertex}
{\cal V}_g=i\sum_{i=1}^5 a_i {\cal A}_i  \,.
\ee
For the external tensor states, we consider a general linear combination of the possible polarizations, so for the $a$-th particle we have
\be
\epsilon^a_{\mu\nu} = P_{a,S} \epsilon^{a,S}_{\mu\nu}+P_{a,V} \epsilon^{a,V}_{\mu\nu}+P_{a,T} \epsilon^{a,T}_{\mu\nu},
\label{eq:generalpol}
\ee
where the polarization tensors are defined as in Section~\ref{sec:eikonlkin} and $P_{a,S}$, $P_{a,V}$, $P_{a,T}$ are some set of coefficients normalized so that $\left\lvert P_{a,S}\right\rvert^2+\left\lvert P_{a,V}\right\rvert^2+\left\lvert P_{a,T}\right\rvert^2=1$, which we may assemble into a unit norm polarization vector
\be
\label{eq:Pa}
{\bf P}_a=\left(\begin{array}{c}
P_{a,S}  \\
 P_{a,V} \\
 P_{a,T} 
\end{array}
\right) \, .
\ee

The $1,3$ and $2,4$ parts of the amplitude factorize making the structure of the eikonal amplitude relatively simple.  The amplitude in the eikonal limit takes the form
\be
{\cal M}_4 = \frac{s^2}{M_{\rm Pl}^{D-2}}\,{\bf P}_{3,4}^{\rm T}\, \hat  {\cal S}(\vec e_1,\vec e_3,- i\vec \partial_b)\otimes\hat {\cal S}(\vec e_2,\vec e_4, i\vec \partial_b)\,{\bf P}_{1,2}
\frac{1}{2\, \pi^\frac{D-2}{2}}\left(\frac{m}{b}\right)^\frac{D-4}{2}K_\frac{D-4}{2}(mb) \, .
\label{eq:generalpoleikamp}
\ee
Here, the polarization vector ${\bf P}_{a,a'} $ is a direct product of the vector of polarization coefficients ${\bf P}_{a,a'} = {\bf P}_a\otimes {\bf P}_{a'}$.
Similarly, $\hat {\cal S}\otimes\hat {\cal S}$ is the tensor product of two copies of the matrix $\hat {\cal S}(\vec e_a,\vec e_{a'}, i\vec \partial_b)$, which is given by
\small
\be
\left(\begin{array}{ccc}
{\cal C}_{SSSS} & -\frac{{\cal C}_{SV}}{m}\vec e_a\cdot\vec \partial_b & -\frac{{\cal C}_{ST}}{m^2}e_a^{ij}\partial_{b^i}\partial_{b^j}\\
-\frac{{\cal C}_{SV}}{m}\vec e_{a'}\cdot\vec \partial_b &{\cal C}_{VV_1}  \vec e_a\cdot\vec e_{a'}+\frac{{\cal C}_{VV_2} }{m^2}\vec e_a\cdot\vec\partial_b \vec e_{a'}\cdot\vec\partial_b&\frac{{\cal C}_{TV} }{m}e_a^{ij}e_{a'}^i\partial_{b^j}+\frac{a_5}{2\sqrt 2m^3}e_a^{ij}e_{a'}^k\partial_{b^i}\partial_{b^j}\partial_{b^k}\\
-\frac{{\cal C}_{ST}}{m^2}e_{a'}^{ij}\partial_{b^i}\partial_{b^j} &\frac{{\cal C}_{TV} }{m}e_{a'}^{ij}e_a^i\partial_{b^j}+\frac{a_5}{2\sqrt 2m^3}e_{a'}^{ij}e_a^k\partial_{b^i}\partial_{b^j}\partial_{b^k} & {a_2\over 2} e_a^{ij} e_{a'}^{ij}+\frac{a_4}{2m^2}e_a^{ij}e_{a'}^{jk}\partial_{b^i}\partial_{b^k}+\frac{a_5}{2m^4}e_a^{ij}e_{a'}^{kl}\partial_{b^i}\partial_{b^j}\partial_{b^k}\partial_{b^l}
\end{array}
\right).
\label{eq:genpolmatrix}
\ee
\normalsize
In the first tensor factor, we send $i\partial_b\to -i\partial_b$ to account for the fact that the internal momentum, $\vec q$, flows out of rather than  into the vertex.
The various coefficients that appear in the matrix, $\hat{\cal S}$, are given by
\begin{align}
\nonumber
{\cal C}_{SSSS} &= \frac{24(D-2)a_1+4(11D-2)a_2-4(5D-2)a_3+\frac{2(12-4D+D^2)}{D-2}a_4+\frac{(D+2)^2}{D-2}a_5}{2^{4}(D-1)}\, ,\\\nonumber
{\cal C}_{SV} &= \frac{1}{8}\sqrt\frac{1}{2(D-2)(D-1)}\left[4(D-2)a_1+8(2D-3)a_2+4(3-2D)a_3+Da_4+(D+2)a_5\right]\, ,\\
{\cal C}_{ST} &=  \frac{1}{8}\sqrt\frac{1}{(D-2)(D-1)}\left[8(D-2)a_2-4(D-2)a_3+4a_4+(D+2)a_5\right]\, ,\\\nonumber
{\cal C}_{VV_1}   &= \frac{(4a_1+12a_2-4a_3+a_4)}{16}\, ,~~~~~~~~~~~~~~{\cal C}_{VV_2}  = \frac{(4a_2-2a_3+a_4+2a_5)}{8}\, , \\\nonumber
{\cal C}_{TV}  &=  \frac{(4a_2-2a_3+a_4)}{4\sqrt 2}\, .
\end{align}

\subsection{Scattering in $D=4$\label{gravfravd4s}}
In order to determine the asymptotic phase shifts we must diagonalize the amplitude~\eqref{eq:generalpoleikamp}. Let us for the time being specialize to the case of $D=4$.  
Since the Gauss--Bonnet combination does not contribute in this case, we set $b_4 = -\frac{a_4}{80}+\frac{a_5}{160} = 0$, which we can think of as fixing $a_4$ in terms of $a_5$.\footnote{In addition to diagonalizing the amplitude order-by-order as we do in this Section, we have explicitly constructed the full $25\times25$ matrix of possible scattering processes, verified that it does not depend on the Gauss--Bonnet combination and then diagonalized it directly. The results are in accord with those reported in this Section.}  

We consider the small $mb$ limit of the amplitude~\eqref{eq:generalpoleikamp}; in this limit, the dominant sector is the tensor sector where the amplitude takes the approximate form
\be
{\cal M}_4 \simeq \frac{s^2}{4M_{\rm Pl}^{D-2}}\frac{a_5^2}{m^8}e_1^{ij}e_2^{kl}e_3^{mn}e_4^{op}\partial_{b^i}\partial_{b^j}\partial_{b^k}\partial_{b^l}\partial_{b^m}\partial_{b^n}\partial_{b^o}\partial_{b^p}\frac{1}{2\pi}K_0(mb).
\ee
Since we are only interested in the $mb\ll 1$ limit of this expression, we can replace the Bessel function with its small argument expansion (which is just the massless propagator) to obtain
\be
{\cal M}_4 \simeq -\frac{s^2}{8\pi M_{\rm Pl}^2}\frac{a_5^2}{m^8}e_1^{ij}e_2^{kl}e_3^{mn}e_4^{op}\partial_{b^i}\partial_{b^j}\partial_{b^k}\partial_{b^l}\partial_{b^m}\partial_{b^n}\partial_{b^o}\partial_{b^p}\log \left(mb\right).
\ee
This is the same expression that appears in the massless case, and it was already argued in~\cite{Camanho:2014apa} that this leads to a time advance for some polarizations. Here we will give a slightly different argument with the same conclusion. In $D=4$ there are two possible tensor polarizations, $\oplus$ and $\otimes$, and their contractions with derivatives with respect to the impact parameter take the form
\begin{align}
D_\oplus &\equiv e_\oplus^{ij}\partial_{b^i}\partial_{b^j} = \frac{1}{\sqrt 2}\left(\partial_{b^1}^2-\partial_{b^2}^2\right)\, ,\\
D_\otimes &\equiv e_\otimes^{ij}\partial_{b^i}\partial_{b^j} = \sqrt 2 \partial_{b^1}\partial_{b^2}.
\end{align}

We choose, without loss of generality, to have $\vec b$ point along the $\hat x$ axis.  Since we will then be setting $\vec b = b \hat x$ after taking derivatives, any expression which has an odd number of $D_\otimes$ operators acting on it will vanish. Note also that the amplitude is symmetric in all the external polarizations, so there are only 3 independent quantities we have to compute:
\begin{align}
D_\oplus D_\oplus D_\oplus D_\oplus \log \left(mb\right) &=D_\otimes D_\otimes D_\otimes D_\otimes \log \left(mb\right)= -\frac{20160}{b^8} \, ,\\
D_\oplus D_\oplus D_\otimes D_\otimes \log \left(mb\right) &=  \frac{20160}{b^8}.
\end{align}
There is then a $4\times 4$ matrix of possible scattering combinations,
\be
{\cal M}_4 = \frac{2520s^2}{\pi M_{\rm Pl}^2}\frac{a_5^2}{(mb)^8} 
\left(\begin{array}{c}
P_{3\oplus}P_{4\oplus}\\
P_{3\oplus}P_{4\otimes}\\
P_{3\otimes}P_{4\oplus}\\
P_{3\otimes}P_{4\otimes}
\end{array}
\right)^{\rm T}
\left(
\begin{array}{cccc}
1 & 0 & 0&  -1\\
0& -1& -1 & 0\\
0& -1 & -1 & 0\\
-1 &  0 & 0 & 1
\end{array}
\right)
\left(\begin{array}{c}
P_{1\oplus}P_{2\oplus}\\
P_{1\oplus}P_{2\otimes}\\
P_{1\otimes}P_{2\oplus}\\
P_{1\otimes}P_{2\otimes}
\end{array}
\right).
\ee
Diagonalizing this matrix is fairly straightforward; it is degenerate and there are only two eigenvalues, leading to the phase shifts
\be
\label{eq:a5const}
\delta(s,b) = \pm \frac{2520s}{\pi M_{\rm Pl}^2}\frac{a_5^2}{(mb)^8}.
\ee
Since the phase shifts come with opposite signs, we see that one linear combination of polarizations will always get a time advance unless we take $a_5 = 0$.\footnote{Here and in what follows we are going to demand that  various coefficients vanish, but what we really mean is that the coefficient of the phase shift must be so small that the time delay is smaller than the inverse cutoff of the effective theory. For massive gravity, the time delays will scale like $p^{\scriptscriptstyle +}\delta t \sim (mb)^{-p}$, with $p$ some power. In order for our calculation to be reliable, at most $p^{\scriptscriptstyle +}$ can be of order the cutoff of the theory, $\Lambda_{\rm c}$. Since $mb\ll 1$, the numerical coefficients have to be quite small to prevent $\delta t\Lambda_{\rm c}$ from growing to be $\sim 1$. For example, a graviton with mass of order Hubble today can have $mb$ as small as $10^{-10}$ and still easily satisfy $b^{-1} \ll \Lambda_{5} = (m^4M_{\rm Pl})^{1/5}$, and so is well within the regime of validity of the EFT. Therefore, the coefficients of cubic structures leading to a time advance have to be also at least ${\cal O}(10^{-10})$, so for practical purposes we call this zero.
} (Note that this also fixes $a_4 = 0$ in $D=4$.)  This sets to zero the Riemann cubed part of the vertex.

After setting $a_5 = 0$, the leading amplitude at small impact parameter is of the form
\begin{align}
{\cal M}_4\simeq -\frac{(2a_2-a_3)^2s^2}{2\pi m^4M_{\rm Pl}^2}
&\left(\frac{1}{\sqrt 6}P_{1S}P_{3T}e_3^{ij}\partial_{b^i}\partial_{b^j}+\frac{1}{\sqrt 6}P_{3S}P_{1T}e_1^{ij}\partial_{b^i}\partial_{b^j}-\frac{1}{4}P_{1V}P_{3V} e_1^i e_3^j\partial_{b^i}\partial_{b^j}
\right)\nn\\
\times&\left(\frac{1}{\sqrt 6}P_{2S}P_{4T}e_4^{kl}\partial_{b^k}\partial_{b^l}+\frac{1}{\sqrt 6}P_{4S}P_{2T}e_2^{kl}\partial_{b^k}\partial_{b^l}-\frac{1}{4}P_{2V}P_{4V} e_2^k e_4^l\partial_{b^k}\partial_{b^l}
\right)\log \left(mb\right)\, . 
\end{align}
Thought of as a matrix, and written out in terms of explicit polarizations, this amplitude is a relatively sparse $25\times25$ matrix. Diagonalizing this matrix to find the eigenvalues of the linear combinations of polarizations which are unchanged by scattering, we find that the phase shifts of the various polarizations are given by
\be
\delta(s,b) = 
\left\{\begin{array}{l}
\pm \frac{(2a_2-a_3)^2s}{\sqrt{2}\pi M_{\rm Pl}^2} \frac{1}{(mb)^4}\, ,\\
\pm \frac{3(2a_2-a_3)^2s}{16\pi M_{\rm Pl}^2} \frac{1}{(mb)^4}\, ,\\
\pm \frac{(2a_2-a_3)^2s}{2\pi M_{\rm Pl}^2} \frac{1}{(mb)^4}\, ,~~~~~~~~{\rm with~multiplicity~2}\, ,\\
\pm \frac{\sqrt{3\over 2}(2a_2-a_3)^2s}{4\pi M_{\rm Pl}^2} \frac{1}{(mb)^4}\, ,~~~~{\rm with~multiplicity~4}\, .\\
\end{array}\right.
\ee
We see that unless we set $a_3 = 2a_2$, some linear combinations of polarizations will acquire a time advance.  This sets to zero the pseudo-linear cubic vertex.

After setting $a_5 = a_4 = 2a_2-a_3 = 0$, the leading contribution to scattering comes from scalar-tensor mixing and the amplitude takes the form
\be
{\cal M}_4 = \frac{a_1^2 s^2}{24\pi m^2 M_{\rm Pl}^2}\left(P_{1V}P_{3S}e_1^i\partial_{b^i}+P_{3V}P_{1S}e_3^i\partial_{b^i}\right)\left(P_{2V}P_{4S}e_2^i\partial_{b^i}+P_{4V}P_{2S}e_4^i\partial_{b^i}\right)\log \left(mb\right).
\ee
We diagonalize this amplitude in the same way as above and find the phase shifts
\be
\delta(s,b) = 
\left\{\begin{array}{l}
\pm\frac{a_1^2 s}{24 \sqrt 2\pi  M_{\rm Pl}^2} \frac{1}{(mb)^2}\, ,\\
\pm\frac{a_1^2 s}{48\pi M_{\rm Pl}^2}\frac{1}{(mb)^2}\, ,~~~~~~~~{\rm with~multiplicity~2}\, .\\
\end{array}\right.
\ee
We see that some of the polarizations will obtain a time advance unless $a_1 = 0$.  This fixes the relative coefficient between the cubic part to the Einstein--Hilbert vertex and the cubic potential term $h_{\mu\nu}^3$.

After making all these parameter choices, the eikonal amplitude is completely diagonal
\begin{align}
{\cal M}_4 = \frac{a_2^2s^2}{16 M_{\rm Pl}^{2}}&\left(P_{1S}P_{3S}+P_{1V}P_{3V} \vec e_2\cdot \vec e_4+2P_{1T}P_{3T}e_2^{ij}e_4^{ij}\right)\nonumber\\
&\times\left(P_{2S}P_{4S}+P_{2V}P_{4V} \vec e_2\cdot \vec e_4+2P_{2T}P_{4T}e_2^{ij}e_4^{ij}\right)\frac{1}{2\pi}K_0(mb),
\end{align}
with positive entries.  Thus all the polarizations will receive asymptotic time delays and there are no further constraints.

Summarizing, the total constraints we find are
\be
a_1 = 0 \, ,~~ a_3 = 2a_2\, ,~~a_4 = a_5 =0 \,.
\label{eq:s2s2D4const}
\ee
Translated into the Lagrangian basis, this corresponds to a cubic Lagrangian of the form:
\be
\label{eq:cubiclinearcombination}
{\cal L}_3 = \frac{a_2}{2}\left(\frac{1}{M_{\rm Pl}}R_{\rm EH}^{(3)} + \frac{m^2}{M_{\rm Pl}}h_{\mu\nu}^3\right) \, .
\ee
Canonically normalized gravity corresponds to $a_2 = 2$.  One interesting thing to note about this combination of terms is that the on-shell cubic amplitude from this linear combination of cubic vertices is the same form as that of the Einstein--Hilbert term in the theory of a massless spin-2, as can be verified by looking at Section \ref{sec:3ptstructures}. In the massive theory, there is a contribution from the Einstein--Hilbert vertex which is proportional to $m^2$, but this is precisely canceled off by this particular choice of $h_{\mu\nu}^3$ coefficient.

\subsection{Amplitude in General $D$}

Diagonalizing the amplitude~\eqref{eq:generalpoleikamp} in general $D$ explicitly is an intricate task. However, it is straightforward to insert the constraints from $D=4$ into the general dimension amplitude and check if the result is sub-luminal. Plugging in the parameter values~\eqref{eq:s2s2D4const}, we obtain
\begin{align}
\label{eq:s2s2Damp}
{\cal M}_4 = \frac{a_2^2s^2}{M_{\rm Pl}^{D-2}}&\left(P_{1S}P_{3S}\frac{D+2}{8(D-1)}+\frac{1}{4}P_{1V}P_{3V} \vec e_2\cdot \vec e_4+\frac{1}{2}P_{1T}P_{3T}e_2^{ij}e_4^{ij}\right)\\\nonumber
&\times\left(P_{2S}P_{4S}\frac{D+2}{8(D-1)}+\frac{1}{4}P_{2V}P_{4V} \vec e_2\cdot \vec e_4+\frac{1}{2}P_{2T}P_{4T}e_2^{ij}e_4^{ij}\right)\frac{1}{2\pi^\frac{D-2}{2}}\left(\frac{m}{b}\right)^\frac{D-4}{2}K_\frac{D-4}{2}(mb).
\end{align}
This amplitude is diagonal and positive, and so the phase shifts are all positive and we see that all of the polarizations experience an asymptotic time delay. There is still the question of whether this is the most general possible amplitude which is consistent with positivity in general dimension.  In order to answer this question, we turn to a slightly different computation; we compute the eikonal amplitude for a massive spin-2 scattering off of a scalar particle. This amplitude is effectively a subsector of~\eqref{eq:generalpoleikamp} and so the constraints in this case must also be satisfied by~\eqref{eq:generalpoleikamp} in order for the theory to experience time delays. We will see that the constraints are the same as~\eqref{eq:s2s2D4const}. Additionally, this calculation is closely related to the Shapiro time delay experienced by a massive graviton propagating in a shockwave background.

\section{Scalar--Spin-2 Eikonal Scattering}
\label{sec:scalarspin2}

We now restrict our attention to the eikonal scattering between a scalar particle, $\phi$, and a massive graviton.  This is effectively a subsector of the previous amplitude where we average over the polarizations of one of the external gravitons so that it acts a scalar source~\cite{Camanho:2014apa}.\footnote{Alternatively, we could imagine sending in a classical coherent state with the scalar polarization: this will cause the final state to also be the scalar polarization, and we will get the same subsector of the amplitude.}
We must again compute the following $t$-channel tree diagram in the eikonal limit, using the kinematics of Section \ref{sec:eikonlkin}:

%\vspace{.4cm}
\begin{center}
\includegraphics[height=1.1in,width=1.7in]{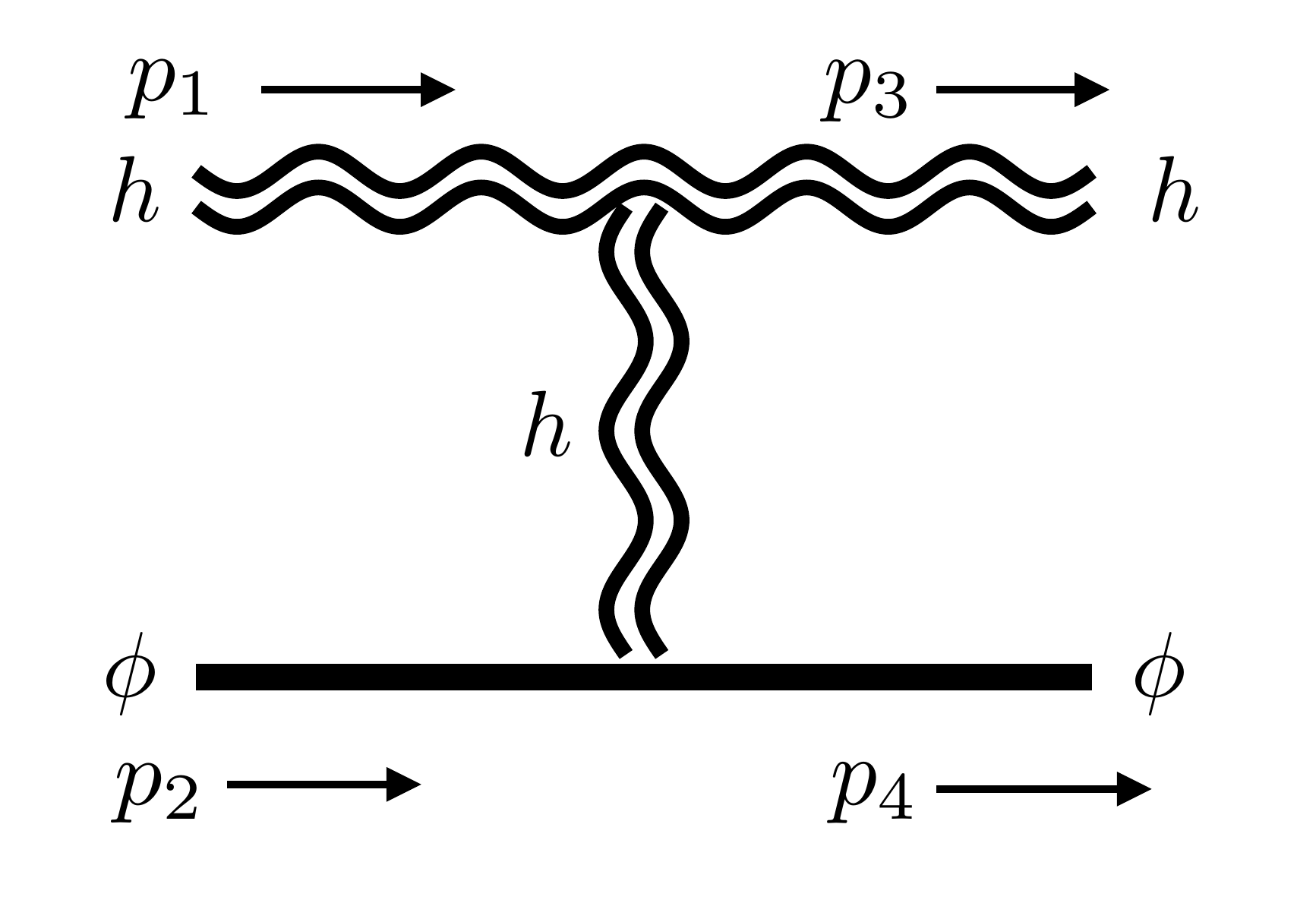}
\end{center}

There are now two types of cubic vertices present in the computation.  In addition to the cubic massive graviton vertex \eqref{eq:spin2vertex} there is also the scalar-scalar-tensor interaction \eqref{eq:scalarscalarspin2vertex},
\be {\cal V}_s=-i\frac{2\,{\cal C}_s}{M_{\rm Pl}^\frac{D-2}{2}}(z_2\cdot p_1)(z_2\cdot p_3) \, ,\ee
where we have allowed for an arbitrary coefficient, ${\cal C}_s$.   The usual canonically-normalized scalar corresponds to ${\cal C}_s=1$.

\subsection{Eikonal Amplitude in General Dimension\label{spin2scalargenssc}}

In the eikonal limit, the 4-point amplitude takes the form of a differential operator acting on a Bessel-$K$ function
\be
\label{eq:generalamplitude}
{\cal M}_4 = \frac{{\cal C}_ss^2}{M_{\rm Pl}^{D-2}}{\bf P}_3^{\rm T}\hat {\cal S}(\vec e_1,\vec e_3, i\vec\partial_b)\, {\bf P}_1
\frac{1}{2\, \pi^\frac{D-2}{2}}\left(\frac{m}{b}\right)^\frac{D-4}{2}K_\frac{D-4}{2}(mb) \, .
\ee
The vector of polarization coefficients ${\bf P}_a$ is defined as above in \eqref{eq:Pa} and the matrix of operators $ \hat {\cal S}(\vec e_1,\vec e_3, i\vec \partial_b)$ is defined as in \eqref{eq:genpolmatrix}.  Note that this amplitude does not depend on the mass of the scalar particle.

Again we must diagonalize this amplitude in order to extract the phase shifts, and we will do so order-by-order in the small parameter, $mb$.  At small $mb$, the leading process in the general amplitude~\eqref{eq:generalamplitude} is pure tensor scattering
\be
{\cal M}_4 \rightarrow P_{1T}P_{3T}\frac{{\cal C}_ss^2}{4\, \pi^\frac{D-2}{2}M_{\rm Pl}^{D-2}}\frac{a_5}{m^4}e_1^{ij}e_3^{kl}\partial_{b^i}\partial_{b^j}\partial_{b^k}\partial_{b^l}\left[\left(\frac{m}{b}\right)^\frac{D-4}{2}K_\frac{D-4}{2}(mb)\right],
\ee
and the tensor polarizations only mix with themselves. Acting with the differential operators, we obtain the expression
\begin{align}
\nonumber
{\cal M}_4\simeq &~ P_{1T}P_{3T}\frac{{\cal C}_ss^2}{4\, \pi^\frac{D-2}{2}M_{\rm Pl}^{D-2}}\frac{a_5}{m^4}\bigg[ 2e_1^{ij}e_3^{ij}\left(\frac{m}{b}\right)^\frac{D}{2}K_\frac{D}{2}(mb)-4 e_1^{ik}e_3^{kj}b^ib^j\left(\frac{m}{b}\right)^\frac{D+2}{2}K_\frac{D+2}{2}(mb)\\
&+e_1^{ij}e_3^{kl} b^ib^jb^kb^l\left(\frac{m}{b}\right)^\frac{D+4}{2}K_\frac{D+4}{2}(mb) \bigg] \,.
\label{eq:TTb4}
\end{align}

Without loss of generality, we now chose the impact parameter to point in the first direction, 
\be 
\vec b=(b,0,0,\cdots).
\ee
With this choice, the various $\oplus$ polarizations mix amongst themselves, but the $\otimes$ polarizations are diagonal and don't mix with the $\oplus$ polarizations. It is therefore convenient to consider the eigenvalues of the $\otimes$ subspace first. Notice that only the $\otimes_{1i}$ polarizations will have a nonzero
$1\,1$ component in $e_1^{ik}e_3^{kj}$, so that only these polarizations receive contributions from the second term in~\eqref{eq:TTb4}. Since none of the $\otimes$ polarizations have a nonzero $1\,1$ component, none of them will receive contributions from $e_1^{ij}e_3^{kl} b^ib^jb^kb^l$.
We can then compute the matrix elements for the various $\otimes$ polarizations as
\begin{align}
T_{\otimes_{1i}} T_{\otimes_{1i}} &= \frac{a_5{\cal C}_ss^2}{4\, \pi^\frac{D-2}{2}M_{\rm Pl}^{D-2}}\left[\frac{2}{m^4}\left(\frac{m}{b}\right)^\frac{D}{2}K_\frac{D}{2}(mb)-\frac{2}{m^2}\left(\frac{m}{b}\right)^\frac{D-2}{2}K_\frac{D+2}{2}(mb)\right]\, ,\\
T_{\otimes_{ab\neq1i}} T_{\otimes_{ab\neq1i}} &= \frac{a_5{\cal C}_ss^2}{4\, \pi^\frac{D-2}{2}M_{\rm Pl}^{D-2}}\frac{2}{m^4}\left(\frac{m}{b}\right)^\frac{D}{2}K_\frac{D}{2}(mb)\, .
\end{align}
Since this matrix is already diagonal, we just have to take the small $b$ limit of each of the diagonal entries. We thus find the following phase shifts from the eigenvalues 
\begin{align}
\delta_{T_{\otimes_{1i}} T_{\otimes_{1i}} }&= -\frac{a_5{\cal C}_s s}{8\, \pi^\frac{D-2}{2}M_{\rm Pl}^{D-2}} \frac{2^\frac{D}{2}(D-1)\Gamma\left[\frac{D}{2}\right]}{m^4 b^D}\, ,\\\
\delta_{T_{\otimes_{ab\neq1i}} T_{\otimes_{ab\neq1i}}} &= \frac{a_5{\cal C}_s s}{8\, \pi^\frac{D-2}{2}M_{\rm Pl}^{D-2}} \frac{2^\frac{D}{2}\Gamma\left[\frac{D}{2}\right]}{m^4 b^D}\, .
\end{align}
Note that these have opposite signs, so at least one of them will lead to a time advance.  Forbidding this, we get our first constraint 
\be a_5 = 0\, .\ee
This is consistent with the general pure spin-2 result~\eqref{eq:a5const} and sets the coefficient of the Riemann cubed vertex to zero.

After setting $a_5$ to zero, the leading-order terms in the amplitude at small impact parameter take the form
\be
\nonumber
{\cal M}^{IJ}_4 \rightarrow
\left(\begin{array}{ccc}
0 & 0 & -\frac{{\cal C}_{ST}}{m^2}e_1^{ij}\partial_{b^i}\partial_{b^j}\\
0 &\frac{{\cal C}_{VV_2} }{m^2}\vec e_1\cdot\vec\partial_b \vec e_3\cdot\vec\partial_b&0\\
-\frac{{\cal C}_{ST}}{m^2}e_3^{ij}\partial_{b^i}\partial_{b^j} &0 & \frac{a_4}{2m^2}e_{1}^{ij}e_3^{jk}\partial_{b^i}\partial_{b^k}\end{array}
\right)\frac{{\cal C}_ss^2}{2\, \pi^\frac{D-2}{2}M_{\rm Pl}^{D-2}}\left(\frac{m}{b}\right)^\frac{D-4}{2}K_\frac{D-4}{2}(mb)\,.
\ee
There are various simplifications: scattering between $\otimes$ polarizations is diagonal and the only mixing between scalar polarizations and tensor polarizations is with the various $\oplus$ polarizations. Using the basis for vector polarizations introduced in Section~\ref{sec:eikonlkin}, the vector polarizations do not mix under scattering either. It is most convenient to consider vector scattering and $\otimes$ tensor scattering. The relevant matrix elements are
\begin{align}
V_1 V_1 &= \frac{{\cal C}_s{\cal C}_{VV_2} s^2}{2\, \pi^\frac{D-2}{2}M_{\rm Pl}^{D-2}}\left[-\frac{1}{m^2}\left(\frac{m}{b}\right)^\frac{D-2}{2}K_\frac{D-2}{2}(mb)+\left(\frac{m}{b}\right)^\frac{D-4}{2}K_\frac{D}{2}(mb)\right] \, ,\\
V_{a\neq 1}V_{a\neq 1} &= -\frac{{\cal C}_ss^2}{2\, \pi^\frac{D-2}{2}M_{\rm Pl}^{D-2}}\frac{{\cal C}_{VV_2} }{m^2}\left(\frac{m}{b}\right)^\frac{D-2}{2}K_\frac{D-2}{2}(mb)\ , \\
T_{\otimes_{1i}} T_{\otimes_{1i}}&=\frac{a_4{\cal C}_ss^2}{4\, \pi^\frac{D-2}{2}M_{\rm Pl}^{D-2}}\left[-\frac{1}{m^2}\left(\frac{m}{b}\right)^\frac{D-2}{2}K_\frac{D-2}{2}(mb)+\frac{1}{2}\left(\frac{m}{b}\right)^\frac{D-4}{2}K_\frac{D}{2}(mb)\right]\, ,\\
T_{\otimes_{ab\neq 1i}}T_{\otimes_{ab\neq 1i}} &= -\frac{a_4{\cal C}_ss^2}{4\, \pi^\frac{D-2}{2}M_{\rm Pl}^{D-2}}\frac{1}{m^2}\left(\frac{m}{b}\right)^\frac{D-2}{2}K_\frac{D-2}{2}(mb)\  .
\end{align}
Let's consider vector scattering first. In the small $mb$ limit, the eigenvalues of that block lead to the phase shifts:
\begin{align}
\delta_{V_1 V_1 } &= \frac{{\cal C}_s s}{4\, \pi^\frac{D-2}{2}M_{\rm Pl}^{D-2}}\frac{{\cal C}_{VV_2} }{m^2} 2^{\frac{D-4}{2}}(D-3)\Gamma\left[\frac{D}{2}-1\right]\frac{1}{b^{D-2}}\, ,\\
\delta_{V_{a\neq 1}V_{a\neq 1}} &= -\frac{{\cal C}_s s}{4\, \pi^\frac{D-2}{2}M_{\rm Pl}^{D-2}}\frac{{\cal C}_{VV_2} }{m^2}2^{\frac{D-4}{2}}\Gamma\left[\frac{D}{2}-1\right]\frac{1}{b^{D-2}}\, .
\end{align}
Independent of the rest of the polarizations, these two polarizations have opposite sign eigenvalues, so we must set ${\cal C}_{VV_2}  = 0$ to avoid a time advance. Since we have already set $a_5 = 0$, ${\cal C}_{VV_2} $  is given by 
${\cal C}_{VV_2}  = {(4a_2-2a_3+a_4)/8}.$
We thus obtain our second constraint:
 \be a_4 = 2a_3-4a_2\, .\ee
 This sets to zero the pseudo-linear cubic vertex.

Next we can consider the $\otimes$ polarizations. These are also diagonal and do not mix with the scalar, so we can read off their eigenvalues as well, leading to the phase shifts
\begin{align}
\delta_{T_{\otimes_{1i}}T_{\otimes_{1i}}} &=\frac{a_4}{m^2}\frac{{\cal C}_s s}{8\, \pi^\frac{D-2}{2}M_{\rm Pl}^{D-2}} 2^\frac{D-6}{2}(D-4)\Gamma\left[\frac{D}{2}-1\right]\frac{1}{b^{D-2}}\, ,\\
\delta_{T_{\otimes_{ab\neq 1i}}T_{\otimes_{ab\neq 1i}} } &= -\frac{a_4}{m^2}\frac{{\cal C}_s s}{8\, \pi^\frac{D-2}{2}M_{\rm Pl}^{D-2}}2^{\frac{D-4}{2}}\Gamma\left[\frac{D}{2}-1\right]\frac{1}{b^{D-2}}\, .
\end{align}
In $D\neq 4$ these two eigenvalues have opposite sign, so one of the two polarizations will have a time advance unless we set $a_4 = 2a_3-4a_2 = 0$, which implies that 
\be a_3 = 2a_2\, .\ee
This sets to zero the cubic vertex coming from the Gauss--Bonnet term.   (In $D=4$ we have no Gauss-Bonnet terms and these two contributions are absent.)   It is then straightforward to check that ${\cal C}_{ST} = 0$ with this parameter choice, so the scattering matrix vanishes at this order and there are no further constraints. 

With these parameter choices, the amplitude~\eqref{eq:genpolmatrix} is very nearly diagonalized. At leading order in small impact parameter, the only contribution is an off-diagonal mixing between scalar and vector polarizations:
\be
{\cal M}_4\simeq -\frac{{\cal C}_ss^2}{4\, \pi^\frac{D-2}{2}M_{\rm Pl}^{D-2}}\frac{a_1}{m}\sqrt\frac{D-2}{2(D-1)}\left(P_{1V}P_{3S} \vec e_1\cdot\vec \partial_b +P_{1S}P_{3V}\vec e_3\cdot\vec \partial_b\right)\left(\frac{m}{b}\right)^\frac{D-4}{2}K_\frac{D-4}{2}(mb)\, .
\ee
Taking the derivative, we find
$
\vec e\cdot\vec \partial_b \left(\frac{m}{b}\right)^\frac{D-4}{2}K_\frac{D-4}{2}(mb) = -\vec e\cdot\vec b \left(\frac{m}{b}\right)^\frac{D-2}{2}K_\frac{D-2}{2}(mb)\, .
$
This is only non-zero for the $V_1$ polarization, so we see that only this polarization mixes with the scalar polarization. It leads to a $2\times2$ matrix to diagonalize which is totally off-diagonal and symmetric. Its eigenvalues are therefore $\pm$ its non-zero entry, so the phase shifts at small $mb$ corresponding to the eigenvalues are
\be
\label{eq:eigen}
\delta(s,b) = \pm a_1\frac{{\cal C}_ss\,2^{(D-11)/2} }{ \pi^\frac{D-2}{2}M_{\rm Pl}^{D-2}}\sqrt\frac{D-2}{D-1} \Gamma\left[{D\over 2}-1 \right] {1\over m b^{D-3}}\, .
\ee
Since they come in opposite signs, we must set $a_1 = 0$. 

To summarize, the total constraints we find are
\be
a_1 = 0 \,, ~~a_3  = 2 a_2\, ,~~a_4 = 0\, , ~~a_5 = 0\, .
\ee
These are in full agreement with those derived in the case of pure massive graviton scattering in $D=4$ in Section \ref{gravfravd4s}, but now in $D>4$ the Gauss--Bonnet cubic vertex is explicitly set to zero by the constraints (rather than being trivial to begin with).

Translated into the Lagrangian basis, this corresponds to a cubic Lagrangian of the form:
\be
\label{eq:cubiclinearcombination}
{\cal L}_3 = \frac{a_2}{2M_{\rm Pl}^\frac{D-2}{2}}R_{\rm EH}^{(3)} + \frac{a_2m^2}{2M_{\rm Pl}^\frac{D-2}{2}}h_{\mu\nu}^3 \, ,
\ee
where the canonically normalized graviton corresponds to $a_2 = 2$.  With this choice of parameters, the amplitude is diagonal:
\be
{\cal M}_4 = \frac{a_2{\cal C}_ss^2}{4\, \pi^\frac{D-2}{2}M_{\rm Pl}^{D-2}}\left(P_{1S}P_{3S}\frac{D+2}{4(D-1)}+\frac{1}{2}P_{1V}P_{3V} \vec e_1\cdot \vec e_3+P_{1T}P_{3T}e_1^{ij}e_3^{ij}\right)\left(\frac{m}{b}\right)^\frac{D-4}{2}K_\frac{D-4}{2}(mb),
\ee
so that by choosing ${\cal C}_{s}a_2 > 0$, everything is positive and all the remaining phase shifts are subluminal. (Canonically-normalized gravity minimally coupled to a canonically-normalized scalar field corresponds to ${\cal C}_s = 1$.)

\subsection{Eikonal Amplitude in $D=4$ and Connection to Shockwaves}
\label{sec:shockwaveamp}
In the context of Einstein gravity, the Shapiro time delay experienced by a particle can be computed from its propagation in a shockwave background described by the Aichelburg--Sexl metric~\cite{Aichelburg:1970dh,Dray:1984ha}. The equivalence of this calculation to the eikonal scattering calculation was shown in~\cite{Kabat:1992tb}. We would like to see that the eikonal and shockwave computations are similarly related in the massive case. We will concentrate on the case of $D=4$ dRGT massive gravity propagating in a shockwave background---as this is perhaps the most phenomenologically interesting situation---but the calculation we describe in this section can easily be generalized to a massive spin-2 theory with arbitrary cubic vertices in arbitrary dimension. (In the same way, the massless spin-2 Shapiro time delay can be computed with higher curvature terms~\cite{Papallo:2015rna,Benakli:2015qlh}.)

On the amplitude side, we first specialize the scalar--spin-2 scattering calculation of Section \ref{spin2scalargenssc} to a theory whose only spin-2 cubic vertices are the cubic Einstein--Hilbert vertex and the cubic potential term $h_{\mu\nu}^3$ in $D=4$.  Taking the graviton to be canonically normalized, this corresponds to the choice of coefficients
\be
a_2 =2\, , ~~~a_3 = 4\, , ~~~a_4 = 0\, , ~~~a_5 = 0\, .
\ee
The only free remaining parameter among the five cubic vertices is $a_1$, which sets the coefficient of the $h_{\mu\nu}^3$ term.  With this choice, the phase shifts for the scalar, vector and tensor modes as calculated in the previous section reduce to 
\begin{align}
\label{eq:eigen1}
\delta_{S}(s,b)&=-(a_1+2)\frac{{\cal C}_s \,s}{16\pi M_{\rm Pl}^2}\log(mb)\, ,\vspace{0.2cm} \\
\label{eq:eigen2}
\delta_{V_1,V_2}(s,b)&=\pm a_1\frac{{\cal C}_s\, s}{8\pi M_{\rm Pl}^2}\frac{1}{\sqrt 3mb}\, , \vspace{0.2cm}\\
\label{eq:eigen3}
\delta_{T_\oplus,T_\otimes}(s,b)&= - \frac{{\cal C}_s \, s}{4\, \pi M_{\rm Pl}^2} \log(mb)  \, ,
\end{align}
for $mb \ll 1$.  Absence of a time advance requires $a_1 =0$, as in the previous section.

We will now see how to reproduce this result directly from the calculation of a massive graviton propagating in a shockwave geometry of the Aichelburg--Sexl type. This calculation has been performed previously in~\cite{Camanho:2016opx}.  However, since our phase shifts differ slightly from theirs, we review the calculation here.

\begin{figure}%[h!]
\begin{center}
\includegraphics[height=3.2in]{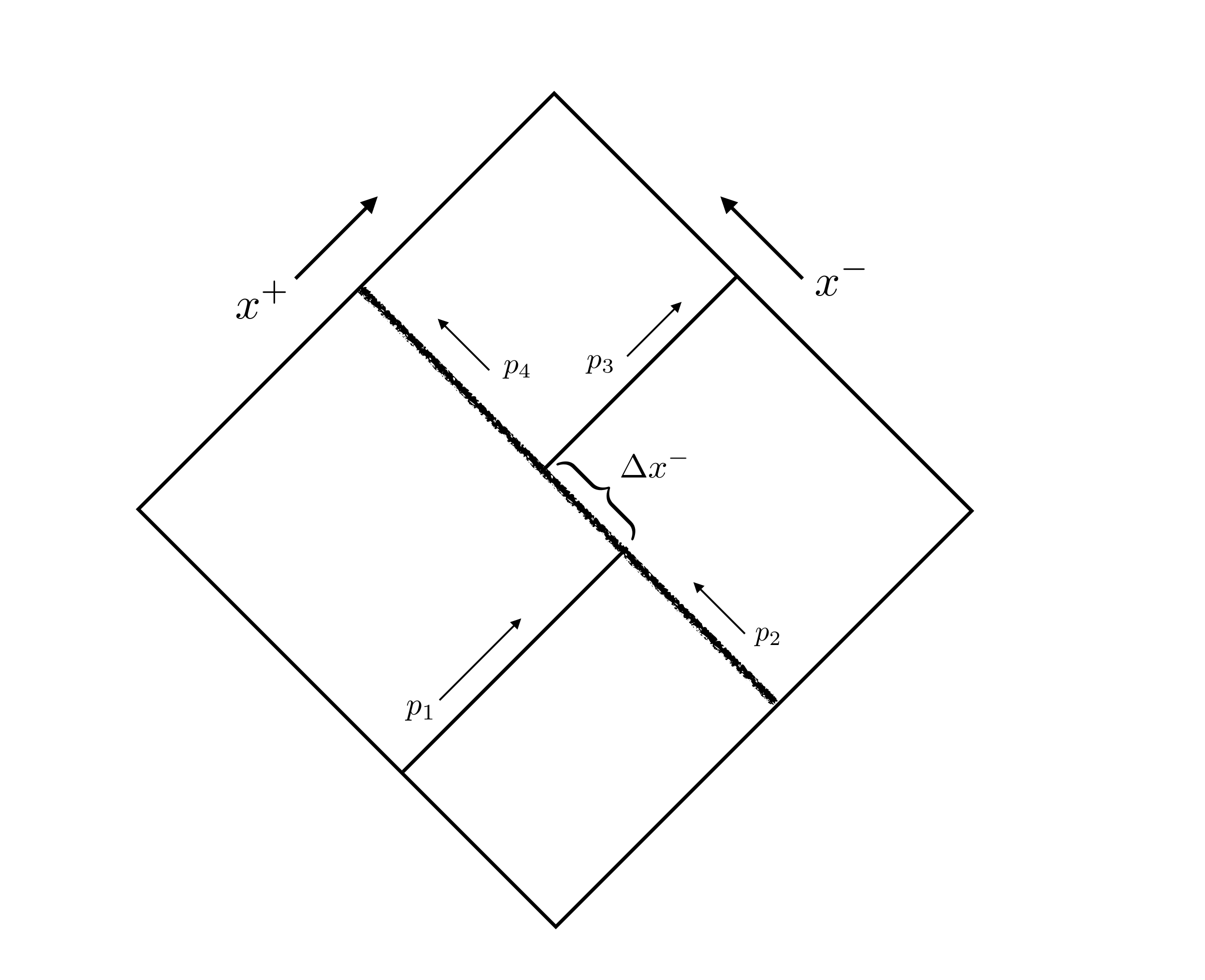}
\caption{\small The shockwave geometry and its connection to scattering.  The shockwave travels in the $x^{\scriptscriptstyle -}$ direction.  It is traversed by a particle traveling in the $x^{\scriptscriptstyle +}$ direction which experiences a time delay $\Delta x^{\scriptscriptstyle -}$ as it crosses the shock.}
\label{shockwavepic}
\end{center}
\end{figure} 

de Rham--Gabadadze--Tolley massive gravity is a two-parameter family of theories. The free parameters are often denoted by $c_3$ and $d_5$ (see, {\it e.g.}~\cite{deRham:2010ik}), where the parameter $c_3$ parametrizes cubic interactions in the potential and is related to the above coefficients by $a_1 = 3(1-4 c_3)$.  The parameter $d_5$ parametrizes quartic  interactions in the potential.  All higher order interactions in the potential are fixed after specifying $c_3$ and $d_5$.  (See Appendix~\ref{app:dRGT} for a review of the structure of dRGT.)

The analogue of the Aichelberg--Sexl metric for a massive spin-2 in lightcone coordinates takes the form
\be
\rd s^2 = -2\rd {x^{\scriptscriptstyle +}} \rd {x^{\scriptscriptstyle -}} + F({x^{\scriptscriptstyle +}},\vec x)\rd {x^{\scriptscriptstyle +}}^2 + \rd\vec x^2,  
\label{eq:massiveaichelburg}
\ee
where the function $F({x^{\scriptscriptstyle +}},\vec x)$ must satisfy a Poisson equation in the transverse variables
\be
(\nabla^2-m^2)F({x^{\scriptscriptstyle +}},\vec x) = -16 \pi G\, T_{\scriptscriptstyle ++} \, ,
\ee
where $T_{\scriptscriptstyle ++}$ is a component of the matter stress tensor $T_{\mu\nu}$ sourcing the shockwave.
Interestingly, this background is a solution to the equations of motion for a massive spin-2 with an arbitrary potential~\cite{Mohseni:2011vv}. This is because the nonlinear terms drop out of the Einstein equations and $F({x^{\scriptscriptstyle +}},\vec x)$ solves, in effect, the linear equations of motion of a massive particle.  This is a particular feature of the metric~\eqref{eq:massiveaichelburg} analogous to the Aichelburg--Sexl solution in General Relativity which also solves both the linear and fully non-linear equations.
We can write the background metric \eqref{eq:massiveaichelburg} in the Kerr--Schild form
\be
\label{eq:background}
\bar{g}_{\mu\nu} = \eta_{\mu\nu}+F({x^{\scriptscriptstyle +}},\vec x)\ell_\mu \ell_\nu \, ,
\ee
where $\ell^\mu$ is a covariantly constant null vector, $\bar\nabla_\mu\ell_\nu = \ell^2  = 0$ chosen to point in the ${x^{\scriptscriptstyle -}}$ direction
\be \ell^\mu=(1,0,\vec 0)\, ,\ee
 and take the stress tensor to be that of a point particle moving at the speed of light with energy $p^{\scriptscriptstyle -}$:\footnote{Although this source is apparently singular at $x^{\scriptscriptstyle +} = 0$---and therefore one might be concerned that the calculation leaves the regime of validity of the EFT---the final answer is sensitive only to the integral across the singularity, which is finite. We can therefore think of $p^{\scriptscriptstyle - }$ as capturing the width of the shock. Alternatively, one can use a smeared source as in~\cite{Camanho:2016opx}; this does not change the answer.
 } 
\be
T_{\mu\nu} = p^{\scriptscriptstyle -}\, \delta({x^{\scriptscriptstyle +}}) \delta(\vec x) \ell_\mu \ell_\nu\, .
\ee
This allows us to solve explicitly for $F$: 
\be
F({x^{\scriptscriptstyle +}},\vec x) =  \frac{p^{\scriptscriptstyle -}}{\pi M_{\rm Pl}^2}  \delta({x^{\scriptscriptstyle +}}) K_0 (m b) \, ,
\ee
where $b = |\vec x|$ is the impact parameter.  As expected, both parameters $c_3$ and $d_5$ drop out of this solution.

To calculate the time delay/advance experienced by a particle in this background \eqref{eq:background}, we consider fluctuations above the background so that the full metric is 
\be
g_{\mu\nu} = \bar g_{\mu\nu} +\frac{2}{M_{\rm Pl}} h_{\mu\nu} \, .
\ee
If we expand the dRGT action~\eqref{eq:drgtaction} around this background, the linearized equations of motion are\footnote{These equations are in agreement with~\cite{Camanho:2016opx} with the conversion $c_3 = (1 - \alpha)/6$. }
\begin{align}
\label{eq:generalEOM}
{\cal E}_{\mu\nu} \equiv& -\bar \nabla^2 h_{\mu\nu}+2 {\bar\nabla}^\lambda{\bar\nabla}_{(\mu}h_{\nu)\rho}
-{\bar\nabla}_\mu{\bar\nabla}_\nu h+{\bar g}_{\mu\nu}\bar \nabla^2 h
-{\bar g}_{\mu\nu}{\bar\nabla}_\lambda{\bar\nabla}_\kappa h^{\lambda\kappa}
+{\bar g}_{\mu\nu}{\bar R}_{\lambda\kappa} h^{\lambda\kappa}-\bar{R} h_{\mu\nu} \nonumber \\
& +m^2\left(h_{\mu\nu}-g_{\mu\nu}h +(\tfrac{1}{2}-3 c_3)F({x^{\scriptscriptstyle +}},\vec x) h \ell_\mu\ell_\nu 
-2(\tfrac{1}{4}-3 c_3)F({x^{\scriptscriptstyle +}},\vec x)\ell_{(\mu} \ell^\lambda h_{\lambda\nu)} \right. \\
&\left. +(1-3 c_3)F({x^{\scriptscriptstyle +}},\vec x) {\bar g}_{\mu\nu} \ell^\lambda\ell^\kappa h_{\lambda\kappa} 
-\tfrac{1}{8}F({x^{\scriptscriptstyle +}},\vec x)^2\ell_\mu\ell_\nu \ell^\lambda\ell^\kappa h_{\lambda\kappa}  \right) =0\, .  \nonumber
\end{align}
 Notably, the parameter $d_5$ has dropped out of these expressions. This is not surprising because we are calculating a propagation effect, which should be sensitive to the leading-order corrections to the quadratic action from the background coming from the cubic interactions.\footnote{In fact, it is possible to obtain these results starting from only the cubic vertices~\eqref{eq:expandeddrgt}. This theory also admits the solution~\eqref{eq:background} and to leading order in $F$, fluctuations about this solution obey an equation equivalent to~\eqref{eq:generalEOM}, making it explicit that the time delay is only sensitive to the cubic terms present in the theory.}
 
The equations of motion contain five on-shell constraints which eliminate five unphysical components of $h_{\mu\nu}$, leaving the five propagating degrees of freedom of the massive spin-2.  The constraints can be expressed as
\be
{\bar \nabla}_\nu {\cal E}^{\nu}_{~\mu} =0 \, , ~~~~ ~~~~~~~~~ 
{\bar \nabla}_\mu{\bar \nabla}_\nu {\cal E}^{\mu\nu} +\frac{m^2}{2} {\cal E}_{\ \mu}^{\mu}=0 \, .
\ee
where indices are raised with the background metric, $\bar g^{\mu\nu}$.
Using four of these constraints, $h_{\scriptscriptstyle ++}$, $h_{\scriptscriptstyle +-}$, $h_{{\scriptscriptstyle +}1}$ and $h_{{\scriptscriptstyle+}2}$ can be solved for algebraically in terms of the remaining components. The remaining constraint eliminates the combination $h_{11}+h_{22}$.  The background spacetime is everywhere Minkowski except along the line ${x^{\scriptscriptstyle +}}=0$, so the physical degrees of freedom satisfy an unsourced wave equation in all directions except the ${x^{\scriptscriptstyle +}}$ direction. We therefore make the ansatz that, for the unconstrained components, the solution is a plane wave moving in the $+{x^{\scriptscriptstyle -}}$ direction times a nontrivial function of $x^{\scriptscriptstyle +}$:
\bea
\begin{array}{lcl}
h_{--}({x^{\scriptscriptstyle -}},{x^{\scriptscriptstyle +}},\vec x)\!\! &\!\! =\!\! &\!\!  H_{\scriptscriptstyle --}({x^{\scriptscriptstyle +}})\, e^{-i(p^{\scriptscriptstyle +} {x^{\scriptscriptstyle -}} +\vec q\cdot\vec x)} \,  , \\
h_{-i}({x^{\scriptscriptstyle -}},{x^{\scriptscriptstyle +}},\vec x)\!\! &\!\! =\!\! &\!\!  H_{{\scriptscriptstyle -}i}({x^{\scriptscriptstyle +}})\, e^{-i(p^{\scriptscriptstyle +} {x^{\scriptscriptstyle -}} +\vec q\cdot\vec x)} \,  , \\
h_{ij}({x^{\scriptscriptstyle -}},{x^{\scriptscriptstyle +}},\vec x) \!\! &\!\! =\!\! & \!\! H_{ij}({x^{\scriptscriptstyle +}})\, e^{-i(p^{\scriptscriptstyle +} {x^{\scriptscriptstyle -}} +\vec q\cdot\vec x)} \,  ,
\end{array}
\eea
where $i,j = 1,2$.  In order to simplify the equations of motion for these modes, it is convenient to perform an additional field redefinition, which sends
\begin{align}
H_{{\scriptscriptstyle-}1}({x^{\scriptscriptstyle +}}) &\mapsto H_{{\scriptscriptstyle -}1}({x^{\scriptscriptstyle +}})+\frac{q_1}{p^{\scriptscriptstyle +}}H_{\scriptscriptstyle --}({x^{\scriptscriptstyle +}})\, ,\\
H_{{\scriptscriptstyle-}2}({x^{\scriptscriptstyle +}}) &\mapsto H_{{\scriptscriptstyle-}2}({x^{\scriptscriptstyle +}})+\frac{q_2}{p^{\scriptscriptstyle +}}H_{\scriptscriptstyle --}({x^{\scriptscriptstyle +}}) \, .
\end{align}
After this redefinition, $H_{{\scriptscriptstyle -}i}$ carries the helicity-1 modes, and $H_{{\scriptscriptstyle --}}$ carries the helicity-0 degree of freedom. The equations of motion satisfied by these fields decouple from the tensor modes, $H_{11}-H_{22}$ and $H_{12}$.  They are given by 
\begin{align}
\nonumber
\left(\partial_+ +ip^{\scriptscriptstyle +}\gamma\right)H_{\scriptscriptstyle --} &= i\frac{p^{\scriptscriptstyle -} p^{\scriptscriptstyle +}}{8\pi M_{\rm Pl}^2} \delta({x^{\scriptscriptstyle +}})\left(8(1-3 c_3)K_0(mb) H_{\scriptscriptstyle --}-4 i (1-4 c_3) \frac{p^{\scriptscriptstyle +} }{m b}K_1(mb) \left(x_1H_{{\scriptscriptstyle-}1}+x_2 H_{{\scriptscriptstyle-}2}\right)\right) \, , \\\nonumber
\left(\partial_+ +ip^{\scriptscriptstyle +}\gamma\right)H_{{\scriptscriptstyle-}1} &= i \frac{p^{\scriptscriptstyle -} p^{\scriptscriptstyle +}}{8\pi M_{\rm Pl}^2} \delta({x^{\scriptscriptstyle +}})\left((5-12 c_3)K_0(mb)\,H_{{\scriptscriptstyle-}1}+3i(1-4 c_3)\frac{m x_1}{p^{\scriptscriptstyle +} b} K_1(mb)\,H_{\scriptscriptstyle --}
\right)\, ,\\
\left(\partial_+ +ip^{\scriptscriptstyle +}\gamma\right)H_{{\scriptscriptstyle -}2} &= i\frac{p^{\scriptscriptstyle -} p^{\scriptscriptstyle +}}{8\pi M_{\rm Pl}^2}\delta({x^{\scriptscriptstyle +}})\left((5-12c_3)K_0(mb)\,H_{{\scriptscriptstyle-}2}+3i(1-4 c_3)\frac{m x_2}{p^{\scriptscriptstyle +}b}K_1(mb)\,H_{\scriptscriptstyle --}
\right) \, ,
\end{align}
where $\gamma \equiv \frac{q^2+m^2}{2(p^{\scriptscriptstyle +})^2}$. 
These equations can be grouped into a matrix equation in the form
\be
\left(\partial_++ip^{\scriptscriptstyle +} \gamma\right)H_I = i\frac{p^{\scriptscriptstyle -} p^{\scriptscriptstyle +}}{8\pi M_{\rm Pl}^2}\delta({x^{\scriptscriptstyle +}}) {\cal M}_{IJ}H_J \, .
\ee
This is a first-order equation, so by diagonalizing the matrix ${\cal M}_{IJ}$, we can directly integrate the resulting equation for the eigenmodes:
\be
\tilde H_I({x^{\scriptscriptstyle +}}) = \tilde H_I({x^{\scriptscriptstyle +}_0}) e^{-ip^{\scriptscriptstyle +}\int_{{x^{\scriptscriptstyle +}_0}}^{x^{\scriptscriptstyle +}} \rd  {\tilde x^{\scriptscriptstyle +}} \left(\gamma -\frac{p^{\scriptscriptstyle -}}{8\pi M_{\rm Pl}^2}\delta( {\tilde x^{\scriptscriptstyle +}}) \lambda_I(b)  \right)},
\ee
where $\lambda_I(b)$ is the eigenvalue of the matrix ${\cal M}_{IJ}$ corresponding to the $I$th eigenmode. We can interpret the $\gamma$ factor in the exponent as the phase shift due to propagation effects of a massive graviton. This is not the piece we are interested in. Instead, we are interested in the anomalous shift of the coordinate ${x^{\scriptscriptstyle -}}$ as we cross the shockwave. In order to isolate this contribution, we can take ${x_0^{\scriptscriptstyle +}}$ and ${x^{\scriptscriptstyle +}}$ to be infinitesimal around ${x^{\scriptscriptstyle +}}=0$. This means we will only pick up the second term in the exponent. We see that the phase shift of the $I$-th polarization due to the shockwave is just
\be
\label{eq:phase}
\delta_I (s,b) =\frac{p^{\scriptscriptstyle -} p^{\scriptscriptstyle +}}{8\pi M_{\rm Pl}^2} \lambda_I (b) =  \frac{s}{16 \pi M_{\rm Pl}^2} \lambda_I(b)  \, .
\ee
The eigenvalues of the matrix ${\cal M}_{IJ}$ are given by
\begin{align}
\lambda_S  &= (5-12 c_3) K_0(mb) \, , \vspace{0.3cm} \\
\lambda_{V_1,V_2}  &= \frac{13-36 c_3}{2} K_0(mb) \pm \frac{1- 4 c_3}{2} \sqrt{9 K_0(mb)^2+48 K_1(mb)^2} \, . 
\end{align}
The eigenvalues for the tensor sector, $H_{11}-H_{22}$ and $H_{12}$, are straightforward to extract and are given by
\be
\lambda_{T_\otimes,T_\oplus} = 4 K_0(mb) \, .
\ee
Converting to the previous coefficient $a_1 = 3(1-4 c_3)$ and taking the limit of impact parameters much smaller than the inverse graviton mass, $mb \ll 1$, the phase shift \eqref{eq:phase} becomes
\begin{align}
\delta_S(s,b) &= - \frac{(a_1+2)\, s}{16 \pi M_{\rm Pl}^2} \log(mb),\\
\delta_{V_1,V_2}(s,b) &= \pm  \frac{a_1\, s}{8 \pi M_{\rm Pl}^2}\frac{1}{\sqrt{3}\, mb},\\
\delta_{T_\otimes,T_\oplus}(s,b) &= - \frac{s}{4\, \pi M_{\rm Pl}^2} \log(mb)\, .
\end{align}
This in perfect agreement with \eqref{eq:eigen1}--\eqref{eq:eigen3} with canonical normalization ${\cal C}_s=1$.  We again conclude that we must have $a_1 =0$ to avoid a time advance.

\section{Discussion and Conclusions}
\label{sec:conclusions}
Demanding that massive spin-2 fields experience an asymptotic time delay is a model-independent constraint that can be placed on generic theories.  
We have found that it completely fixes the cubic structure of the spin-2 self interactions.  If cubic vertices are present, they must appear in the combination~\eqref{eq:cubiclinearcombination} in order for the EFT description of an isolated massive spin-2 to be valid. This constraint is independent of the number of dimensions (for $D > 3$). In the case that other cubic structures are present, new physics must appear at the low scale, $m$, the spin-2 mass. This is somewhat surprising, as perturbative unitarity does not break down until a parametrically higher scale. This is the power of the eikonal scattering techniques we have employed---the choice of kinematics provides a complementary picture to analyticity arguments and allows us to see that new degrees of freedom must enter earlier than other probes would suggest they should.\footnote{This is essentially because at a given impact parameter, $b$, eikonal scattering is sensitive to the presence of particles which have masses $\sim b^{-1}$.} 

It is important to keep in mind the assumptions to which the constraints derived from the eikonal scattering are subject.   Any of these may be violated in any given situation or application.  These assumptions include:

\begin{itemize}

\item That the eikonal approximation is valid, {\it i.e.}, that summing over only ladder type diagrams is a consistent limit of the full scattering amplitude.  The validity of the eikonal approximation as applied to higher spin particles is still not a settled issue, and there may be subtleties that could invalidate it.  It is thought that the validity of the eikonal approximation requires that the spin of the exchanged particle in the rungs of the ladder is $J\geq 2$. The essential requirement fulfilled by particles with $J\geq 2$ is that the eikonal phase shift, which scales as $\delta \sim s^{J-1}$, must grow with $s$.  A physical argument for this requirement is given in~\cite{Camanho:2014apa}. The failure of the eikonal approximation for lower-spin particles is described in~\cite{Tiktopoulos:1971hi,Cheng:1987ga,Kabat:1992pz}.   There is an additional subtlety even in the $J\geq 2$ case; the eikonal graphs re-sum into a phase, but if similar cancellations do not occur for subleading graphs, the eikonal approximation will break down~\cite{Kabat:1992pz}. It is unknown if this happens to all orders, but the leading corrections to the graviton eikonal amplitude also appear to re-sum into a subleading phase~\cite{Akhoury:2013yua}.

\item That the absence of asymptotic time advances is a fundamental requirement of a UV theory.  As far as we know, there is no direct derivation of the absence of asymptotic time advances in the $S$-matrix as a consequence of more fundamental $S$-matrix notions such as analyticity or locality, or quantum field theory postulates such as space-like commutativity of local operators.  The presence of superluminality of this kind may not necessarily lead to any microcausality, acausality or consistency issues \cite{Babichev:2007dw,Geroch:2010da,Burrage:2011cr}.

\item That flat space is a solution out to length scales $\gg m^{-1}$.  In most cosmological applications of massive gravity\footnote{See \cite{Hinterbichler:2017sbd} for a status report.} the horizon size is of order $m^{-1}$, and so there is no notion of a flat space $S$-matrix at scales larger than the Compton wavelength of the graviton.  In this case, the bounds derived don't directly apply.

\end{itemize}

The most direct application of our results is to constrain effective theories of an isolated massive spin-2 field.  The most interesting of these theories is dRGT massive gravity. dRGT massive gravity is a two-parameter family of theories and the positivity of the eikonal phase completely fixes one of these parameters, leaving a one-parameter family of theories consistent with the constraint. Another independent scattering constraint comes from forward dispersion relations~\cite{Adams:2006sv,deRham:2017zjm}, which in~\cite{Cheung:2016yqr} were applied to dRGT.  There they found a blob-like two dimensional compact subregion of the parameter space consistent with these constraints.  As was noted in~\cite{Camanho:2016opx} there is a nontrivial overlap of these two regions, identifying a small line in parameter space consistent with both constraints.  Our results are also applicable to the pseudo-linear theory of a massive spin-2 \cite{Folkerts:2011ev,Hinterbichler:2013eza}.  In this case, there is no region of parameter space consistent with dispersion relation constraints~\cite{Bonifacio:2016wcb} and similarly our results constrain the cubic terms in this theory to vanish (as the Einstein--Hilbert vertex is not part of the theory).  Within the context of massive gravity and related theories, a natural direction to pursue is to extend these eikonal constraints to the case of bimetric \cite{Hassan:2011zd} or multi-metric theories \cite{Hinterbichler:2012cn}, where both massive and massless spin-2 particles are present, or to massive scalar tensor theories \cite{DAmico:2012hia,Gabadadze:2012tr} where scalars are present.

In this paper, we have only considered the leading contribution to eikonal scattering of a massive spin-2. At this order, the phase shift is only sensitive to the on-shell cubic vertices in the theory. It is expected that sub-leading eikonal graphs should be sensitive to the higher-order interactions in the theory. Some work has been done on sub-leading corrections to the eikonal amplitude in the context of General Relativity~\cite{Akhoury:2013yua,Bjerrum-Bohr:2016hpa}, but it remains a largely unexplored subject. It would also be interesting to understand the connection between positivity of eikonal scattering and positivity of forward scattering~\cite{Adams:2006sv,deRham:2017zjm} or positivity constraints which come from the conformal bootstrap~\cite{Hartman:2015lfa,Hartman:2016dxc}. Two interesting test cases are the galileon~\cite{Nicolis:2008in} and the shift symmetric scalar with a ``wrong sign" $(\partial\phi)^4$ interaction. In both theories, there are no on-shell three-point functions. This guarantees that the phase shift in the leading eikonal approximation will vanish. This is consistent with the explicit leading-order computation of~\cite{Creminelli:2014zxa} for the galileon.\footnote{Note that it is important to integrate the phase shift from the asymptotic past to the asymptotic future in order to get zero. In~\cite{Goon:2016une} a similar computation is done from the origin outward, which yields a nonzero time advance.}  The time advance computation was done to sub-leading order in~\cite{Creminelli:2014zxa}, where there is a non-zero time advance. Presumably this computation would be captured by sub-leading graphs in the eikonal approximation where quartic vertices are important. Similarly, the sign of the $(\partial\phi)^4$ operator in the shift-symmetric case should be constrained by some subleading eikonal amplitude.

The constraints on the interaction of massive spin-2s should also provide some insights into the interactions of higher-spin particles more generally. For example, in Kaluza--Klein theories where massive spin-2 states appear, it would be interesting to understand if each of the massive spin-2's separately has a positive Shapiro time delay or whether cancellations between them make the theory subluminal. Similarly, the leading Regge trajectory of the open bosonic string has a massive spin-2 excitation. In our language, its on-shell cubic amplitude is given by~\cite{Sagnotti:2010at}
\be
{\cal A}_{\rm on-shell} \propto \frac{40}{3}{\cal B}_1+\frac{14}{3}{\cal B}_2-\frac{2{}}{15}{\cal B}_4+\frac{2}{9}{\cal B}_5,
\ee
where the ${\cal B}_i$ are the Lagrangian amplitudes given above in Section \ref{sec:3ptstructures} and the spin-2 two mass is $m^2=1/\alpha'$, with $\alpha'$ the Regge slope (note that the pseudo-linear term ${\cal B}_3$ does not appear here). Taken by itself, this implies that the spin-2 excitation will experience an asymptotic time advance. Presumably, the full tree-level Shapiro time delay in string theory is positive, which implies that there should be tree-level contributions from all the higher-spin intermediate states on the same Regge trajectory which make the time delay positive for all polarizations, along the lines of what happens for the massless spin-2 mode~\cite{DAppollonio:2015fly}. It would be interesting to understand how this works in detail.  Such considerations also apply to large $N$ confining gauge theories, and are complimentary to other $S$-matrix bootstrap constraints on the possible spectra of interacting higher spin particles~\cite{Caron-Huot:2016icg}.

\vspace{-.4cm}
\paragraph{Acknowledgements:} We would like to thank James Bonifacio, Clifford Cheung, Garrett Goon, Daniel Kabat, Juan Maldacena, Rakibur Rahman and Glenn Starkman for helpful conversations and correspondence.  RAR is supported by DOE grant DE-SC0011941  and AJ and RAR are supported by NASA grant NNX16AB27G.  The authors would like to thank the Sitka Sound Science Center for their hospitality while some of this work was completed.

\appendix

\section{Eikonal Resummation with Spin}
\label{app:eikonal}

Here we exhibit the resummation of ladder diagrams into an eikonal phase in the case of spinning particles. Consider a generic theory which has a particle, $A$, with mass $M_A$, a particle, $B$, with mass $M_B$, and a particle, $C$, with mass $m$.  Some of the particles may be the same, and all may have arbitrary spin.  There are on-shell non-trivial 3-point vertices $AAC$ and $BBC$.   

We're interested in the elastic 4-particle process $AB\rightarrow AB$, with the kinematics of Section~\ref{sec:eikonlkin}.  
The leading eikonal approximation sums up---in a certain approximation---all ladder diagrams where $A$ interacts with $B$ only through the exchange of $C$s which take the form
\begin{center}
\includegraphics[height=2in,width=6.0in]{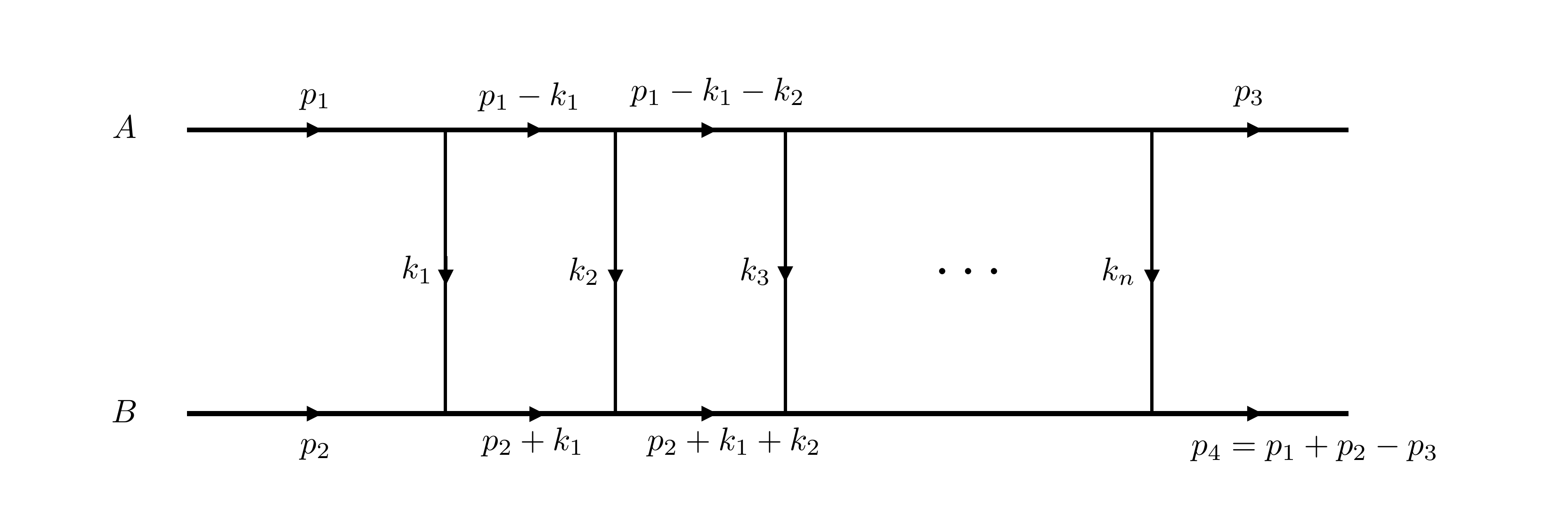}
\end{center}
along with all diagrams where the rungs are crossed.
The $0$-loop ladder is  the $t$-channel tree level exchange diagram.  At $1$-loop order there are two possible diagrams: a box and a cross.  A $n$-loop order, there are $n!$ possible diagrams, corresponding to all possible crosses of the ladder which are in one-to-one correspondence with the permutations of the rungs of the ladder.

Each ladder diagram is computed using the following rules.  For an $n-1$ loop ladder diagram, which has $n$ rungs:
\begin{itemize}
\item  Label the momenta along the rungs of the ladder as $k_1,k_2,\cdots,k_n$.   
\item Integrate over all the undetermined rung momenta and compensate for the extra free momentum by inserting a factor of $(2\pi)^D\delta^D(p_3-p_1+\sum_{i=i}^n k_i)$ which fixes the total momentum transfer to be the difference between $p_3$ and $p_1$.
\item Treat the rung propagator denominators with particle $C$ exactly: 
\be{-i\over k_i^2+m^2-i\epsilon}\, .\ee
\item For the propagator denominators of $A$ on the top rails, use the approximation 
\be {-i\over (p_1+k)^2+M_A^2-i\epsilon}\rightarrow {-i\over 2p_1\cdot k-i\epsilon},\ee 
and for the propagator denominators of $B$ on the bottom rails, use the approximation 
\be {-i\over (p_2+k)^2+M_B^2-i\epsilon}\rightarrow {-i\over 2p_2\cdot k-i\epsilon},\ee
where in each case $k$ is the appropriate sum of the rung momenta.  

Furthermore, breaking up $k^\mu=\left(k^-,k^+,\kb\right)$, take the large $p^{\scriptscriptstyle +},p^{\scriptscriptstyle -}$ limit in the dot products, which leads to 
\be p_1\cdot k\rightarrow -p^{\scriptscriptstyle +} k^-,\ \  \ p_2\cdot k\rightarrow -p^{\scriptscriptstyle -} k^+. \ee
\item For all propagator numerators, take $k^\mu = 0$.
\item For the $AAC$ vertex factors, we take the on-shell cubic Feynman rules with momenta $\{p_1,-(p_1-k),-k\}$.  Similarly, for the $BBC$ vertex factors, take the on-shell cubic Feynman rules with momenta $\{p_2,-(p_2+k),k\}$.  

\end{itemize}

Using these rules, we will see that the ladder diagrams sum up into an exponential of the tree level diagram in impact parameter space. The claim\footnote{As far as we are aware, there is no proof of this claim, and it is thought to fail in some cases \cite{Tiktopoulos:1971hi,Cheng:1987ga,Kabat:1992pz}.} of the eikonal approximation is that the full $2\to 2$ amplitude organizes into an exponential whose argument has a natural expansion in powers of $p^{\scriptscriptstyle +},p^{\scriptscriptstyle -}$, and that the leading exponent is captured by the summation of ladder diagrams using the above rules and expanding to leading order.

\subsection{Computing the Diagrams}

We first compute a generic ladder diagram using the above eikonal rules.  We will allow all the particles to be spinning and/or massive, and so we introduce the following generic indices for the various particles:
\begin{itemize}
\item  Particle $A$ has Lorentz labels $\IA,\JA,\ldots$ and polarization labels $i,j,\ldots$.  The polarization tensors are $\epsilon^i_{\IA}$.  The propagator numerators are $N_{\IA\JA}$.
\item Particle $B$ has Lorentz labels $\IB,\JB,\ldots$ and polarization labels $\ti,\tj,\ldots$.  The polarization tensors are $\epsilon^\ti_{\IB}$.  The propagator numerators are $N_{\IB\JB}$.
\item Particle $C$ has Lorentz labels $I,J,\ldots$.  The propagator numerators are $N_{IJ}$.
\end{itemize}
Here each of the Lorentz labels is a multi-index collectively labelling whatever Lorentz representation the field transforms in, for example if particle $A$ is a spin-$s$ then ${\cal I}$ is a string of $s$ symmetric Lorentz indices, ${\cal I} \equiv \mu_{1}\cdots\mu_{s_s}$.
The Feynman rules for the cubic vertices are
\begin{itemize}
\item $AAC$ vertex: $V_{(A)}^{\IA\JA, I}(p,k)$.
\item $BBC$ vertex: $V_{(A)}^{\IB\JB, I}(p,k)$.
\end{itemize}

With all this, we can use the Feynman rules to write down the expression for all the $n-1$ loop ladder diagrams,
\begin{align}
 i{\cal M}_{n-1}^{i\ti,j\tj}=&~
\int {\rd^Dk_1\over (2\pi)^D}\cdots  {\rd^Dk_n\over (2\pi)^D} {-iN_{I_1J_1}\over k_1^2+m^2}\cdots {-iN_{I_nJ_n}\over k_n^2+m^2}  \,(2\pi)^D\delta^D\left(p_3-p_1+\sum_{i=i}^n k_i\right) \nn\\
&\times {\epsilon}^i_{\IA_1}(p_1) V_{(A)}^{\IA_1 \JA_1,I_1}(p_1,-k_1) V_{(A)}^{\IA_2 \JA_2,I_2}(p_1,-k_2) \cdots V_{(A)}^{\IA_n \JA_n,I_n}(p_1,-k_n) {\epsilon^\ast}^j_{\JA_n}(p_1) \nn\\
&\times  {-iN_{\JA_1 \IA_2}(p_1)\over -2p_1\cdot k_1-i\epsilon}{-iN_{\JA_2 \IA_3}(p_1)\over -2p_1\cdot (k_1+k_2)-i\epsilon}\cdots {-iN_{\JA_{n-1} \IA_n}(p_1)\over -2p_1\cdot (k_1+k_2+\cdots+k_{n-1})-i\epsilon} \nn\\
& \times \sum_{\rm perms} {\epsilon}^\ti_{\IB_1}(p_2) V_{(B)}^{\IB_1 \JB_1,J_1}(p_2,k_1) V_{(B)}^{\IB_2 \JB_2,J_2}(p_2,k_2) \cdots V_{(B)}^{\IB_n \JB_n,J_n}(p_2,k_n) {\epsilon^\ast}^\tj_{\JB_n}(p_2) \nn\\
& \times {-iN_{\JB_1 \IB_2}(p_2)\over 2p_2\cdot k_1-i\epsilon}{-iN_{\JB_2 \IB_3}(p_2)\over 2p_2\cdot (k_1+k_2)-i\epsilon}\cdots {-iN_{\JB_{n-1} \IB_n}(p_2)\over 2p_2\cdot (k_1+k_2+\cdots+k_{n-1})-i\epsilon}\, .
\end{align}
The sum $\sum_{\rm perms} $ is over all permutations of $\{k_1,k_2,\cdots,k_n\}$, and accounts for all the $n!$ ladder diagrams with all the various crossings.

Replacing the dot products in the denominators according to the eikonal rules, $p_1\cdot k\rightarrow -p^{\scriptscriptstyle +} k^-$,  $p_2\cdot k\rightarrow -p^{\scriptscriptstyle -} k^+$, we have
\begin{align}
i{\cal M}_{n-1}=&~
{1\over 4^{n-1}}\int {\rd^Dk_1\over (2\pi)^D}\cdots  {\rd^Dk_n\over (2\pi)^D} {-iN_{I_1J_1}\over k_1^2+m^2}\cdots {-iN_{I_nJ_n}\over k_n^2+m^2}  \, (2\pi)^D\delta^D\left(p_3-p_1+\sum_{i=i}^n k_i\right) \nn\\
&\times  \epsilon^i_{\IA_1}(p_1) V_{(A)}^{\IA_1 \JA_1,I_1}(p_1,-k_1) V_{(A)}^{\IA_2 \JA_2,I_2}(p_1,-k_2) \cdots V_{(A)}^{\IA_n \JA_n,I_n}(p_1,-k_n) {\epsilon^\ast}^j_{\JA_n}(p_1)  \nn\\
&\times  {-iN_{\JA_1 \IA_2}(p_1) \over p^{\scriptscriptstyle +} k^-_1-i\epsilon}{-iN_{\JA_2 \IA_3}(p_1) \over p^{\scriptscriptstyle +} (k^-_1+k^-_2) -i\epsilon}\cdots {-iN_{\JA_{n-1} \IA_n}(p_1)\over p^{\scriptscriptstyle +} (k^-_1+k^-_2+\cdots+k^-_{n-1} )-i\epsilon} \nn\\
& \times \sum_{\rm perms}  {\epsilon}^\ti_{\IB_1}(p_2) V_{(B)}^{\IB_1 \JB_1,J_1}(p_2,k_1) V_{(B)}^{\IB_2 \JB_2,J_2}(p_2,k_2) \cdots V_{(B)}^{\IB_n \JB_n,J_n}(p_2,k_n) {\epsilon^\ast}^\tj_{\JB_n}(p_2)  \nn\\
& \times{-iN_{\JB_1 \IB_2}(p_2)\over  -p^{\scriptscriptstyle -} k^+_1-i\epsilon}{-iN_{\JB_2 \IB_3}(p_2)\over  -p^{\scriptscriptstyle -} (k^+_1+k^+_2)-i\epsilon}\cdots {-iN_{\JB_{n-1} \IB_n}(p_2)\over -p^{\scriptscriptstyle -} (k^+_1+k^+_2+\cdots+k^+_{n-1})-i\epsilon} \\\nonumber  
=&~{1\over 4^{n-1}}\int {\rd^{D-2}\kb_1\over (2\pi)^{D-2}}\cdots  {\rd^{D-2}\kb_n\over (2\pi)^{D-2}}\int {\rd k^+_1\over 2\pi}
\cdots {\rd k^+_n\over 2\pi} \int {\rd k^-_1\over 2\pi}
\cdots {\rd k^-_n\over 2\pi} \nn\\
& \times {-iN_{I_1J_1}\over -2k^+_1 k^-_1+\kb_1^2+m^2}\cdots {-iN_{I_nJ_n}\over -2k^+_n k^-_n+\kb_n^2+m^2} \nn\\
& \times  \, (2\pi)^D\delta^{D-2}\left(-\qb+\sum_{i=i}^n \kb_i\right)\delta\left(\sum_{i=i}^n k^+_i\right)\delta\left(\sum_{i=i}^n k^-_i\right) \nn\\
& \times   \epsilon^i_{\IA_1}(p_1) V_{(A)}^{\IA_1 \JA_1,I_1}(p_1,-k_1) \cdots V_{(A)}^{\IA_n \JA_n,I_n}(p_1,-k_n) {\epsilon^\ast}^j_{\JA_n}(p_1) N_{\JA_1 \IA_2}(p_1)\cdots N_{\JA_{n-1} \IA_n}(p_1)  \nn\\
&\times  {1\over p^{\scriptscriptstyle +} k^-_1-i\epsilon}{1\over p^{\scriptscriptstyle +} (k^-_1+k^-_2) -i\epsilon}\cdots {1\over p^{\scriptscriptstyle +} (k^-_1+k^-_2+\cdots+k^-_{n-1} )-i\epsilon} \nn\\
& \times {\epsilon}^\ti_{\IB_1}(p_2) V_{(B)}^{\IB_1 \JB_1,J_1}(p_2,k_1)  \cdots V_{(B)}^{\IB_n \JB_n,J_n}(p_2,k_n) {\epsilon^\ast}^\tj_{\JB_n}(p_2) N_{\JB_1 \IB_2}(p_2)\cdots N_{\JB_{n-1} \IB_n}(p_2) \nn\\
&\times  \sum_{\rm perms}  {1\over  p^{\scriptscriptstyle -} k^+_1+i\epsilon}{1\over  p^{\scriptscriptstyle -} (k^+_1+k^+_2)+i\epsilon}\cdots {1\over p^{\scriptscriptstyle -} (k^+_1+k^+_2+\cdots+k^+_{n-1})+i\epsilon}\, . \label{eqrefinta1}
\end{align}
Note that we have pulled the $B$ vertices and numerators outside the sum over permutation, since their product is permutation-invariant when combined with all the other stuff.

We now make use of the following delta function identity (see the appendix of \cite{Saotome:2012vy} for a proof)
\be
\lim_{\epsilon\rightarrow 0}\delta(x_1+x_2+\cdots+x_n)\sum_{{\rm perms}}{1\over x_1\pm i\epsilon}{1\over x_1+x_2\pm i\epsilon}\cdots {1\over x_1+x_2+\cdots x_{n-1}\pm i\epsilon}=(\mp 2\pi i)^{n-1}\delta(x_1)\delta(x_2)\cdots \delta(x_n)\,. \label{crazydelta}
\ee
Here, the sum over permutations means that we sum over all $n!$ permutations of the set $\{x_1,x_2,\cdots,x_n\}$.  For example, the first two instances of \eqref{crazydelta} read
\bea  &&n=1,\ \ \  \delta(x_1)=\delta(x_1)\, , \\
&&  n=2,\ \ \  \delta(x_1+x_2)\left({1\over x_1\pm i\epsilon}+{1\over x_2\pm i\epsilon}\right)=\mp 2\pi i\delta(x_1)\delta(x_2)\,.
\eea
Using \eqref{crazydelta} on the $k^+$ variables in \eqref{eqrefinta1}, we have
\begin{align}
i{\cal M}_{n-1}=&~{1\over 4^{n-1}}\int {\rd^{D-2}\kb_1\over (2\pi)^{D-2}}\cdots  {\rd^{D-2}\kb_n\over (2\pi)^{D-2}}\int {\rd k^+_1\over 2\pi}
\cdots {\rd k^+_n\over 2\pi} \int {\rd k^-_1\over 2\pi}
\cdots {\rd k^-_n\over 2\pi} \nn\\
& \times {-iN_{I_1J_1}\over -2k^+_1 k^-_1+\kb_1^2+m^2}\cdots {-iN_{I_nJ_n}\over -2k^+_n k^-_n+\kb_n^2+m^2}  \, (2\pi)^D \delta^{D-2}\left(-\qb+\sum_{i=i}^n \kb_i\right)\delta\left(\sum_{i=i}^n k^-_i\right) \nn\\
& \times   \epsilon^i_{\IA_1}(p_1) V_{(A)}^{\IA_1 \JA_1,I_1}(p_1,-k_1) \cdots V_{(A)}^{\IA_n \JA_n,I_n}(p_1,-k_n) {\epsilon^\ast}^j_{\JA_n}(p_1) N_{\JA_1 \IA_2}(p_1)\cdots N_{\JA_{n-1} \IA_n}(p_1)  \nn\\
&\times {1\over p^{\scriptscriptstyle +} k^-_1-i\epsilon}{1\over p^{\scriptscriptstyle +} (k^-_1+k^-_2) -i\epsilon}\cdots {1\over p^{\scriptscriptstyle +} (k^-_1+k^-_2+\cdots+k^-_{n-1} )-i\epsilon} \nn\\
&\times {\epsilon}^\ti_{\IB_1}(p_2) V_{(B)}^{\IB_1 \JB_1,J_1}(p_2,k_1)  \cdots V_{(B)}^{\IB_n \JB_n,J_n}(p_2,k_n) {\epsilon^\ast}^\tj_{\JB_n}(p_2) N_{\JB_1 \IB_2}(p_2)\cdots N_{\JB_{n-1} \IB_n}(p_2) \nn\\
&\times{(-2\pi i)^{n-1}\over |p^{\scriptscriptstyle -}|^{n-1}} \delta(k^+_1)\cdots \delta(k^+_n) \, . 
\end{align}
We can now use the delta functions in $k^+$ to do the $k^+$ integrals,
\begin{align}
i{\cal M}_{n-1}=&~{(-2\pi i)^{n-1}\over |p^{\scriptscriptstyle -}|^{n-1}} {1\over 4^{n-1}}{1\over (2\pi)^n}\int {\rd^{D-2}\kb_1\over (2\pi)^{D-2}}\cdots  {\rd^{D-2}\kb_n\over (2\pi)^{D-2}} \int {\rd k^-_1\over 2\pi}
\cdots {\rd k^-_n\over 2\pi} \nn\\
& \times {-iN_{I_1J_1}\over \kb_1^2+m^2}\cdots {-iN_{I_nJ_n}\over \kb_n^2+m^2}  \, (2\pi)^D\delta^{D-2}\left(-\qb+\sum_{i=i}^n \kb_i\right)\delta\left(\sum_{i=i}^n k^-_i\right) \nn\\
&\times   \epsilon^i_{\IA_1}(p_1) V_{(A)}^{\IA_1 \JA_1,I_1}(p_1,-k_1) \cdots V_{(A)}^{\IA_n \JA_n,I_n}(p_1,-k_n) {\epsilon^\ast}^j_{\JA_n}(p_1) N_{\JA_1 \IA_2}(p_1)\cdots N_{\JA_{n-1} \IA_n}(p_1) \nn\\
&\times \left. {\epsilon}^\ti_{\IB_1}(p_2) V_{(B)}^{\IB_1 \JB_1,J_1}(p_2,k_1)  \cdots V_{(B)}^{\IB_n \JB_n,J_n}(p_2,k_n) {\epsilon^\ast}^\tj_{\JB_n}(p_2) N_{\JB_1 \IB_2}(p_2)\cdots N_{\JB_{n-1} \IB_n}(p_2) \right|_{k^+=0}\nn\\
&\times {1\over p^{\scriptscriptstyle +} k^-_1-i\epsilon}{1\over p^{\scriptscriptstyle +} (k^-_1+k^-_2) -i\epsilon}\cdots {1\over p^{\scriptscriptstyle +} (k^-_1+k^-_2+\cdots+k^-_{n-1} )-i\epsilon}\, . 
\end{align}
The integral measure is symmetric under the interchange of any of the $k^-$ variables.  We may therefore symmetrize the entire integrand over the $k^-$ variables,
\begin{align}
i{\cal M}_{n-1}=&~{(-2\pi i)^{n-1}\over |p^{\scriptscriptstyle -}|^{n-1}} {1\over 4^{n-1}}{1\over (2\pi)^n}\int {\rd^{D-2}\kb_1\over (2\pi)^{D-2}}\cdots  {\rd^{D-2}\kb_n\over (2\pi)^{D-2}} \int {\rd k^-_1\over 2\pi}
\cdots {\rd k^-_n\over 2\pi} \nn\\
& \times {-iN_{I_1J_1}\over \kb_1^2+m^2}\cdots {-iN_{I_nJ_n}\over \kb_n^2+m^2}   \, (2\pi)^D\delta^{D-2}\left(-\qb+\sum_{i=i}^n \kb_i\right)\delta\left(\sum_{i=i}^n k^-_i\right) \nn\\
&\times   \epsilon^i_{\IA_1}(p_1) V_{(A)}^{\IA_1 \JA_1,I_1}(p_1,-k_1) \cdots V_{(A)}^{\IA_n \JA_n,I_n}(p_1,-k_n) {\epsilon^\ast}^j_{\JA_n}(p_1) N_{\JA_1 \IA_2}(p_1)\cdots N_{\JA_{n-1} \IA_n}(p_1) \nn\\
&\times \left. {\epsilon}^\ti_{\IB_1}(p_2) V_{(B)}^{\IB_1 \JB_1,J_1}(p_2,k_1)  \cdots V_{(B)}^{\IB_n \JB_n,J_n}(p_2,k_n) {\epsilon^\ast}^\tj_{\JB_n}(p_2) N_{\JB_1 \IB_2}(p_2)\cdots N_{\JB_{n-1} \IB_n}(p_2) \right|_{k^+=0}\nn\\
&\times {1\over n!}\sum_{\rm perms}{1\over p^{\scriptscriptstyle +} k^-_1-i\epsilon}{1\over p^{\scriptscriptstyle +} (k^-_1+k^-_2) -i\epsilon}\cdots {1\over p^{\scriptscriptstyle +} (k^-_1+k^-_2+\cdots+k^-_{n-1} )-i\epsilon}\, .
\end{align}
Now we use the delta identity \eqref{crazydelta} again on the $k^-$ variables, giving
\begin{align}
i{\cal M}_{n-1}=&~{(-2\pi i)^{n-1}\over |p^{\scriptscriptstyle -}|^{n-1}} {1\over 4^{n-1}}{1\over (2\pi)^n}\int {\rd^{D-2}\kb_1\over (2\pi)^{D-2}}\cdots  {\rd^{D-2}\kb_n\over (2\pi)^{D-2}} \int {dk^-_1\over 2\pi}
\cdots {dk^-_n\over 2\pi} \nn\\
& \times {-iN_{I_1J_1}\over \kb_1^2+m^2}\cdots {-iN_{I_nJ_n}\over \kb_n^2+m^2}  \, (2\pi)^D\delta^{D-2}\left(-\qb+\sum_{i=i}^n \kb_i\right) \nn\\
&\times   \epsilon^i_{\IA_1}(p_1) V_{(A)}^{\IA_1 \JA_1,I_1}(p_1,-k_1) \cdots V_{(A)}^{\IA_n \JA_n,I_n}(p_1,-k_n) {\epsilon^\ast}^j_{\JA_n}(p_1) N_{\JA_1 \IA_2}(p_1)\cdots N_{\JA_{n-1} \IA_n}(p_1) \nn\\
&\times \left. {\epsilon}^\ti_{\IB_1}(p_2) V_{(B)}^{\IB_1 \JB_1,J_1}(p_2,k_1)  \cdots V_{(B)}^{\IB_n \JB_n,J_n}(p_2,k_n) {\epsilon^\ast}^\tj_{\JB_n}(p_2) N_{\JB_1 \IB_2}(p_2)\cdots N_{\JB_{n-1} \IB_n}(p_2) \right|_{k^+=0}\nn\\
&\times {1\over n!}{(2\pi i)^{n-1}\over |p^{\scriptscriptstyle +}|^{n-1}}\delta(k^-_1)\cdots \delta(k^-_n) \, , 
\end{align}
and then use the delta functions to do the $k^-$ integrals,
\begin{align}
i{\cal M}_{n-1}=&~{(-2\pi i)^{n-1}\over |p^{\scriptscriptstyle -}|^{n-1}} {(2\pi i)^{n-1}\over |p^{\scriptscriptstyle +}|^{n-1}} {1\over n!}{1\over 4^{n-1}}{1\over (2\pi)^n}{1\over (2\pi)^n}\int {\rd^{D-2}\kb_1\over (2\pi)^{D-2}}\cdots  {\rd^{D-2}\kb_n\over (2\pi)^{D-2}} \nn\\
& \times {-iN_{I_1J_1}\over \kb_1^2+m^2}\cdots {-iN_{I_nJ_n}\over \kb_n^2+m^2}  \nn\\
&\times   \epsilon^i_{\IA_1}(p_1) V_{(A)}^{\IA_1 \JA_1,I_1}(p_1,-k_1) \cdots V_{(A)}^{\IA_n \JA_n,I_n}(p_1,-k_n) {\epsilon^\ast}^j_{\JA_n}(p_1) N_{\JA_1 \IA_2}(p_1)\cdots N_{\JA_{n-1} \IA_n}(p_1) \nn\\
&\times \left. {\epsilon}^\ti_{\IB_1}(p_2) V_{(B)}^{\IB_1 \JB_1,J_1}(p_2,k_1)  \cdots V_{(B)}^{\IB_n \JB_n,J_n}(p_2,k_n) {\epsilon^\ast}^\tj_{\JB_n}(p_2) N_{\JB_1 \IB_2}(p_2)\cdots N_{\JB_{n-1} \IB_n}(p_2) \right|_{k^+,k^-=0}\nn\\
 & \times (2\pi)^D\delta^{D-2}\left(-\qb+\sum_{i=i}^n \kb_i\right)\, . 
\end{align}
We'll now use the fact that the sum over polarizations gives a propagator numerator, for both the $A$ and $B$ particles,
\begin{align} 
{\epsilon^\ast}^i_{\IA}(p)\epsilon^i_{\JA}(p)&=N_{\IA \JA}(p)\, ,\nn\\
{\epsilon^\ast}^\ti_{\IB}(p)\epsilon^\ti_{\JB}(p)&=N_{\IB \JB}(p),
\end{align}
to replace all instances of $N_{\IA \JA}(p)$ and $N_{\IB \JB}(p)$ in the above.  
 Now we can group together the numerators and vertices.  Define 
 \be {\cal V}^{i\ti,j\tj}(p^{\scriptscriptstyle +},p^{\scriptscriptstyle -},\kb_i)\equiv \left. \epsilon^i_{\IA}(p_1) V_{(A)}^{\IA \JA,I}(p_1,-k_i){\epsilon^\ast}^j_{\JA}(p_1) N_{IJ} \epsilon^\ti_{\IB}(p_2) V_{(B)}^{\IB\JB,J}(p_2,k_i){\epsilon^\ast}^\tj_{\JB}(p_2) \right|_{k^+,k^-=0} , \ \  i=1,\cdots,n.\ee
With this the amplitude becomes
\begin{align}
i{\cal M}^{i\ti,j\tj}_{n-1}=&~ (2\pi)^{D-2}{1\over |p^{\scriptscriptstyle -}p^{\scriptscriptstyle +}|^{n-1}}  {1\over n!}{ 1\over 4^{n-1}}  \int {\rd^{D-2}\kb_1\over (2\pi)^{D-2}}\cdots  {\rd^{D-2}\kb_n\over (2\pi)^{D-2}}  {-i\over \kb_1^2+m^2}\cdots {-i\over \kb_n^2+m^2} \nn\\
&\times {\cal V}^{i\ti,}_{\ \  i_2\ti_2}(p^{\scriptscriptstyle +},p^{\scriptscriptstyle -},\kb_1){\cal V}^{i_2\ti_2,}_{\ \ \ \ i_3\ti_2}(p^{\scriptscriptstyle +},p^{\scriptscriptstyle -},\kb_2)\cdots {\cal V}^{i_n\ti_2,j\tj}(p^{\scriptscriptstyle +},p^{\scriptscriptstyle -},\kb_n)  \, \delta^{D-2}\left(-\qb+\sum_{i=i}^n \kb_i\right) \,.\nn\\
\end{align}
we can write the $(D-2)$-dimensional delta function in terms of an integral over ${\bf b}$, 
\be 
\delta^{D-2}\left(-\qb+\sum_{i=i}^n \kb_i\right)=\int {\rd^{D-2}{\bf b}\over (2\pi)^{D-2}} e^{-i{\bf b}\cdot \left(-\qb+\sum_{i=i}^n \kb_i\right)},\ee
and write the $(n-1)$-loop amplitude as
\be
i{\cal M}^{i\ti,j\tj}_{n-1}=4{|p^{\scriptscriptstyle -}p^{\scriptscriptstyle +}|}   \int {\rd^{D-2}{\bf b}}\, e^{i{\bf b}\cdot \qb} {1\over n!}\left[ {1\over 4|p^{\scriptscriptstyle -}p^{\scriptscriptstyle +}|} {  }  \int {\rd^{D-2}\kb \over (2\pi)^{D-2}}{\cal V}^{i\ti,j\tj}(p^{\scriptscriptstyle +},p^{\scriptscriptstyle -},\kb){-i\over \kb^2+m^2}e^{-i{\bf b}\cdot \kb}\right]^n \, .
\ee
with matrix multiplication implied for ${\cal V}$.

\subsection{Summing All Loops}
Now that we have a relatively simple expression for a the $(n-1)$-loop diagrams, we can see how to sum up the different loop orders. Defining the eikonal phase,
\begin{align} 
\delta^{i\ti,j\tj}({\bf b})&=-{1\over 4|p^{\scriptscriptstyle -}p^{\scriptscriptstyle +}|} { }  \int {\rd^{D-2}\kb \over (2\pi)^{D-2}}{\cal V}^{i\ti,j\tj}(p^{\scriptscriptstyle +},p^{\scriptscriptstyle -},\kb){1\over \kb^2+m^2}e^{-i{\bf b}\cdot \kb}\, \nn\\
&=-{1\over 4|p^{\scriptscriptstyle -}p^{\scriptscriptstyle +}|} { }{\cal V}^{i\ti,j\tj}(p^{\scriptscriptstyle +},p^{\scriptscriptstyle -},i{\partial_{\bf b}})  \int {\rd^{D-2}\kb \over (2\pi)^{D-2}}{1\over \kb^2+m^2}e^{-i{\bf b}\cdot \kb}\, \nn\\
&=   -{1\over 4|p^{\scriptscriptstyle -}p^{\scriptscriptstyle +}|} { }{\cal V}^{i\ti,j\tj}(p^{\scriptscriptstyle +},p^{\scriptscriptstyle -},i{\partial_{\bf b}})  \left[ \frac{1}{2\, \pi^{\frac{D-2}{2}}}\left(\frac{m}{b}\right)^\frac{D-4}{2}K_\frac{D-4}{2}(mb)\right],
\end{align}
the full amplitude is now seen to exponentiate,
\be   i{\cal M}^{i\ti,j\tj}=\sum_{n=1}^\infty i{\cal M}_{n-1}=4{|p^{\scriptscriptstyle -}p^{\scriptscriptstyle +}|}   \int {\rd^{D-2}{\bf b}}\, e^{i{\bf b}\cdot \qb} \left(e^{i \delta^{}({\bf b})}-1\right)^{i\ti,j\tj}\,.\ee
with matrix exponentiation implied for ${\delta}^{i\ti,j\tj}$.

Note that the tree level diagram is
\be
{\cal M}^{i\ti,j\tj}_{0}=4{\lvert p^{\scriptscriptstyle -}p^{\scriptscriptstyle +}\rvert}   \int {\rd^{D-2}{\bf b}}\,e^{i{\bf b}\cdot \qb}\delta^{i\ti,j\tj}({\bf b}) \, ,
\ee
so inverse Fourier transforming, we can write the eikonal phase as the Fourier transform of the tree level diagram,
\be  \delta^{i\ti,j\tj}({\bf b})= {1\over 4|p^{\scriptscriptstyle -}p^{\scriptscriptstyle +}|}  \int {\rd^{D-2}\qb\over (2\pi)^{D-2}}\,  e^{-i{\bf b}\cdot \qb} {\cal M}^{i\ti,j\tj}_{0}(\qb)\,,
\ee
which is the $2\to2$ scattering amplitude in impact parameter space.

The eikonal phase is a matrix, and diagonalizing this matrix gives the eigenstates which propagate with a definite phase.  For each such state, the associated eigenvalue, $\delta$, is then 
related to the time delay of propagation of that state by
\be \Delta x^-={1\over |p^{\scriptscriptstyle -}|}\delta\,. \ee

\section{Cubic Vertices of dRGT}
\label{app:dRGT}
One of the applications of our analysis is to constrain the possible parameters in nonlinear massive gravity which are consistent with positivity of the eikonal amplitude. Therefore, it is worth being explicit about the cubic vertices that appear in the dRGT theory~\cite{deRham:2010kj} and what the constraints are. Here we specialize to $D=4$. The dRGT theory is a 2-parameter family which can be written as
\begin{align}
\label{eq:drgtaction}
S =M_{\rm Pl}^2\int\rd^4x\sqrt{-g}\bigg(\frac{R}{2}+\frac{m^2}{2}\Big[&\left([{\cal K}]^2-[{\cal K}^2]\right)+\alpha_3\left([{\cal K}]^3-3[{\cal K}][{\cal K}^2]+2[{\cal K}^3]\right)\\\nonumber
&+\alpha_4\left([{\cal K}]^4-6[{\cal K}]^2[{\cal K}^2]+8[{\cal K}][{\cal K}^3]+3[{\cal K}^2]^2-6[{\cal K}^4]\right)\Big]\,
\bigg)\, .
\end{align}
The tensor ${\cal K}$ is defined to be (with the definition $g_{\mu\nu} = \eta_{\mu\nu} + h_{\mu\nu}$)
\be
{\cal K}^\mu_{~\nu} = \delta^\mu_\nu -\sqrt{g^{\mu\alpha} \eta_{\alpha\nu}} = -\sum_{n=1}^\infty \frac{(2n)!}{(1-2n)(n!)^2 4^n}(H^n)^\mu_{\ \nu},
\ee
with $H^\mu_{\ \nu} = g^{\mu\alpha}h_{\alpha\nu}$. 
The relation between the parameters $\alpha_3$, $\alpha_4$ here and the parameters $c_3$, $d_5$ used in Section~\ref{sec:shockwaveamp} is
\be
\alpha_3 = -2c_3\, ,~~~~~~~~\alpha_4 = -4d_5.
\ee

In order to see what cubic terms are present we expand the action~\eqref{eq:drgtaction} out to cubic order 
\be
{\cal L}_h = M_{\rm Pl}^2\left(\frac{1}{8}h{\cal E}h - \frac{m^2}{8}(h_{\mu\nu}^2-h^2)+R_{\rm EH}^{(3)}[h]+\frac{m^2}{16}\left((3+2\alpha_3)h_{\mu\nu}^3-(4+3\alpha_3)h h_{\mu\nu}^2+(1+\alpha_3)h^3\right)\right)+{\cal O}\left(h^4\right)\,,
\ee
where  $R_{\rm EH}^{(3)}[h]$ is the cubic term coming from the Einstein--Hilbert action and $\frac{1}{2}h{\cal E}h$ is the usual graviton kinetic term
\be
\frac{1}{2}h^{\mu\nu}{\cal E}_{\mu\nu\alpha\beta}h^{\alpha\beta} = -\frac{1}{2}\partial_\mu h_{\alpha\beta}\partial^\mu h^{\alpha\beta}+\partial_\mu h_{\alpha\beta}\partial^\alpha h^{\mu\beta}- \partial_\mu h^{\mu\nu}\partial_\nu h+\frac{1}{2}\partial_\mu h\partial^\mu h\, .
\ee
This action is not canonically-normalized, so we must redefine $h_{\mu\nu}\mapsto \frac{2}{M_{\rm Pl}}h_{\mu\nu}$ so that we have
\be
{\cal L}_{2h} = \frac{1}{2}h{\cal E}h - \frac{m^2}{2}(h_{\mu\nu}^2-h^2)+\frac{1}{M_{\rm Pl}}R_{\rm EH}^{(3)}[2h]+\frac{m^2}{2M_{\rm Pl}}\left((3+2\alpha_3)h_{\mu\nu}^3-(4+3\alpha_3)h h_{\mu\nu}^2+(1+\alpha_3)h^3\right)\,.
\label{eq:expandeddrgt}
\ee
Unsurprisingly, the quartic dRGT potential and the corresponding coefficient $\alpha_4$ does not contribute to the action up to cubic order~\eqref{eq:expandeddrgt}. We can now read off the coefficients of the cubic terms in the Lagrangian basis:
\begin{align}
b_1 &= \frac{3}{2}(3+2\alpha_3)\, ,\\
b_2 & = 1\, .
\end{align}
It is then straightforward to translate these to the structures basis using the formulae in Section~\ref{sec:3ptstructures}
\begin{align}
a_1 &= 3(1+2\alpha_3)\, ,\\
a_2 &= 2\, ,\\
a_3 &= 4\, .
\end{align}
Both the shockwave analysis and the scattering computation  tell us that we need to fix the coefficient of the $h^3$ vertex in terms of the Einstein--Hilbert cubic term. Specifically, the special point in parameter space is $a_1 = 0$. This corresponds to 
\be \alpha_3 = -\frac{1}{2}.\ee
 This fixes the cubic dRGT action to take the form
\be
{\cal L}_{2h} = \frac{1}{2}h{\cal E}h - \frac{m^2}{2}(h_{\mu\nu}^2-h^2)+\frac{1}{M_{\rm Pl}}R_{\rm EH}^{(3)}[2h]+\frac{m^2}{M_{\rm Pl}}\left(h_{\mu\nu}^3-\frac{5}{4}h h_{\mu\nu}^2+\frac{1}{4}h^3\right)\, .
\ee
Having fixed the parameter $\alpha_3$, we are left with only the freedom to rescale $m, M_{\rm Pl}$ and choose $\alpha_4$. 

As was noted in~\cite{Camanho:2016opx}, part of this reduced parameter space is still consistent with the region of parameter space identified in~\cite{Cheung:2016yqr} by demanding consistency with forward dispersion relations following from $S$-matrix analyticity~\cite{Adams:2006sv,deRham:2017zjm}.

\renewcommand{\em}{}
\bibliographystyle{utphys}
\addcontentsline{toc}{section}{References}
\bibliography{spin2superluminality16}

\providecommand{\href}[2]{#2}\begingroup\raggedright\begin{thebibliography}{10}

\bibitem{ArkaniHamed:2002sp}
N.~Arkani-Hamed, H.~Georgi, and M.~D. Schwartz, ``{Effective field theory for
  massive gravitons and gravity in theory space},''
  \href{http://dx.doi.org/10.1016/S0003-4916(03)00068-X}{{\em Annals Phys.}
  {\bf 305} (2003)  96--118},
\href{http://arxiv.org/abs/hep-th/0210184}{{\tt arXiv:hep-th/0210184
  [hep-th]}}.
%%CITATION = HEP-TH/0210184;%%.

\bibitem{Creminelli:2005qk}
P.~Creminelli, A.~Nicolis, M.~Papucci, and E.~Trincherini, ``{Ghosts in massive
  gravity},'' \href{http://dx.doi.org/10.1088/1126-6708/2005/09/003}{{\em JHEP}
  {\bf 09} (2005)  003},
\href{http://arxiv.org/abs/hep-th/0505147}{{\tt arXiv:hep-th/0505147
  [hep-th]}}.
%%CITATION = HEP-TH/0505147;%%.

\bibitem{Goon:2014paa}
G.~Goon, K.~Hinterbichler, A.~Joyce, and M.~Trodden, ``{Einstein Gravity,
  Massive Gravity, Multi-Gravity and Nonlinear Realizations},''
  \href{http://dx.doi.org/10.1007/JHEP07(2015)101}{{\em JHEP} {\bf 07} (2015)
  101},
\href{http://arxiv.org/abs/1412.6098}{{\tt arXiv:1412.6098 [hep-th]}}.
%%CITATION = ARXIV:1412.6098;%%.

\bibitem{Torabian:2017bqu}
M.~Torabian, ``{Towards a Lorentz Invariant UV Completion for Massive Gravity:
  dRGT Theory from Spontaneous Symmetry Breaking},''
\href{http://arxiv.org/abs/1707.04403}{{\tt arXiv:1707.04403 [hep-th]}}.
%%CITATION = ARXIV:1707.04403;%%.

\bibitem{Porrati:2001db}
M.~Porrati, ``{Higgs phenomenon for 4-D gravity in anti-de Sitter space},''
  \href{http://dx.doi.org/10.1088/1126-6708/2002/04/058}{{\em JHEP} {\bf 04}
  (2002)  058},
\href{http://arxiv.org/abs/hep-th/0112166}{{\tt arXiv:hep-th/0112166
  [hep-th]}}.
%%CITATION = HEP-TH/0112166;%%.

\bibitem{Gabadadze:2017jom}
G.~Gabadadze, ``{A Scale-up of Lambda 3},''
\href{http://arxiv.org/abs/1707.01739}{{\tt arXiv:1707.01739 [hep-th]}}.
%%CITATION = ARXIV:1707.01739;%%.

\bibitem{Adams:2006sv}
A.~Adams, N.~Arkani-Hamed, S.~Dubovsky, A.~Nicolis, and R.~Rattazzi,
  ``{Causality, analyticity and an IR obstruction to UV completion},''
  \href{http://dx.doi.org/10.1088/1126-6708/2006/10/014}{{\em JHEP} {\bf 10}
  (2006)  014},
\href{http://arxiv.org/abs/hep-th/0602178}{{\tt arXiv:hep-th/0602178
  [hep-th]}}.
%%CITATION = HEP-TH/0602178;%%.

\bibitem{deRham:2017avq}
C.~de~Rham, S.~Melville, A.~J. Tolley, and S.-Y. Zhou, ``{Positivity Bounds for
  Scalar Theories},''
\href{http://arxiv.org/abs/1702.06134}{{\tt arXiv:1702.06134 [hep-th]}}.
%%CITATION = ARXIV:1702.06134;%%.

\bibitem{deRham:2017zjm}
C.~de~Rham, S.~Melville, A.~J. Tolley, and S.-Y. Zhou, ``{UV complete me:
  Positivity Bounds for Particles with Spin},''
\href{http://arxiv.org/abs/1706.02712}{{\tt arXiv:1706.02712 [hep-th]}}.
%%CITATION = ARXIV:1706.02712;%%.

\bibitem{Cheung:2016yqr}
C.~Cheung and G.~N. Remmen, ``{Positive Signs in Massive Gravity},''
  \href{http://dx.doi.org/10.1007/JHEP04(2016)002}{{\em JHEP} {\bf 04} (2016)
  002},
\href{http://arxiv.org/abs/1601.04068}{{\tt arXiv:1601.04068 [hep-th]}}.
%%CITATION = ARXIV:1601.04068;%%.

\bibitem{Bonifacio:2016wcb}
J.~Bonifacio, K.~Hinterbichler, and R.~A. Rosen, ``{Positivity constraints for
  pseudolinear massive spin-2 and vector Galileons},''
  \href{http://dx.doi.org/10.1103/PhysRevD.94.104001}{{\em Phys. Rev.} {\bf
  D94} (2016) no.~10, 104001},
\href{http://arxiv.org/abs/1607.06084}{{\tt arXiv:1607.06084 [hep-th]}}.
%%CITATION = ARXIV:1607.06084;%%.

\bibitem{Velo:1969bt}
G.~Velo and D.~Zwanziger, ``{Propagation and quantization of Rarita-Schwinger
  waves in an external electromagnetic potential},''
\href{http://dx.doi.org/10.1103/PhysRev.186.1337}{{\em Phys. Rev.} {\bf 186}
  (1969)  1337--1341}.
%%CITATION = PHRVA,186,1337;%%.

\bibitem{Velo:1970ur}
G.~Velo and D.~Zwanziger, ``{Noncausality and other defects of interaction
  lagrangians for particles with spin one and higher},''
\href{http://dx.doi.org/10.1103/PhysRev.188.2218}{{\em Phys. Rev.} {\bf 188}
  (1969)  2218--2222}.
%%CITATION = PHRVA,188,2218;%%.

\bibitem{Deser:2012qx}
S.~Deser and A.~Waldron, ``{Acausality of Massive Gravity},''
  \href{http://dx.doi.org/10.1103/PhysRevLett.110.111101}{{\em Phys. Rev.
  Lett.} {\bf 110} (2013) no.~11, 111101},
\href{http://arxiv.org/abs/1212.5835}{{\tt arXiv:1212.5835 [hep-th]}}.
%%CITATION = ARXIV:1212.5835;%%.

\bibitem{Deser:2014fta}
S.~Deser, K.~Izumi, Y.~C. Ong, and A.~Waldron, ``{Problems of massive
  gravities},'' \href{http://dx.doi.org/10.1142/S0217732315400064}{{\em Mod.
  Phys. Lett.} {\bf A30} (2015)  1540006},
\href{http://arxiv.org/abs/1410.2289}{{\tt arXiv:1410.2289 [hep-th]}}.
%%CITATION = ARXIV:1410.2289;%%.

\bibitem{Deser:2014hga}
S.~Deser, M.~Sandora, A.~Waldron, and G.~Zahariade, ``{Covariant constraints
  for generic massive gravity and analysis of its characteristics},''
  \href{http://dx.doi.org/10.1103/PhysRevD.90.104043}{{\em Phys. Rev.} {\bf
  D90} (2014) no.~10, 104043},
\href{http://arxiv.org/abs/1408.0561}{{\tt arXiv:1408.0561 [hep-th]}}.
%%CITATION = ARXIV:1408.0561;%%.

\bibitem{Deser:2015wta}
S.~Deser, A.~Waldron, and G.~Zahariade, ``{Propagation peculiarities of mean
  field massive gravity},''
  \href{http://dx.doi.org/10.1016/j.physletb.2015.07.055}{{\em Phys. Lett.}
  {\bf B749} (2015)  144--148},
\href{http://arxiv.org/abs/1504.02919}{{\tt arXiv:1504.02919 [hep-th]}}.
%%CITATION = ARXIV:1504.02919;%%.

\bibitem{Burrage:2011cr}
C.~Burrage, C.~de~Rham, L.~Heisenberg, and A.~J. Tolley, ``{Chronology
  Protection in Galileon Models and Massive Gravity},''
  \href{http://dx.doi.org/10.1088/1475-7516/2012/07/004}{{\em JCAP} {\bf 1207}
  (2012)  004},
\href{http://arxiv.org/abs/1111.5549}{{\tt arXiv:1111.5549 [hep-th]}}.
%%CITATION = ARXIV:1111.5549;%%.

\bibitem{Hassan:2017ugh}
S.~F. Hassan and M.~Kocic, ``{On the local structure of spacetime in ghost-free
  bimetric theory and massive gravity},''
\href{http://arxiv.org/abs/1706.07806}{{\tt arXiv:1706.07806 [hep-th]}}.
%%CITATION = ARXIV:1706.07806;%%.

\bibitem{Goon:2016une}
G.~Goon and K.~Hinterbichler, ``{Superluminality, black holes and EFT},''
  \href{http://dx.doi.org/10.1007/JHEP02(2017)134}{{\em JHEP} {\bf 02} (2017)
  134},
\href{http://arxiv.org/abs/1609.00723}{{\tt arXiv:1609.00723 [hep-th]}}.
%%CITATION = ARXIV:1609.00723;%%.

\bibitem{tHooft:1987vrq}
G.~'t~Hooft, ``{Graviton Dominance in Ultrahigh-Energy Scattering},''
\href{http://dx.doi.org/10.1016/0370-2693(87)90159-6}{{\em Phys. Lett.} {\bf
  B198} (1987)  61--63}.
%%CITATION = PHLTA,B198,61;%%.

\bibitem{Kabat:1992tb}
D.~N. Kabat and M.~Ortiz, ``{Eikonal quantum gravity and Planckian
  scattering},'' \href{http://dx.doi.org/10.1016/0550-3213(92)90627-N}{{\em
  Nucl. Phys.} {\bf B388} (1992)  570--592},
\href{http://arxiv.org/abs/hep-th/9203082}{{\tt arXiv:hep-th/9203082
  [hep-th]}}.
%%CITATION = HEP-TH/9203082;%%.

\bibitem{Camanho:2014apa}
X.~O. Camanho, J.~D. Edelstein, J.~Maldacena, and A.~Zhiboedov, ``{Causality
  Constraints on Corrections to the Graviton Three-Point Coupling},''
  \href{http://dx.doi.org/10.1007/JHEP02(2016)020}{{\em JHEP} {\bf 02} (2016)
  020},
\href{http://arxiv.org/abs/1407.5597}{{\tt arXiv:1407.5597 [hep-th]}}.
%%CITATION = ARXIV:1407.5597;%%.

\bibitem{Zwiebach:1985uq}
B.~Zwiebach, ``{Curvature Squared Terms and String Theories},''
\href{http://dx.doi.org/10.1016/0370-2693(85)91616-8}{{\em Phys. Lett.} {\bf
  156B} (1985)  315--317}.
%%CITATION = PHLTA,156B,315;%%.

\bibitem{DAppollonio:2015fly}
G.~D'Appollonio, P.~Di~Vecchia, R.~Russo, and G.~Veneziano, ``{Regge behavior
  saves String Theory from causality violations},''
  \href{http://dx.doi.org/10.1007/JHEP05(2015)144}{{\em JHEP} {\bf 05} (2015)
  144},
\href{http://arxiv.org/abs/1502.01254}{{\tt arXiv:1502.01254 [hep-th]}}.
%%CITATION = ARXIV:1502.01254;%%.

\bibitem{Edelstein:2016nml}
J.~D. Edelstein, G.~Giribet, C.~Gomez, E.~Kilicarslan, M.~Leoni, and B.~Tekin,
  ``{Causality in 3D Massive Gravity Theories},''
  \href{http://dx.doi.org/10.1103/PhysRevD.95.104016}{{\em Phys. Rev.} {\bf
  D95} (2017) no.~10, 104016},
\href{http://arxiv.org/abs/1602.03376}{{\tt arXiv:1602.03376 [hep-th]}}.
%%CITATION = ARXIV:1602.03376;%%.

\bibitem{deRham:2010ik}
C.~de~Rham and G.~Gabadadze, ``{Generalization of the Fierz-Pauli Action},''
  \href{http://dx.doi.org/10.1103/PhysRevD.82.044020}{{\em Phys. Rev.} {\bf
  D82} (2010)  044020},
\href{http://arxiv.org/abs/1007.0443}{{\tt arXiv:1007.0443 [hep-th]}}.
%%CITATION = ARXIV:1007.0443;%%.

\bibitem{deRham:2010kj}
C.~de~Rham, G.~Gabadadze, and A.~J. Tolley, ``{Resummation of Massive
  Gravity},'' \href{http://dx.doi.org/10.1103/PhysRevLett.106.231101}{{\em
  Phys. Rev. Lett.} {\bf 106} (2011)  231101},
\href{http://arxiv.org/abs/1011.1232}{{\tt arXiv:1011.1232 [hep-th]}}.
%%CITATION = ARXIV:1011.1232;%%.

\bibitem{Hinterbichler:2011tt}
K.~Hinterbichler, ``{Theoretical Aspects of Massive Gravity},''
  \href{http://dx.doi.org/10.1103/RevModPhys.84.671}{{\em Rev. Mod. Phys.} {\bf
  84} (2012)  671--710},
\href{http://arxiv.org/abs/1105.3735}{{\tt arXiv:1105.3735 [hep-th]}}.
%%CITATION = ARXIV:1105.3735;%%.

\bibitem{deRham:2014zqa}
C.~de~Rham, ``{Massive Gravity},''
  \href{http://dx.doi.org/10.12942/lrr-2014-7}{{\em Living Rev. Rel.} {\bf 17}
  (2014)  7},
\href{http://arxiv.org/abs/1401.4173}{{\tt arXiv:1401.4173 [hep-th]}}.
%%CITATION = ARXIV:1401.4173;%%.

\bibitem{Folkerts:2011ev}
S.~Folkerts, A.~Pritzel, and N.~Wintergerst, ``{On ghosts in theories of
  self-interacting massive spin-2 particles},''
\href{http://arxiv.org/abs/1107.3157}{{\tt arXiv:1107.3157 [hep-th]}}.
%%CITATION = ARXIV:1107.3157;%%.

\bibitem{Hinterbichler:2013eza}
K.~Hinterbichler, ``{Ghost-Free Derivative Interactions for a Massive
  Graviton},'' \href{http://dx.doi.org/10.1007/JHEP10(2013)102}{{\em JHEP} {\bf
  10} (2013)  102},
\href{http://arxiv.org/abs/1305.7227}{{\tt arXiv:1305.7227 [hep-th]}}.
%%CITATION = ARXIV:1305.7227;%%.

\bibitem{Camanho:2016opx}
X.~O. Camanho, G.~Lucena~Gomez, and R.~Rahman, ``{Causality Constraints on
  Massive Gravity},''
\href{http://arxiv.org/abs/1610.02033}{{\tt arXiv:1610.02033 [hep-th]}}.
%%CITATION = ARXIV:1610.02033;%%.

\bibitem{Dray:1984ha}
T.~Dray and G.~'t~Hooft, ``{The Gravitational Shock Wave of a Massless
  Particle},''
\href{http://dx.doi.org/10.1016/0550-3213(85)90525-5}{{\em Nucl. Phys.} {\bf
  B253} (1985)  173--188}.
%%CITATION = NUPHA,B253,173;%%.

\bibitem{Tipler:1980aq}
R.~Penrose, {\em {``On Schwarzschild Causality A Problem for Lorentz Covariant
  General Relativity" in {\it Essays in General Relativity. A Festschrift for
  Abraham Taub}}}.
\newblock
1980.
\newblock
%%CITATION = INSPIRE-162229;%%.

\bibitem{Olum:1998mu}
K.~D. Olum, ``{Superluminal travel requires negative energies},''
  \href{http://dx.doi.org/10.1103/PhysRevLett.81.3567}{{\em Phys. Rev. Lett.}
  {\bf 81} (1998)  3567--3570},
\href{http://arxiv.org/abs/gr-qc/9805003}{{\tt arXiv:gr-qc/9805003 [gr-qc]}}.
%%CITATION = GR-QC/9805003;%%.

\bibitem{Gao:2000ga}
S.~Gao and R.~M. Wald, ``{Theorems on gravitational time delay and related
  issues},'' \href{http://dx.doi.org/10.1088/0264-9381/17/24/305}{{\em Class.
  Quant. Grav.} {\bf 17} (2000)  4999--5008},
\href{http://arxiv.org/abs/gr-qc/0007021}{{\tt arXiv:gr-qc/0007021 [gr-qc]}}.
%%CITATION = GR-QC/0007021;%%.

\bibitem{Babichev:2007dw}
E.~Babichev, V.~Mukhanov, and A.~Vikman, ``{k-Essence, superluminal
  propagation, causality and emergent geometry},''
  \href{http://dx.doi.org/10.1088/1126-6708/2008/02/101}{{\em JHEP} {\bf 02}
  (2008)  101},
\href{http://arxiv.org/abs/0708.0561}{{\tt arXiv:0708.0561 [hep-th]}}.
%%CITATION = ARXIV:0708.0561;%%.

\bibitem{Geroch:2010da}
R.~Geroch, ``{Faster Than Light?},'' {\em AMS/IP Stud. Adv. Math.} {\bf 49}
  (2011)  59--70,
\href{http://arxiv.org/abs/1005.1614}{{\tt arXiv:1005.1614 [gr-qc]}}.
%%CITATION = ARXIV:1005.1614;%%.

\bibitem{Papallo:2015rna}
G.~Papallo and H.~S. Reall, ``{Graviton time delay and a speed limit for small
  black holes in Einstein-Gauss-Bonnet theory},''
  \href{http://dx.doi.org/10.1007/JHEP11(2015)109}{{\em JHEP} {\bf 11} (2015)
  109},
\href{http://arxiv.org/abs/1508.05303}{{\tt arXiv:1508.05303 [gr-qc]}}.
%%CITATION = ARXIV:1508.05303;%%.

\bibitem{Gott:1990zr}
J.~R. Gott, III, ``{Closed timelike curves produced by pairs of moving cosmic
  strings: Exact solutions},''
\href{http://dx.doi.org/10.1103/PhysRevLett.66.1126}{{\em Phys. Rev. Lett.}
  {\bf 66} (1991)  1126--1129}.
%%CITATION = PRLTA,66,1126;%%.

\bibitem{Carroll:1991nr}
S.~M. Carroll, E.~Farhi, and A.~H. Guth, ``{An Obstacle to building a time
  machine},'' \href{http://dx.doi.org/10.1103/PhysRevLett.68.263}{{\em Phys.
  Rev. Lett.} {\bf 68} (1992)  263--266}.
[Erratum: Phys. Rev. Lett.68,3368(1992)].
%%CITATION = PRLTA,68,263;%%.

\bibitem{Dubovsky:2007ac}
S.~Dubovsky, A.~Nicolis, E.~Trincherini, and G.~Villadoro, ``{Microcausality in
  curved space-time},''
  \href{http://dx.doi.org/10.1103/PhysRevD.77.084016}{{\em Phys. Rev.} {\bf
  D77} (2008)  084016},
\href{http://arxiv.org/abs/0709.1483}{{\tt arXiv:0709.1483 [hep-th]}}.
%%CITATION = ARXIV:0709.1483;%%.

\bibitem{Carroll:2004st}
S.~M. Carroll, {\em {Spacetime and geometry: An introduction to general
  relativity}}.
\newblock 2004.
\newblock
\url{http://www.slac.stanford.edu/spires/find/books/www?cl=QC6:C37:2004}.
\newblock
%%CITATION = INSPIRE-650093;%%.

\bibitem{Cheng:1969eh}
H.~Cheng and T.~T. Wu, ``{High-energy elastic scattering in quantum
  electrodynamics},''
\href{http://dx.doi.org/10.1103/PhysRevLett.22.666}{{\em Phys. Rev. Lett.} {\bf
  22} (1969)  666}.
%%CITATION = PRLTA,22,666;%%.

\bibitem{Levy:1969cr}
M.~Levy and J.~Sucher, ``{Eikonal approximation in quantum field theory},''
\href{http://dx.doi.org/10.1103/PhysRev.186.1656}{{\em Phys. Rev.} {\bf 186}
  (1969)  1656--1670}.
%%CITATION = PHRVA,186,1656;%%.

\bibitem{Abarbanel:1969ek}
H.~D.~I. Abarbanel and C.~Itzykson, ``{Relativistic eikonal expansion},''
\href{http://dx.doi.org/10.1103/PhysRevLett.23.53}{{\em Phys. Rev. Lett.} {\bf
  23} (1969)  53}.
%%CITATION = PRLTA,23,53;%%.

\bibitem{Giudice:2001ce}
G.~F. Giudice, R.~Rattazzi, and J.~D. Wells, ``{Transplanckian collisions at
  the LHC and beyond},''
  \href{http://dx.doi.org/10.1016/S0550-3213(02)00142-6}{{\em Nucl. Phys.} {\bf
  B630} (2002)  293--325},
\href{http://arxiv.org/abs/hep-ph/0112161}{{\tt arXiv:hep-ph/0112161
  [hep-ph]}}.
%%CITATION = HEP-PH/0112161;%%.

\bibitem{Giddings:2011xs}
S.~B. Giddings, ``{The gravitational S-matrix: Erice lectures},''
  \href{http://dx.doi.org/10.1142/9789814522489_0005}{{\em Subnucl. Ser.} {\bf
  48} (2013)  93--147},
\href{http://arxiv.org/abs/1105.2036}{{\tt arXiv:1105.2036 [hep-th]}}.
%%CITATION = ARXIV:1105.2036;%%.

\bibitem{Tiktopoulos:1971hi}
G.~Tiktopoulos and S.~B. Treiman, ``{Relativistic eikonal approximation},''
\href{http://dx.doi.org/10.1103/PhysRevD.3.1037}{{\em Phys. Rev.} {\bf D3}
  (1971)  1037--1040}.
%%CITATION = PHRVA,D3,1037;%%.

\bibitem{Cheng:1987ga}
H.~Cheng and T.~T. Wu, {\em {Expanding Protons: Scattering At High-Energies}}.
\newblock
1987.
\newblock
%%CITATION = INSPIRE-256205;%%.

\bibitem{Kabat:1992pz}
D.~N. Kabat, ``{Validity of the Eikonal approximation},'' {\em Comments Nucl.
  Part. Phys.} {\bf 20} (1992) no.~6, 325--335,
\href{http://arxiv.org/abs/hep-th/9204103}{{\tt arXiv:hep-th/9204103
  [hep-th]}}.
%%CITATION = HEP-TH/9204103;%%.

\bibitem{Akhoury:2013yua}
R.~Akhoury, R.~Saotome, and G.~Sterman, ``{High Energy Scattering in
  Perturbative Quantum Gravity at Next to Leading Power},''
\href{http://arxiv.org/abs/1308.5204}{{\tt arXiv:1308.5204 [hep-th]}}.
%%CITATION = ARXIV:1308.5204;%%.

\bibitem{Bjerrum-Bohr:2016hpa}
N.~E.~J. Bjerrum-Bohr, J.~F. Donoghue, B.~R. Holstein, L.~Plante, and
  P.~Vanhove, ``{Light-like Scattering in Quantum Gravity},''
  \href{http://dx.doi.org/10.1007/JHEP11(2016)117}{{\em JHEP} {\bf 11} (2016)
  117},
\href{http://arxiv.org/abs/1609.07477}{{\tt arXiv:1609.07477 [hep-th]}}.
%%CITATION = ARXIV:1609.07477;%%.

\bibitem{Saotome:2012vy}
R.~Saotome and R.~Akhoury, ``{Relationship Between Gravity and Gauge Scattering
  in the High Energy Limit},''
  \href{http://dx.doi.org/10.1007/JHEP01(2013)123}{{\em JHEP} {\bf 01} (2013)
  123},
\href{http://arxiv.org/abs/1210.8111}{{\tt arXiv:1210.8111 [hep-th]}}.
%%CITATION = ARXIV:1210.8111;%%.

\bibitem{Aichelburg:1970dh}
P.~C. Aichelburg and R.~U. Sexl, ``{On the Gravitational field of a massless
  particle},''
\href{http://dx.doi.org/10.1007/BF00758149}{{\em Gen. Rel. Grav.} {\bf 2}
  (1971)  303--312}.
%%CITATION = GRGVA,2,303;%%.

\bibitem{Britto:2005fq}
R.~Britto, F.~Cachazo, B.~Feng, and E.~Witten, ``{Direct proof of tree-level
  recursion relation in Yang-Mills theory},''
  \href{http://dx.doi.org/10.1103/PhysRevLett.94.181602}{{\em Phys. Rev. Lett.}
  {\bf 94} (2005)  181602},
\href{http://arxiv.org/abs/hep-th/0501052}{{\tt arXiv:hep-th/0501052
  [hep-th]}}.
%%CITATION = HEP-TH/0501052;%%.

\bibitem{Benincasa:2007xk}
P.~Benincasa and F.~Cachazo, ``{Consistency Conditions on the S-Matrix of
  Massless Particles},''
\href{http://arxiv.org/abs/0705.4305}{{\tt arXiv:0705.4305 [hep-th]}}.
%%CITATION = ARXIV:0705.4305;%%.

\bibitem{Schuster:2008nh}
P.~C. Schuster and N.~Toro, ``{Constructing the Tree-Level Yang-Mills S-Matrix
  Using Complex Factorization},''
  \href{http://dx.doi.org/10.1088/1126-6708/2009/06/079}{{\em JHEP} {\bf 06}
  (2009)  079},
\href{http://arxiv.org/abs/0811.3207}{{\tt arXiv:0811.3207 [hep-th]}}.
%%CITATION = ARXIV:0811.3207;%%.

\bibitem{Benakli:2015qlh}
K.~Benakli, S.~Chapman, L.~Darmé, and Y.~Oz, ``{Superluminal graviton
  propagation},'' \href{http://dx.doi.org/10.1103/PhysRevD.94.084026}{{\em
  Phys. Rev.} {\bf D94} (2016) no.~8, 084026},
\href{http://arxiv.org/abs/1512.07245}{{\tt arXiv:1512.07245 [hep-th]}}.
%%CITATION = ARXIV:1512.07245;%%.

\bibitem{Costa:2011mg}
M.~S. Costa, J.~Penedones, D.~Poland, and S.~Rychkov, ``{Spinning Conformal
  Correlators},'' \href{http://dx.doi.org/10.1007/JHEP11(2011)071}{{\em JHEP}
  {\bf 11} (2011)  071},
\href{http://arxiv.org/abs/1107.3554}{{\tt arXiv:1107.3554 [hep-th]}}.
%%CITATION = ARXIV:1107.3554;%%.

\bibitem{deRham:2014naa}
C.~de~Rham, L.~Heisenberg, and R.~H. Ribeiro, ``{On couplings to matter in
  massive (bi-)gravity},''
  \href{http://dx.doi.org/10.1088/0264-9381/32/3/035022}{{\em Class. Quant.
  Grav.} {\bf 32} (2015)  035022},
\href{http://arxiv.org/abs/1408.1678}{{\tt arXiv:1408.1678 [hep-th]}}.
%%CITATION = ARXIV:1408.1678;%%.

\bibitem{Hinterbichler:2015yaa}
K.~Hinterbichler and R.~A. Rosen, ``{Note on ghost-free matter couplings in
  massive gravity and multigravity},''
  \href{http://dx.doi.org/10.1103/PhysRevD.92.024030}{{\em Phys. Rev.} {\bf
  D92} (2015) no.~2, 024030},
\href{http://arxiv.org/abs/1503.06796}{{\tt arXiv:1503.06796 [hep-th]}}.
%%CITATION = ARXIV:1503.06796;%%.

\bibitem{Mohseni:2011vv}
M.~Mohseni, ``{Exact plane gravitational waves in the de Rham-Gabadadze-Tolley
  model of massive gravity},''
  \href{http://dx.doi.org/10.1103/PhysRevD.84.064026}{{\em Phys. Rev.} {\bf
  D84} (2011)  064026},
\href{http://arxiv.org/abs/1109.4713}{{\tt arXiv:1109.4713 [hep-th]}}.
%%CITATION = ARXIV:1109.4713;%%.

\bibitem{Hinterbichler:2017sbd}
K.~Hinterbichler, ``{Cosmology of Massive Gravity and its Extensions},'' in
  {\em {Proceedings, 51st Rencontres de Moriond, Cosmology session: La Thuile,
  Italy, March 19-26, 2016}}, pp.~223--232.
\newblock 2016.
\newblock \href{http://arxiv.org/abs/1701.02873}{{\tt arXiv:1701.02873
  [astro-ph.CO]}}.
\newblock
\url{https://inspirehep.net/record/1508592/files/arXiv:1701.02873.pdf}.
\newblock
%%CITATION = ARXIV:1701.02873;%%.

\bibitem{Hassan:2011zd}
S.~F. Hassan and R.~A. Rosen, ``{Bimetric Gravity from Ghost-free Massive
  Gravity},'' \href{http://dx.doi.org/10.1007/JHEP02(2012)126}{{\em JHEP} {\bf
  02} (2012)  126},
\href{http://arxiv.org/abs/1109.3515}{{\tt arXiv:1109.3515 [hep-th]}}.
%%CITATION = ARXIV:1109.3515;%%.

\bibitem{Hinterbichler:2012cn}
K.~Hinterbichler and R.~A. Rosen, ``{Interacting Spin-2 Fields},''
  \href{http://dx.doi.org/10.1007/JHEP07(2012)047}{{\em JHEP} {\bf 07} (2012)
  047},
\href{http://arxiv.org/abs/1203.5783}{{\tt arXiv:1203.5783 [hep-th]}}.
%%CITATION = ARXIV:1203.5783;%%.

\bibitem{DAmico:2012hia}
G.~D'Amico, G.~Gabadadze, L.~Hui, and D.~Pirtskhalava, ``{Quasidilaton: Theory
  and cosmology},'' \href{http://dx.doi.org/10.1103/PhysRevD.87.064037}{{\em
  Phys. Rev.} {\bf D87} (2013)  064037},
\href{http://arxiv.org/abs/1206.4253}{{\tt arXiv:1206.4253 [hep-th]}}.
%%CITATION = ARXIV:1206.4253;%%.

\bibitem{Gabadadze:2012tr}
G.~Gabadadze, K.~Hinterbichler, J.~Khoury, D.~Pirtskhalava, and M.~Trodden,
  ``{A Covariant Master Theory for Novel Galilean Invariant Models and Massive
  Gravity},'' \href{http://dx.doi.org/10.1103/PhysRevD.86.124004}{{\em Phys.
  Rev.} {\bf D86} (2012)  124004},
\href{http://arxiv.org/abs/1208.5773}{{\tt arXiv:1208.5773 [hep-th]}}.
%%CITATION = ARXIV:1208.5773;%%.

\bibitem{Hartman:2015lfa}
T.~Hartman, S.~Jain, and S.~Kundu, ``{Causality Constraints in Conformal Field
  Theory},'' \href{http://dx.doi.org/10.1007/JHEP05(2016)099}{{\em JHEP} {\bf
  05} (2016)  099},
\href{http://arxiv.org/abs/1509.00014}{{\tt arXiv:1509.00014 [hep-th]}}.
%%CITATION = ARXIV:1509.00014;%%.

\bibitem{Hartman:2016dxc}
T.~Hartman, S.~Jain, and S.~Kundu, ``{A New Spin on Causality Constraints},''
  \href{http://dx.doi.org/10.1007/JHEP10(2016)141}{{\em JHEP} {\bf 10} (2016)
  141},
\href{http://arxiv.org/abs/1601.07904}{{\tt arXiv:1601.07904 [hep-th]}}.
%%CITATION = ARXIV:1601.07904;%%.

\bibitem{Nicolis:2008in}
A.~Nicolis, R.~Rattazzi, and E.~Trincherini, ``{The Galileon as a local
  modification of gravity},''
  \href{http://dx.doi.org/10.1103/PhysRevD.79.064036}{{\em Phys. Rev.} {\bf
  D79} (2009)  064036},
\href{http://arxiv.org/abs/0811.2197}{{\tt arXiv:0811.2197 [hep-th]}}.
%%CITATION = ARXIV:0811.2197;%%.

\bibitem{Creminelli:2014zxa}
P.~Creminelli, M.~Serone, G.~Trevisan, and E.~Trincherini, ``{Inequivalence of
  Coset Constructions for Spacetime Symmetries},''
  \href{http://dx.doi.org/10.1007/JHEP02(2015)037}{{\em JHEP} {\bf 02} (2015)
  037},
\href{http://arxiv.org/abs/1403.3095}{{\tt arXiv:1403.3095 [hep-th]}}.
%%CITATION = ARXIV:1403.3095;%%.

\bibitem{Sagnotti:2010at}
A.~Sagnotti and M.~Taronna, ``{String Lessons for Higher-Spin Interactions},''
  \href{http://dx.doi.org/10.1016/j.nuclphysb.2010.08.019}{{\em Nucl. Phys.}
  {\bf B842} (2011)  299--361},
\href{http://arxiv.org/abs/1006.5242}{{\tt arXiv:1006.5242 [hep-th]}}.
%%CITATION = ARXIV:1006.5242;%%.

\bibitem{Caron-Huot:2016icg}
S.~Caron-Huot, Z.~Komargodski, A.~Sever, and A.~Zhiboedov, ``{Strings from
  Massive Higher Spins: The Asymptotic Uniqueness of the Veneziano
  Amplitude},''
\href{http://arxiv.org/abs/1607.04253}{{\tt arXiv:1607.04253 [hep-th]}}.
%%CITATION = ARXIV:1607.04253;%%.

\end{thebibliography}\endgroup

\end{document}